\documentclass[a4paper,11pt]{article}
\pdfoutput=1 

\usepackage{jheppub} 
\makeatletter
\DeclareRobustCommand*{\bfseries}{%
  \not@math@alphabet\bfseries\mathbf
  \fontseries\bfdefault\selectfont
  \boldmath
}
\makeatother

\usepackage{calc}
\usepackage{rotating}
\usepackage[english]{babel}
\usepackage{hyperref}
\usepackage{graphicx}
\usepackage{subfig}
\usepackage{float}
\usepackage{amsmath,bm}
\usepackage{amssymb}
\usepackage{amsthm}
\usepackage{latexsym}
\usepackage{dcolumn}

\newcommand{\newc}{\newcommand*}

\long\def\begincomment#1\endcomment{%
        \begingroup\sf\baselineskip12pt#1\endgroup}

\newc{\etal}{\textrm{et al.}} 
\newc{\eg}{\textrm{e.g.}} 
\newc{\ie}{\textrm{i.e.}}
\newc{\etc}{\textrm{etc}}
\newc\vs{\textrm{vs.}}
\newc{\cl}{\rm {C.L.}}

\newc{\ev}{\ensuremath{\,\mathrm{eV}}}
\newc{\kev}{\ensuremath{\,\mathrm{keV}}}
\newc{\mev}{\ensuremath{\,\mathrm{MeV}}}
\newc{\gev}{\ensuremath{\,\mathrm{GeV}}}
\newc{\tev}{\ensuremath{\,\mathrm{TeV}}}
\newc{\MeV}{\mev} 
\newc{\TeV}{\tev}
\newc{\invpb}{\ensuremath{/\text{pb}}}
\newc{\invfb}{\ensuremath{/\text{fb}}}
\newc\nb{\ensuremath{\,\mathrm{nb}}} \newc\pb{\ensuremath{\,\mathrm{pb}}} \newc\fb{\ensuremath{\,\mathrm{fb}}}
\newc\pc{\ensuremath{\,\mathrm{pc}}}
\newc\kpc{\ensuremath{\,\mathrm{kpc}}}
\newc\mpc{\ensuremath{\,\mathrm{Mpc}}}
\newc\ps{\ensuremath{\,\mathrm{ps}}} 
\newc\cmeter{\ensuremath{\,\mathrm{cm}}} 
\newc\meter{\ensuremath{\,\mathrm{m}}} 
\newc\kmeter{\ensuremath{\,\mathrm{km}}}
\newc\second{\ensuremath{\,\mathrm{s}}}
\newc\msecond{\ensuremath{\,\mathrm{ms}}}
\newc\nsecond{\ensuremath{\,\mathrm{ns}}}
\newc\psecond{\ensuremath{\,\mathrm{ps}}}

\newc{\chisqmin}{\ensuremath{\chi^2_{\mathrm{min}}}}
\newc{\Delchisq}{\ensuremath{\Delta\chi^2}}
\newc{\chisq}{\ensuremath{\chi^2}}
\newc{\like}{\ensuremath{\mathcal{L}}}

\newc\lsim{\ensuremath{\mathrel{\rlap{\lower4pt\hbox{\hskip1pt$\sim$}}\raise1pt\hbox{$<$}}}}
\newc\gsim{\ensuremath{\mathrel{\rlap{\lower4pt\hbox{\hskip1pt$\sim$}}\raise1pt\hbox{$>$}}}}
\newc{\VEV}[1]{\ensuremath{\langle #1 \rangle}}
\newc{\dl}{\ensuremath{\stackrel{\leftarrow}{D}}}
\newc{\dr}{\ensuremath{\stackrel{\rightarrow}{D}}}

\newc{\bcenter}{\begin{center}}    \newc{\ecenter}{\end{center}}
\newc{\bfl}{\begin{flushleft}}    \newc{\efl}{\end{flushleft}}
\newc{\bfr}{\begin{flushright}}    \newc{\efr}{\end{flushright}}

\newc{\bi}{\begin{itemize}}
\newc{\ei}{\end{itemize}}
\newc{\bed}{\begin{description}}
\newc{\eed}{\end{description}}
\newc{\ben}{\begin{enumerate}}
\newc{\een}{\end{enumerate}}

\newc{\be}{\begin{equation}}
\newc{\ee}{\end{equation}}
\newc{\bea}{\begin{eqnarray}}
\newc{\eea}{\end{eqnarray}}
\newc{\bfle}{\begin{flalign}}
\newc{\efle}{\end{flalign}}
\newc{\ra}{\rightarrow}

\newc{\alphas}{\ensuremath{\alpha_s}}
\newc{\alphatwo}{\ensuremath{\alpha_2}}
\newc{\alphaone}{\ensuremath{\alpha_1}}
\newc{\alphai}[1]{\ensuremath{\alpha_{#1}}}
\newc{\alphaem}{\ensuremath{\alpha_{\mathrm{em}}}}
\newc{\alphaeff}{\ensuremath{\alpha_{\mathrm{eff}}}}
\newc{\sineff}{\ensuremath{\sin^2 \theta_{\mathrm{eff}}}}
\newc{\sinsqeff}{\ensuremath{\sin^2 \theta_{\mathrm{eff}}}}
\newc{\dalphahad}{\ensuremath{\Delta \alpha_{\mathrm{had}}}}
\newc{\yt}{\ensuremath{h_t}} \newc{\yb}{\ensuremath{h_b}} \newc{\ytau}{\ensuremath{h_{\tau}}}
\newc\mz{\ensuremath{m_Z}} 
\newc\mw{\ensuremath{m_W}}
\newc\mZ{\mz}        \newc\mW{\mw}
\newc\mhsm{\ensuremath{ m_{H_{\mathrm{SM}}}}}
\newc{\mtop}{\ensuremath{ m_t}}               \newc{\mtpole}{\ensuremath{ M_t}}
\newc{\mbottom}{\ensuremath{ m_b}} 
\newc{\mtau}{\ensuremath{ m_{\tau}}}
\newc{\mt}{\mtpole}
\newc{\mb}{\mbottom} 
\newc{\rtwogg}{\ensuremath{R_{h_2}(\gamma\gamma)}}
\newc{\rtwozz}{\ensuremath{R_{h_2}(ZZ)}}
\newc{\ronegg}{\ensuremath{R_{h_1}(\gamma\gamma)}}
\newc{\ronezz}{\ensuremath{R_{h_1}(ZZ)}}
\newc{\rsiggg}{\ensuremath{R_{h_\textrm{sig}}(\gamma\gamma)}}
\newc{\rsigzz}{\ensuremath{R_{h_\textrm{sig}}(ZZ)}}
\newc{\llbar}{\ensuremath{\ell\bar{\ell}}}
\newc{\tauptaum}{\ensuremath{ \tau^+\tau^-}}
\newc{\qqbar}{\ensuremath{ q\bar{q}}} \newc{\ppbar}{\ensuremath{ p\bar{p}}}
\newc{\bbbar}{\ensuremath{ b\bar{b}}} \newc{\ttbar}{\ensuremath{ t\bar{t}}}
\newc{\ffbar}{\ensuremath{ f\bar{f}}} \newc{\tautaubar}{\ensuremath{ \tau\bar{\tau}}}

\newc{\mchi}{\ensuremath{m_\neutone}}
\newc{\squark}{\ensuremath{\tilde{q}}}
\newc{\slepton}{\ensuremath{\tilde{l}}}
\newc{\gluino}{\ensuremath{\tilde{g}}} 
\newc{\mgluino}{\ensuremath{{m_{\gluino}}}}
\newc{\wino}{\ensuremath{\tilde{W}}} 
\newc{\mwino}{\ensuremath{{m_{\wino}}}}
\newc{\tone}{\ensuremath{{\tilde{t}_1}}}
\newc{\Hone}{\ensuremath{{\tilde{H}_{1}}}}
\newc{\Htwo}{\ensuremath{{\tilde{H}_{2}}}}
\newc{\Hhtwo}{\ensuremath{{H_{2}}}}
\newc{\qli}{\ensuremath{{\tilde{Q}_{i}}}}
\newc{\uri}{\ensuremath{{\tilde{u}_{i}}}}
\newc{\dri}{\ensuremath{{\tilde{d}_{i}}}}
\newc{\lli}{\ensuremath{{\tilde{L}_{i}}}}
\newc{\eri}{\ensuremath{{\tilde{e}_{i}}}}

\newc{\sthw}{\ensuremath{ \sin\theta_W}}              \newc{\cthw}{\ensuremath{\cos\theta_W}}
\newc{\tanthw}{\ensuremath{ \tan\theta_W}}              \newc{\cotthw}{\ensuremath{\cot\theta_W}}
\newc{\ssqthw}{\ensuremath{\sin^2 \theta_W}}
\newc{\msbar}{\ensuremath{\overline{MS}}} \newc{\drbar}{\ensuremath{\overline{DR}}}
\newc{\mtmtsmmsbar}{\ensuremath{ m_t(m_t)^{\msbar}_{{\mathrm{SM}}}}}
\newc{\mtmtsmdrbar}{\ensuremath{ m_t(m_t)^{\drbar}_{{\mathrm{SM}}}}}
\newc{\mtmtmssmdrbar}{\ensuremath{ m_t(m_t)^{\drbar}_{{\mathrm{SUSY}}}}}
\newc{\mbmbmsbar}{\ensuremath{ m_b(m_b)^{\msbar} }}
\newc{\mbmbsmmsbar}{\ensuremath{ m_b(m_b)^{\msbar}_{{\mathrm{SM}}}}}
\newc{\mbmzsmmsbar}{\ensuremath{ m_b(\mz)^{\msbar}_{{\mathrm{SM}}}}}
\newc{\mbmzsmdrbar}{\ensuremath{ m_b(\mz)^{\drbar}_{{\mathrm{SM}}}}}
\newc{\mbmzmssmdrbar}{\ensuremath{ m_b(\mz)^{\drbar}_{{\mathrm{SUSY}}}}}
\newc{\mtaumzsmmsbar}{\ensuremath{ m_{\tau}(\mz)^{\msbar}_{{\mathrm{SM}}}}}
\newc{\mtaumzsmdrbar}{\ensuremath{ m_{\tau}(\mz)^{\drbar}_{{\mathrm{SM}}}}}
\newc{\mtaumzmssmdrbar}{\ensuremath{ m_{\tau}(\mz)^{\drbar}_{{\mathrm{SUSY}}}}}
\newc{\alphasmzms}{\ensuremath{\alpha_s(M_Z)^{\overline{MS}}}}
\newc{\alphaimzms}[1]{\ensuremath{\alpha_{#1}(M_Z)^{\overline{MS}}}}
\newc{\alphaemmz}{\ensuremath{\alpha_{\mathrm{em}}(M_Z)^{\overline{MS}}}}

\newc{\mzero}{\ensuremath{{m_0}}}
\newc{\mhalf}{\ensuremath{ m_{1/2}}}
\newc{\tanb}{\ensuremath{\tan\beta}}
\newc{\azero}{\ensuremath{ A_0}}
\newc{\signmu}{\ensuremath{\rm{sgn}\,\mu}}
\newc{\atau}{\ensuremath{{A_{\tau}}}}
\newc{\mueff}{\ensuremath{\mu_{\rm{eff}}}}
\newc{\lam}{\ensuremath{{\lambda}}}
\newc{\kap}{\ensuremath{{\kappa}}}
\newc{\alam}{\ensuremath{{A_{\lambda}}}}
\newc{\akap}{\ensuremath{{A_{\kappa}}}}
\newc{\hs}{\ensuremath{ H_s}}      
\newc{\mhs}{\ensuremath{ m_{H_s}}} 
\newc{\mgut}{\ensuremath{ M_{\rm GUT}}}
\newc{\gut}{\ensuremath{{\rm GUT}}}
\newc{\mplanck}{\ensuremath{ M_{\rm P}}}      \newc{\mpl}{\ensuremath{ M_{\rm Pl}}}
\newc{\msusy}{\ensuremath{ M_{\rm SUSY}}}      \newc{\ms}{\ensuremath{ M_{\rm S}}}
\newc{\mew}{\ensuremath{ M_{\rm EW}}}  
 \newc{\hu}{\ensuremath{ H_u}}       \newc{\hd}{\ensuremath{ H_d}}
 \newc{\mhu}{\ensuremath{ m_{H_u}}}       \newc{\mhd}{\ensuremath{ m_{H_d}}}
 \newc{\mhuew}{\ensuremath{ m^{\ast}_{H_u}}}       \newc{\mhdew}{\ensuremath{ m^{\ast}_{H_d}}}
 \newc{\mhuewsq}{\ensuremath{ m^{\ast\, 2}_{H_u}}}       \newc{\mhdewsq}{\ensuremath{ m^{\ast\, 2}_{H_d}}}
 \newc{\mhl}{\ensuremath{m_\hl}} 
 \newc{\mhone}{\ensuremath{m_{h_1}}} 
 \newc{\mhtwo}{\ensuremath{m_{h_2}}} 
 \newc{\mhi}{\ensuremath{m_{\tilde{h}}}} 
 \newc{\mul}{\ensuremath{m_{\tilde{u}_L}}} 
 \newc{\mtone}{\ensuremath{m_{\tilde{t}_1}}} 
 \newc{\ma}{\ensuremath{m_A}} 
 \newc{\mH}{\ensuremath{m_H}} 
 \newc{\maone}{\ensuremath{m_{a_1}}} 
 \newc{\matwo}{\ensuremath{m_{a_2}}}
 \newc{\hone}{\ensuremath{h_1}}
 \newc{\htwo}{\ensuremath{h_2}}
 \newc{\aone}{\ensuremath{a_1}}
 \newc{\atwo}{\ensuremath{a_2}}
 \newc{\mqthree}{\ensuremath{m_{\tilde{Q}_3}^2}}
 \newc{\muthree}{\ensuremath{m_{\tilde{u}_3}^2}}
 \newc{\mql}{\ensuremath{m_{\tilde{Q}}}}
 \newc{\mqlij}{\ensuremath{(m_{\tilde{Q}}^2)_{ij}}}
 \newc{\mur}{\ensuremath{m_{\tilde{u}}}}
 \newc{\mdr}{\ensuremath{m_{\tilde{d}}}}
 \newc{\murij}{\ensuremath{(m_{\tilde{u}}^2)_{ij}}}
 \newc{\md}{\ensuremath{m_{\tilde{D}}}}
 \newc{\muu}{\ensuremath{m_{\tilde{U}}}}
 \newc{\mdrij}{\ensuremath{(m_{\tilde{d}}^2)_{ij}}}
 \newc{\mll}{\ensuremath{m_{\tilde{L}}}}
 \newc{\mllij}{\ensuremath{m_{\tilde{L}_{ij}}}}
 \newc{\mer}{\ensuremath{m_{\tilde{e}}}}
 \newc{\merij}{\ensuremath{m_{\tilde{e}_{ij}}}}
 \newc{\ts}{\ensuremath{T_{SUSY}}}

\newc{\sigsip}{\ensuremath{\sigma^{\rm SI}_{p}}}	\newc{\sigsin}{\ensuremath{\sigma^{\rm SI}_{n}}}
\newc{\sigsdp}{\ensuremath{\sigma^{\rm SD}_{p}}}	\newc{\sigsdn}{\ensuremath{\sigma^{\rm SD}_{n}}}
\newc{\sigsi}{\ensuremath{\sigma^{\rm SI}}}	\newc{\sigsd}{\ensuremath{\sigma^{\rm SD}}}
\newc{\abund}{\ensuremath{ \Omega h^2}}
\newc{\omegadm}{\ensuremath{ \Omega_{{\rm DM}}}}     \newc{\abunddm}{\ensuremath{ \Omega_{{\rm DM}} h^2}} 
\newc{\omegam}{\ensuremath{ \Omega_{{\rm m}}}}       \newc{\abundm}{\ensuremath{ \Omega_{{\rm m}} h^2}}
\newc{\omegab}{\ensuremath{ \Omega_{{\rm b}}}}	\newc{\abundb}{\ensuremath{ \Omega_{{\rm b}} h^2}}
\newc{\omegatot}{\ensuremath{ \Omega_{{\rm TOT}}}}
\newc{\omegacdm}{\ensuremath{ \Omega_{{\rm CDM}}}}   \newc{\abundcdm}{\ensuremath{ \Omega_{{\rm CDM}} h^2}}
\newc{\omegalambda}{\ensuremath{ \Omega_{\Lambda}}} \newc{\abundlambda}{\ensuremath{ \Omega_{\Lambda} h^2}}
\newc{\omegarad}{\ensuremath{ \Omega_{{\rm rad}}}}  \newc{\abundrad}{\ensuremath{ \Omega_{{\rm rad}} h^2}}
\newc{\rhocrit}{\ensuremath{ \rho_{\rm crit}}}
\newc{\rhochi}{\ensuremath{ \rho_{\chi}}}
\newc{\abunchi}{\ensuremath{\Omega_\chi h^2}}
\newc{\abundlsp}{\ensuremath{\Omega_{\rm LSP}h^2}}
\newcommand*{\abundchi}{\ensuremath{\Omega_\chi h^2}}

\newc{\amu}{\ensuremath{ a_{\mu}}}        \newc{\amususy}{\ensuremath{ a_{\mu}^{\mathrm{SUSY}}}}
\newc{\amuexpt}{\ensuremath{ a_{\mu}^{\mathrm{expt}}}}        \newc{\amusm}{\ensuremath{ a_{\mu}^{\mathrm{SM}}}}
\newc\deltaamu{\ensuremath{\Delta a_{\mu}}} \newc{\deltaamususy}{\ensuremath{\delta a_{\mu}^{\mathrm{SUSY}}}}
\newc\gmtwo{\ensuremath{ (g-2)_{\mu}}} 
\newc{\deltagmtwomususy}{\ensuremath{\delta\left(g-2\right)_{\mu}^{\mathrm{SUSY}}}}
\newc{\deltagmtwomu}{\ensuremath{\delta\left(g-2\right)_{\mu}}}
\newc\BR{\ensuremath{\rm BR}}
\newc\bsgamma{\ensuremath{ b\rightarrow s \gamma }}
\newc\bxsgamma{\ensuremath{\overline{B}\rightarrow X_{s}\gamma}}
\newc\brbsgamma{\ensuremath{\BR\left(\bsgamma\right)}}
\newc\brbxsgamma{\ensuremath{\BR\left(\bxsgamma\right)}}
\newc\bsmumu{\ensuremath{B_s\to\mu^+\mu^-}}
\newc\bdmumu{\ensuremath{B_d\to\mu^+\mu^-}}
\newc\brbsmumu{\ensuremath{\BR\left(B_s\to\mu^+\mu^-\right)}}
\newc\brbdmumu{\ensuremath{\BR\left(B_d\to\mu^+\mu^-\right)}}
\newc\bdmmumu{\ensuremath{\overline{B}_d\to\mu^+\mu^-}}
\newc\bbbarmix{\ensuremath{\overline{B}_s\mbox{-}B_s}}      
\newc\delmbs{\ensuremath{\Delta M_{B_s}}}
\newc\thc{\ensuremath{t\to h c}}
\newc\thu{\ensuremath{t\to h u}}
\newc{\butaunu}{\ensuremath{B_u \rightarrow \tau \nu}}
\newc{\brbutaunu}{\ensuremath{\BR\left(B_u \rightarrow \tau \nu\right)}}

\newcommand*{\reffig}[1]{Fig.~\ref{#1}}
 
        \newcommand*{\refeq}[1]{Eq.~(\ref{#1})}

\newcommand*{\neutone}{\ensuremath{\tilde{\chi}^0_1}}
\newcommand*{\neuttwo}{\ensuremath{\tilde{{\chi}}^0_2}}

\newcommand*{\charone}{\ensuremath{\tilde{{\chi}}^{\pm}_1}}

\newcommand*{\stau}{\ensuremath{\tilde{\tau}}}

\newcommand*{\eight}{\ensuremath{\sqrt{s}=8\tev}}

\newcommand*{\dsusy}{DarkSUSY}
\newcommand*{\higgsbounds}{H{\scriptsize IGGS}B{\scriptsize OUNDS}}
\newcommand*{\higgssignals}{H{\scriptsize IGGS}S{\scriptsize IGNALS}}
\newcommand*{\feynhiggs}{\texttt{FeynHiggs}}
\newcommand*{\spheno}{\texttt{SPheno}}

\newcommand*{\multinest}{MultiNest}

\newcommand*{\susyflav}{\text{SUSY\textunderscore FLAVOR}}

\newcommand*{\stauc}{\text{\stau-coannihilation}}

\let\oldcite\cite
\renewcommand*{\cite}{~\oldcite}

\newcommand*{\hl}{\ensuremath{h}}

\newcommand*{\kk}{\ensuremath{\bar{K^0}-K^0}}

\newcommand*{\bdbd}{\ensuremath{\bar{B_d^0}-B_d^0}}
\newcommand*{\bsbs}{\ensuremath{\bar{B_s^0}-B_s^0}}
\newcommand*{\dullab}{\ensuremath{(\delta_{12}^u)_{LL}}}
\newcommand*{\dullac}{\ensuremath{(\delta_{13}^u)_{LL}}}
\newcommand*{\dullbc}{\ensuremath{(\delta_{23}^u)_{LL}}}

\newcommand*{\durrab}{\ensuremath{(\delta_{12}^u)_{RR}}}
\newcommand*{\durrac}{\ensuremath{(\delta_{13}^u)_{RR}}}
\newcommand*{\durrbc}{\ensuremath{(\delta_{23}^u)_{RR}}}

\newcommand*{\dulrab}{\ensuremath{(\delta_{12}^u)_{LR}}}
\newcommand*{\dulrac}{\ensuremath{(\delta_{13}^u)_{LR}}}
\newcommand*{\dulrbc}{\ensuremath{(\delta_{23}^u)_{LR}}}
\newcommand*{\durlab}{\ensuremath{(\delta_{21}^u)_{LR}}}
\newcommand*{\durlac}{\ensuremath{(\delta_{31}^u)_{LR}}}
\newcommand*{\durlbc}{\ensuremath{(\delta_{32}^u)_{LR}}}

\restylefloat{figure}

\title{Phenomenology of SUSY with General Flavour Violation} 

\author{Kamila Kowalska} 

\affiliation{National Centre for Nuclear Research,
  Ho{\. z}a 69, 00-681 Warsaw, Poland} 

\emailAdd{Kamila.Kowalska@fuw.edu.pl}

\abstract{We discuss the consequences of relaxing the Minimal Flavour Violation assumption in the up-squark sector on the phenomenology of SUSY models. We study the impact of the off-diagonal entries in the soft SUSY-breaking matrices on the mass of the lightest Higgs scalar and we derive the approximate analytical formulae that quantify this effect. We show that \mhl\ can be enhanced by up to $13-14\gev$ in the case of the phenomenological MSSM with the inverted hierarchy of masses in the squark sector and zero stop mixing, and up to $4-5\gev$ in GUT-constrained scenarios where the magnitude of the enhancement is mitigated by renormalization group effects. We also perform a global analysis of an inverted hierarchy GFV scenario, taking into account the experimental bounds from the measurements of relic density, EW precision observables and $B$-physics. We show that the allowed parameter space of the model is strongly constrained by \mw, \sineff\ and \brbsmumu, requiring $\mzero(3)<1500\gev$ and $\mhalf<1800\gev$, as well as a large non-zero (2,3) entry in the up-squark trilinear matrix. }

\begin{document}
\maketitle
\flushbottom

\section{Introduction}\label{sec:intro}

The discovery of the Higgs boson\cite{Chatrchyan:2012ufa,Aad:2012tfa} was an unquestionable and historical success of the LHC 8\tev\ run. With a mass of around 126\gev\cite{CMS-PAS-HIG-13-005}, the new scalar is consistent with the predictions of both the Standard Model (SM) and its minimal supersymmetric extension. While in the former case the Higgs mass remains a free parameter of the theory, in the latter it is totally determined by the gauge and soft supersymmetry-breaking (SSB) sectors. In the Minimal Supersymmetric Standard Model (MSSM) the enhancement from the tree-level value, necessary to obtain \mhl\ in agreement with the experimental data, can be achieved through radiative corrections to the scalar potential. Those, to be large enough, require either relatively heavy stops or almost maximal mixing in the stop sector\cite{Haber:1996fp}. 

The phenomenology of the SUSY landscape after the Higgs discovery has been widely studied in the literature, both in the context of GUT-constrained scenarios, as well as of models described by a set of supersymmetric parameters defined at the electro-weak symmetry breaking (EWSB) scale (for a non-comprehensive list of articles see for example\cite{Fowlie:2012im,Akula:2012kk,Beskidt:2012sk,Buchmueller:2012hv,Altmannshofer:2012ks,Strege:2012bt,Cabrera:2012vu,Kowalska:2013hha,Badziak:2012rf,Arbey:2012dq}).
The vast majority of those analyses were performed under the assumption of Minimal Flavour Violation (MFV). In this framework the only source of flavour mixing in the sfermion sector are the CKM and PMNS matrices, so the supersymmetric contributions the the Flavour Changing Neutral Currents (FCNC) transitions are suppressed via the super-GIM mechanism, in analogy to what happens in the Standard Model. The MFV assumption can be realised in two different ways: a) the trilinear terms and soft masses are aligned with the corresponding Yukawa matrices, thus becoming nearly diagonal after rotation to the SCKM-basis; b) diagonal entries of soft mass matrices are degenerate so that the off-diagonal elements in the SCKM-basis are automatically zero. Moreover, if sfermions are heavy enough, SUSY loop contributions to the FCNC will be naturally suppressed regardless the SSB structure. The last assumption, combined with the non-observation of SUSY particles at the LHC and the fact that the third generation of colour sfermions should not be too heavy in order to keep the fine tuning of the model reasonably low\cite{Ellis:1986yg,Barbieri:1987fn}, seems to somehow favour the pattern of so-called inverted hierarchy (IH)\cite{Cohen:1996vb}, where the first two generations of sfermions are significantly heavier than the third one.

However, in the most general case, the soft masses and trilinear terms do not need to be constrained in any way and can be treated as additional free parameters of the model. This General Flavour Violation (GVF) is a source of what is called the FCNC supersymmetric problem, as the unrestricted off-diagonal entries of SSB matrices can lead to disastrously large SUSY contributions to FCNC processes. Note that even in a relatively simple case when the soft masses and trilinears are diagonal in the interaction basis though not proportional to the unity matrix, after the rotation of the fermion fields to the physical basis the non-diagonal SSB terms  arise, proportional to the mass splitting between the diagonal entries. This effect is particularly important in the case of the soft mass \mqlij, where the mass splitting impact can be enhanced by the corresponding CKM matrix elements.

The FCNC processes are not the only area where a possible discrepancy between the GFV structure of the soft-SUSY breaking sector and the experimental data can manifest. Firstly, it was observed in Ref.\cite{Heinemeyer:2004by} that the one-loop flavour violating corrections to the mass of the $W$ boson and the value of \sineff\ can be very large and therefore provide severe limits on the allowed parameter space of a supersymmetric model. Secondly, the authors of Ref.\cite{Herrmann:2011xe} showed that the annihilation cross-section of the neutralino can be enhanced if the squark mass splitting is increased by GFV effects, and that new annihilation channels can open due to the presence of a flavour-mixing coupling between squarks and neutralinos. Finally, also the Higgs sector can be affected by the presence of GFV soft SUSY-breaking terms. 
It was shown in Refs.\cite{Heinemeyer:2004by,AranaCatania:2011ak} (recently updated in\cite{Arana-Catania:2014ooa}) that the corresponding radiative corrections to the lightest CP-even scalar mass due to flavour mixings between the second and third squark generation can be either moderate and positive (up to $2-4\gev$) or large and negative, leading to a reduction of \mhl\ well below the LEP limit. The latter effect was also observed in\cite{Cao:2006xb}.

In this study we pursue the question as to what extent the Higgs boson mass can be actually enhanced by the non-diagonal SSB terms without introducing heavy stops or maximal stop mixing. 
We derive approximate formulae that allow to quantify this effect analytically. We then show that the strongest Higgs mass enhancement can be achieved through non-zero (2,3), (3,2), (1,3) and (3,1) entries of the up-squark trilinear coupling for a particular choice of SSB mass matrices, namely 	
in an inverted hierarchy scenario where the first two generations of sfermions are significantly heavier than the third one. The size of the GFV contribution can reach $13-14\gev$ in the case of the phenomenological MSSM, and up to $3-5\gev$ for the hierarchical model defined at the GUT scale. We also show that the enhancement in  \mhl\ can be significantly reduced by introducing non-zero stop mixing.

We also analyse the consistency of a GUT-constrained inverted hierarchy scenario with other experimental data, in particular the measurement of relic density by PLANCK\cite{Ade:2013zuv}, constraints from B-physics and EW precision observables\cite{Beringer:1900zz}. 
We emphasise that the latter set of constraints is of key importance as GFV effects due to mass splitting between the first/second and third generation give rise to large one-loop corrections to \mw\ and \sineff. As a consequence, a part of the parameter space corresponding to the universal mass of the third generation squarks heavier than $1500\gev$ is strongly disfavoured. 
Finally, we point out a tension between the measurement of \brbsmumu\ and the GFV assumption, which can be significantly reduced if a large non-zero (2,3) entry in the up-squark trilinear matrix is present.

This paper is organised as follows. In Sec.\ref{sec:higgs} we discuss the enhancement of the Higgs boson mass due to GFV effects and we derive approximate formulae to quantify their impact. In Sec.\ref{sec:fcnc} we calculate the limits from the FCNC processes, renormalization of the CKM matrix, and vacuum stability on the relevant GFV parameters. In Sec.\ref{sec:guthiggs} we analyse the corresponding effects in the models defined at the \gut\ scale and show that the Higgs mass enhancement is reduced due to the effects of RGEs. In Sec.\ref{sec:scans} we present the results of a global analysis that combines GFV effects in the Higgs sector with other experimental data. We summarise our findings in Sec.\ref{sec:concl}.

\section{The Higgs boson mass in the GFV MSSM}\label{sec:higgs}

We will start this section with a brief review of the notation we will be using throughout the paper.
In the interaction basis, the R-parity conserving superpotential of the MSSM is given by
\be
W = \epsilon_{ab}[H_2^bQ_i^{a}(Y_u)_{ij}\bar{U}_j+H_1^bQ_i^{a}(Y_d)_{ij}\bar{D}_j+H_1^bL_i^{a}(Y_e)_{ij}\bar{E}_j-\mu H_1^aH_2^b],
\ee
where $a,b=1,2$ indicate $SU(2)$ indices and $i,j=1,2,3$ are generation indices. 
The soft SUSY-breaking part of the lagrangian can be written as
\bea
\like_{\textrm{soft}}&=&\tilde{Q}^\dagger_i\mqlij\tilde{Q}_j+\tilde{U}^\dagger_i\murij\tilde{U}_j+\tilde{D}^\dagger_i\mdrij\tilde{D}_j+\tilde{L}^\dagger_i\mllij\tilde{L}_j+\tilde{E}^\dagger_i\merij\tilde{E}_j+,\nonumber\\
&+&\mhd^2H_1^*H_2+\mhu^2H_2^*H_2+(m_3^2\epsilon_{ab}H_1^aH_2^b+h.c.)\nonumber\\
&+&\epsilon_{ab}((T_u)_{ij}H_2^b\tilde{Q}_i^{a}\tilde{U}^*_j+(T_d)_{ij}H_1^b\tilde{Q}_i^{a}\tilde{D}^*_j+(T_e)_{ij}H_1^b\tilde{L}_i^{a}\tilde{E}^*_j)+\textrm{h.c.}.
\eea
The rotation of the quark fields into the basis in which they are diagonal - the so called super-CKM basis - means that also the squarks need to be rotated accordingly (since in this paper we are only interested in the effects coming from the inter-generation mixing in the quark/squark sector, we will treat neutrinos as massless, henceforth assuming that there is no tree-level mixing between the leptons). Defining the quark rotation matrices $V$ as
\be
Q_{\textrm{sCKM}}=\left( \begin{array}{c} V^u_L U \\ V^d_L D\end{array}\right),\qquad \bar{U}_{\textrm{sCKM}}=V^u_R\bar{U},\qquad \bar{D}_{\textrm{sCKM}}=V^d_R\bar{D}
\ee
and demanding them to diagonalise the Yukawa matrices
\be
Y_u^{\textrm{diag}}=V^u_R Y_u^T V^{u\dagger}_L,\qquad Y_d^{\textrm{diag}}=V^d_R Y_d^T V^{d\dagger}_L,
\ee
the soft SUSY-breaking mass matrices and the trilinear terms in the super-CKM basis take the form:
\bea\label{sckm_mass}
&(\mql^2)_{LL}=V_L^d\mql^2 V_L^{d\dagger},\quad (\muu^2)_{RR}=V_R^u\mur^2 V_R^{u\dagger},\quad (\md^2)_{RR}=V_R^d\mdr^2 V_R^{d\dagger},\nonumber\\
&(\muu^2)_{LR}=\frac{v_2}{\sqrt{2}}V^u_R (T_u)V_L^{u\dagger}, \quad (\md^2)_{LR}=\frac{v_1}{\sqrt{2}}V^d_R (T_d)V_L^{d\dagger}.
\eea
The $6\times 6$ mass matrices for the up and down squarks are then constructed as (we follow here the notation of the SLHA2\cite{Allanach:2008qq}):
\bea\label{supmatrix}
\mathcal{M}^2_{\tilde{u}}&=&\left( \begin{array}{cc} K(\mql^2)_{LL}K^{\dagger}+(m_u^{\textrm{diag}})^2-\frac{\cos2\beta}{6}(\mz^2-4\mw^2)\mathbb{I} & (\muu^2)_{LR}^{\dagger}-m_u^{\textrm{diag}}\mu^*\cot\beta \\ (\muu^2)_{LR}-m_u^{\textrm{diag}}\mu\cot\beta & (\muu^2)_{RR}+(m_u^{\textrm{diag}})^2+\frac{2\cos2\beta}{3}\mz^2\ssqthw\mathbb{I}   \end{array}\right),\nonumber\\
\mathcal{M}^2_{\tilde{d}}&=&\left( \begin{array}{cc} (\mql^2)_{LL}+(m_d^{\textrm{diag}})^2-\frac{\cos2\beta}{6}(\mz^2+2\mw^2)\mathbb{I} & (\md^2)_{LR}^{\dagger}-m_d^{\textrm{diag}}\mu^*\tan\beta\\ (\md^2)_{LR}-m_d^{\textrm{diag}}\mu\tan\beta &(\md^2)_{RR}+(m_d^{\textrm{diag}})^2-\frac{\cos2\beta}{3}\mz^2\ssqthw\mathbb{I} \end{array}.\right)
\eea
In the above $m_u^{\textrm{diag}}=\frac{v_2}{\sqrt{2}}Y_u^{\textrm{diag}}$ and $m_d^{\textrm{diag}}=\frac{v_1}{\sqrt{2}}Y_d^{\textrm{diag}}$ are the diagonal matrices of quark masses, $\theta_W$ is the Weinberg angle, $\tan\beta$ is the ratio of the Higgs doublets' vacuum expectation values (VEV), $\tan\beta\equiv\frac{v_2}{v_1}$, and $\mathbb{I}$ denotes the unity matrix in the generation space. The CKM matrix is defined as $K=V_L^{u\dagger}V^d_L$.
Note that the left-handed blocks in $\mathcal{M}^2_{\tilde{u}}$ and $\mathcal{M}^2_{\tilde{d}}$ can not be simultaneously diagonal in the SCKM-basis if any mass splitting in $(m_Q^2)_{LL}$ is present. That is an important issue and we will come back to it in Sec.\ref{sec:scans}.

It is convenient to parametrise the non-diagonal entries of the squark mass matrices given in \refeq{sckm_mass} in terms of dimensionless parameters $(\delta_{ij})_{AB}$, normalised to the geometrical average of the diagonal elements:
\be\label{deltas_def}
(\delta^{u}_{ij})_{AB}=\frac{(\muu^2)^{ij}_{AB}}{\sqrt{(\muu^2)^{ii}_{BB}(\muu^2)^{jj}_{AA}}}, \qquad (\delta^{d}_{ij})_{AB}=\frac{(\md^2)^{ij}_{AB}}{\sqrt{(\md^2)^{ii}_{BB}(\md^2)^{jj}_{AA}}},
\ee
where for consistency we defined $(\muu^2)_{LL}=K(\mql^2)_{LL}K^{\dagger}$ and $(\md^2)_{LL}=(\mql^2)_{LL}$.

\bigskip
We can now proceed to discuss the dependence of the lightest Higgs boson mass on the size of parameters $(\delta_{ij})_{AB}$. The full one-loop corrections to the masses of the Higgs scalars in the GFV framework have been calculated by several groups in the diagrammatic approach and implemented in the publicly available numerical codes, \feynhiggs\cite{feynhiggs:99,feynhiggs:00,feynhiggs:03,feynhiggs:06} and \spheno\cite{Porod:2003um,Porod:2011nf}. In the numerical analysis throughout the paper we will use \spheno\_v.3.2.4.  

To have a grasp of possible effects that can arise after taking into account the non-zero values of parameters $(\delta_{ij})_{AB}$, we will start with deriving  approximate analytical formulae for the GFV corrections to the Higgs boson mass. In what follows we adopt the procedure proposed in Section 6 of Ref.\cite{Haber:1993an}, based on the effective potential technique\cite{Ellis:1990nz,Ellis:1991zd}, as well as the subsequent results obtained in Ref.\cite{Haber:1996fp} where the effects of the mixing and non-degeneracy of the soft masses in the stop sector have been thoroughly analysed.

In order to derive relatively simple expressions that would parametrise the GFV effects, in the following we will limit ourselves to analysing the contribution from the up-squark sector only, which is known to be the dominating one\footnote{In principle the contributions from the bottom and tau sectors can become significant for large \tanb. However, by choosing $A_{b,\tau}=0$ and small $\mu$ their impact can be strongly reduced.}. 
Let us further assume that the diagonal elements of the up squark mass matrix $\mathcal{M}^2_{\tilde{u}}$ are degenerate and equal to the common mass $\tilde{m}^2$. We also neglect the terms of the order of $\mz^2$ and all the quarks masses but \mtop. The matrix (\ref{supmatrix}) takes then the form:
\be\label{mu2approx}
\mathcal{M}^2_{\tilde{u}}=\tilde{m}^2\left( \begin{array}{ccc|ccc} 1 & \dullab &\dullac & 0 & \durlab &\durlac \\ \dullab &1 & \dullbc &\dulrab  & 0 &\durlbc\\ \dullac & \dullbc& 1+\frac{\mtop^2}{\tilde{m}^2} & \dulrac &\dulrbc & \frac{\langle v_2\rangle\tilde{X}_t}{\sqrt{2}\tilde{m}^2} \\ \hline 0 & \dulrab &\dulrac & 1 & \durrab &\durrac \\ \durlab& 0  &\dulrbc & \durrab & 1 & \durrbc \\ \durlac & \durlbc &\frac{\langle v_2\rangle\tilde{X}_t}{\sqrt{2}\tilde{m}^2}& \durrac &\durrbc & 1+\frac{\mtop^2}{\tilde{m}^2} \end{array}\right),
\ee
where we defined $\tilde{X}_t=(T_u)_{33}-Y_t\mu\cot\beta$ in analogy the the common mixing parameter $X_t$. Note that the MFV case corresponds to $(\delta_{ij})_{AB}=0$.

We will now discuss the GFV corrections to the Higgs boson mass, assuming various structures of the mass matrix (\ref{mu2approx}). Here we present only the final results. The details of calculation can be found in Appendix \ref{sec:appen}.

\paragraph{ Case 1:} No GFV and $\tilde{X}_t\ne0$.

We will start with the well known scenario of non-zero mixing in the stop sector, which we will use as a reference case while studying the GFV effects. The masses of physical stops are given by $\tilde{m}_{t_1}^2=\tilde{m}^2+m_t^2-\tilde{X}_t\frac{v_2}{\sqrt{2}}$, $\tilde{m}_{t_2}^2=\tilde{m}^2+m_t^2+\tilde{X}_t\frac{v_2}{\sqrt{2}}$. The one-loop correction to \mhl\ reads:

\be
\Delta\mhl^2=\frac{3}{8\pi^2 v^2}Y_t^4v_2^4\ln\frac{\tilde{m}^2}{\mtop^2}+\frac{3v^4_2}{8\pi^2v^2}\left[\frac{\tilde{X}_t^2}{\tilde{m}^2}\left(Y_t^2-\frac{\tilde{X}_t^2}{12\tilde{m}^2}\right)\right].
\ee
If the trilinear term $T^u_{33}$ is proportional to the top Yukawa coupling, $T_{33}^u=Y_tA_{33}^u$, the above expression takes the usual form given in\cite{Haber:1996fp}. 

\paragraph{ Case 2:} $\dulrbc\ne0$ or $\dulrac\ne0$ .

The presence of non-zero off-diagonal term \dulrbc\ in the mass matrix  $(\muu^2)_{LR}$ induces chirality flipping mixing between stop $\tilde{t}_L$ and scharm $\tilde{c}_R$, leading to the following mass eigenstates: $m^2_{\tilde{t}_R}=\mtop^2+\tilde{m}^2$, $m^2_{\tilde{c}_L}=\tilde{m}^2$, $m^2_{1}=\frac{1}{2}\mtop^2+\tilde{m}^2-\frac{1}{2}\sqrt{\mtop^4+4\tilde{m}^4\dulrbc^2}$, $m^2_{2}=\frac{1}{2}\mtop^2+\tilde{m}^2+\frac{1}{2}\sqrt{\mtop^4+4\tilde{m}^4\dulrbc^2}$, where we assumed that $\tilde{X}_t=0$. Exactely the same effect arises in the case of $\tilde{u}_R$ - $\tilde{t}_L$ mixing, with \dulrbc\ replaced by \dulrac. 

If $\dulrbc>\frac{\mtop^2}{2\tilde{m}^2}$ (which corresponds to $\dulrbc>0.06$ for $\tilde{m}=0.5\tev$) the masses of the mixed eigenstates reduce to $m^2_{\tilde{t}_1}=\tilde{m}^2+\frac{1}{2}\mtop^2+\tilde{m}^2\dulrbc$, $m^2_{\tilde{c}_1}=\tilde{m}^2+\frac{1}{2}\mtop^2-\tilde{m}^2\dulrbc$. Since \dulrbc\ depends linearly on the VEV $v_2$ (see \refeq{sckm_mass}), its correction  to the Higgs boson mass is similar to the one generated by the stop mixing term $\tilde{X}_t$,
\be\label{case2}
\Delta\mhl^2(\dulrbc)=\frac{3}{4\pi^2}\left[\tilde{m}^2\dulrbc^2\left(\frac{1}{2}Y_t^2\sin^2\beta-\frac{\tilde{m}^2\dulrbc^2}{6v^2}\right)\right].
\ee
It is straightforward to check that the maximal effect is expected when $\dulrbc\simeq \sqrt{3/2}vY_t\sin\beta/\tilde{m}$.

\paragraph{ Case 3:} $\dulrab\ne0$

The presence of this term induces the mixing between $\tilde{u}_R$ and $\tilde{c}_L$, while the stops remain degenerate. The masses of the mixed eigenstates are given by $m^2_{1}=\tilde{m}^2+\tilde{m}^2\dulrab$ and $m^2_{2}=\tilde{m}^2-\tilde{m}^2\dulrab$. The terms proportional to $v_2^2$ are suppressed by the corresponding Yukawa couplings and have been neglected. On the other hand, since at this point we do not make any assumption about the size of the non-diagonal term \dulrab, its contribution to the Higgs boson mass should be taken into account and gives:
\bea\label{case3}
\Delta\mhl^2(\dulrab)=-\frac{3}{24\pi^2v^2}\tilde{m}^4\dulrab^4.
\eea
Notice that, unlike the case of $\dulrac$ or $\dulrbc$ driven mixing, this contribution is always negative. 

\begin{table}[t]
\begin{center}
\begin{tabular}{|c|c|c|c|c|c|c|c|c|c|c|c|}
\hline 
    & $M_1$ & $M_2$ & $M_3$ & $A_{t,b,\tau}$ & $\mu$ & \ma & \tanb & $m_{\tilde{L},\tilde{e}}$ & $m_{\tilde{Q}_{1,2},\tilde{u}_{1,2},\tilde{d}}$ & $m_{\tilde{Q}_3,\tilde{u}_3}$\\
\hline 
BP1 &  126 & 233 & 670 & 0 & 242 & 1860 & 20 & 5950 & 560 & 530\\
BP2 &  126 & 233 & 670 & 0 & 362 & 1712 & 30 & 1455 & 945 & 850 \\
BP3 &  126 & 233 & 670 & 0 & 462 & 1512 & 30 & 2500 & 1300 & 1200 \\
\hline 
\end{tabular}
\caption{\footnotesize Input parameters for benchmark points BP1 - BP3 at \msusy. The masses and trilinear terms are in \gev.}
\label{tab:BP_23}
\end{center}
\end{table}

\begin{figure}[t]
\centering
\subfloat[]{
\label{fig:a}
\includegraphics[width=0.45\textwidth]{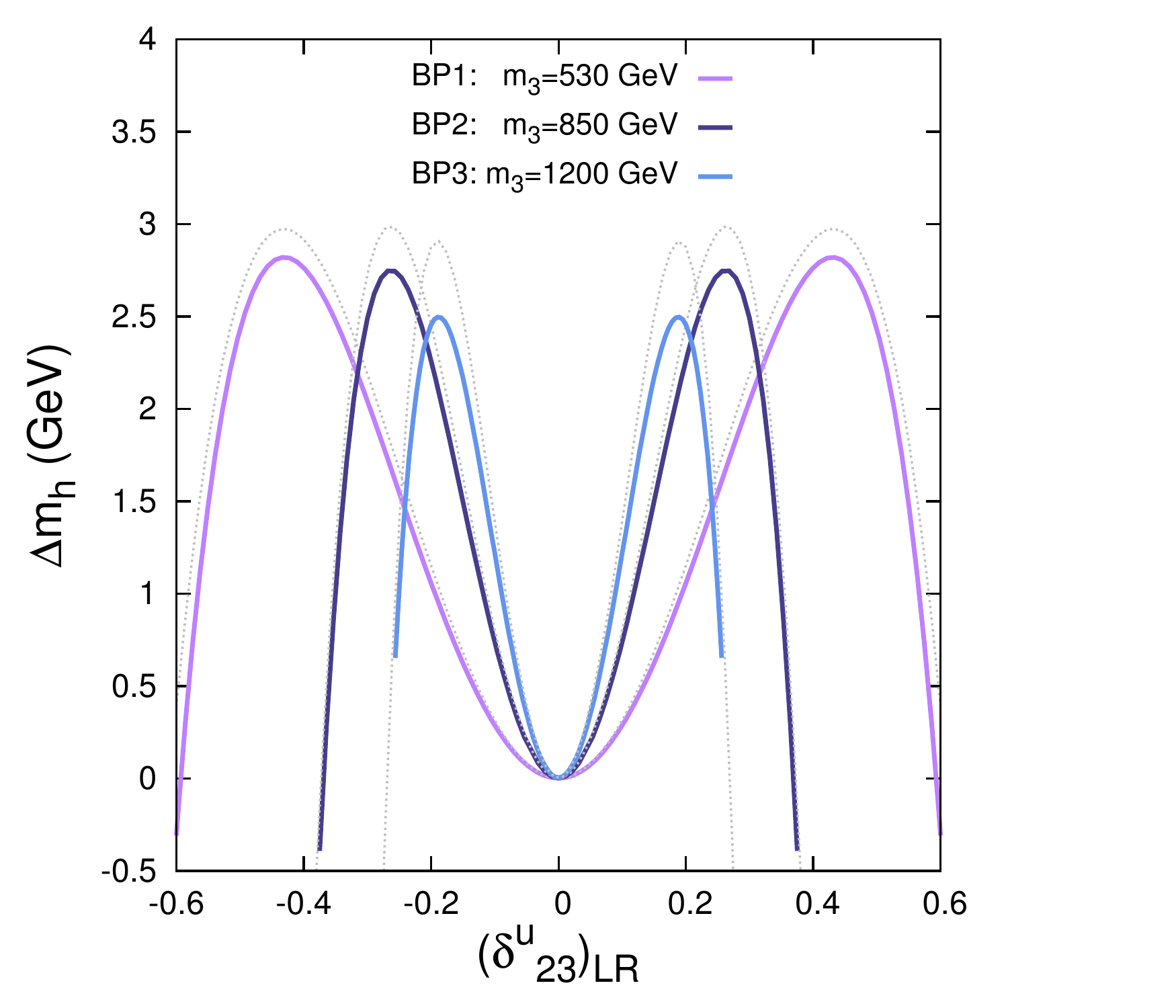}}
\subfloat[]{
\label{fig:b}
\includegraphics[width=0.45\textwidth]{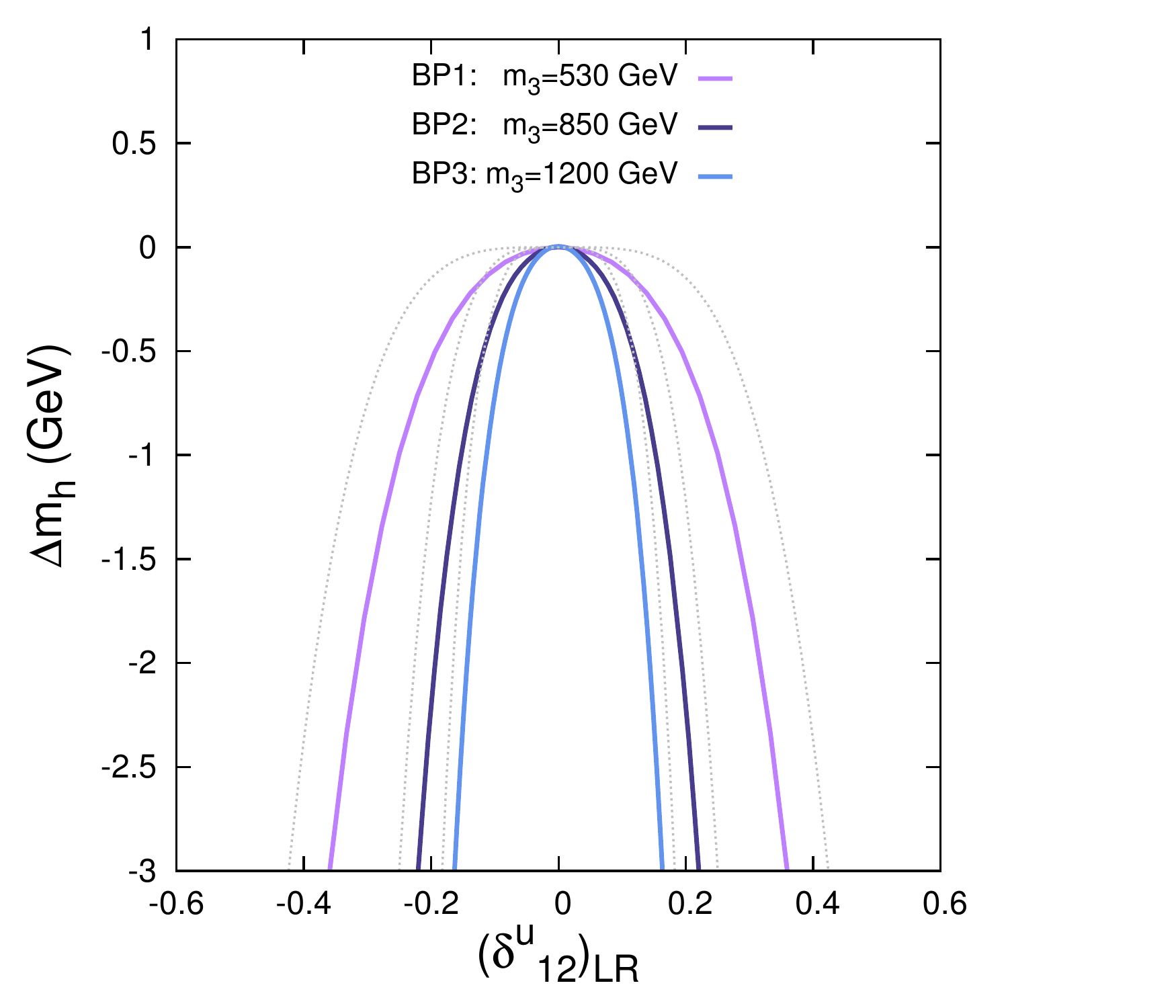}}\\
\vspace{-5mm}
\subfloat[]{
\label{fig:c}
\includegraphics[width=0.45\textwidth]{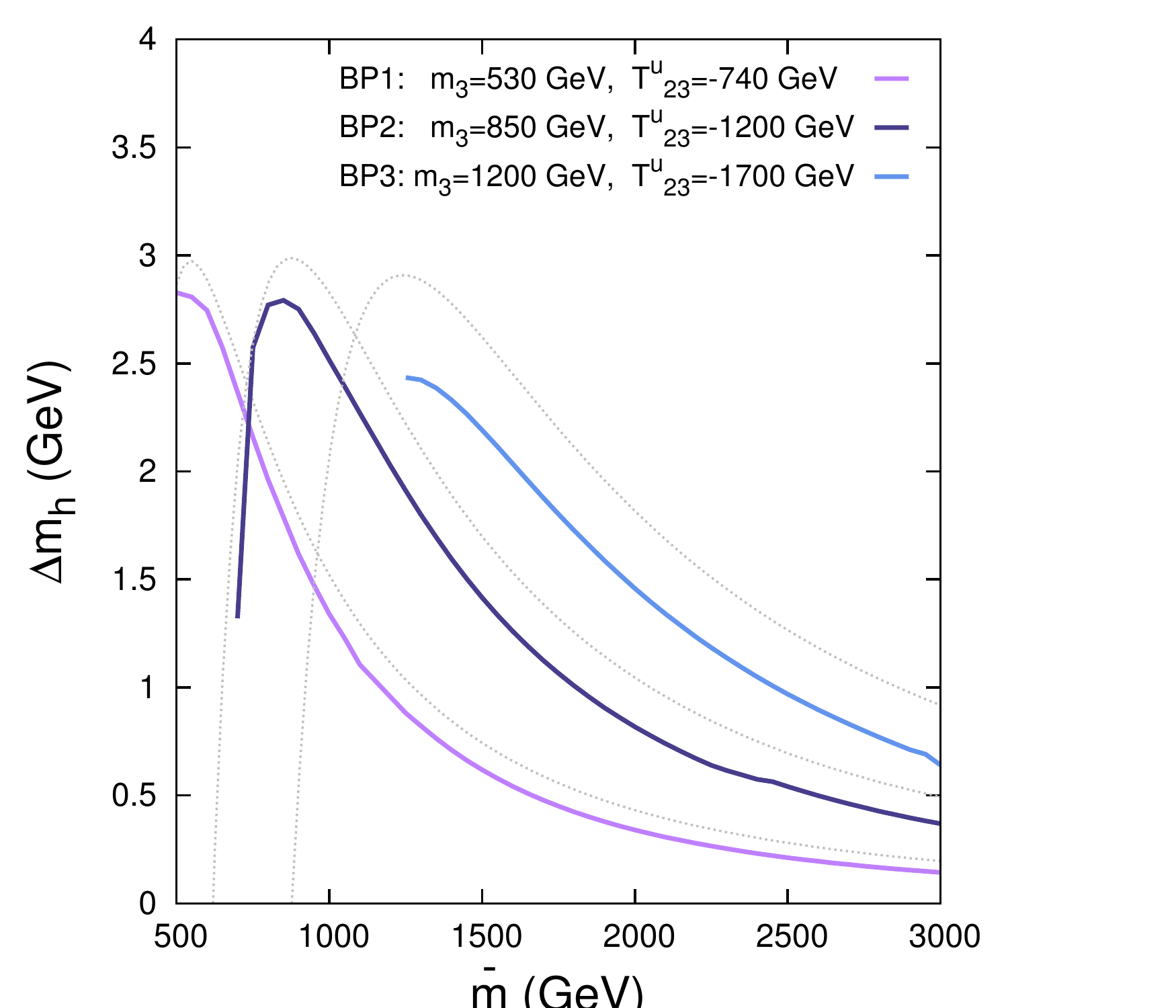}}
\caption{\footnotesize The enhancement of the Higgs boson mass \mhl\ due to flavour violating parameters \protect\subref{fig:a} \dulrbc\ and \protect\subref {fig:b} \dulrab\ for the benchmark points defined in Table~\ref{tab:BP_23}. \protect\subref{fig:c} Dependence of $\Delta\mhl^2$ on the supersymmetric mass scale $\tilde{m}$ for fixed values of $T^u_{23}$. Thick solid lines indicate the results form \spheno\ while the thin ones show the predictions of analytical formulae (\ref{case2}) and (\ref{case3}).}
\label{fig:case23}
\end{figure}

To illustrate the above discussion, we calculated the GFV-corrected Higgs boson mass with \spheno\_v.3.2.4 and compared it with the analytical formulae given in \refeq{case2} and \refeq{case3}. Our three benchmark points at the scale $\msusy=\sqrt{\tilde{m}_{t_1}\tilde{m}_{t_2}}$ are defined in Table~\ref{tab:BP_23}. Gaugino masses, $\mu$ and trilinear terms  were chosen to suppress the corrections not related to parameters $(\delta^u)_{LR}$, in particular those induced by mixing in the sbottom sector.\footnote{The corresponding corrections to the Higgs boson mass squared, calculated using the formulae given in\cite{Haber:1996fp}, do not exceed $1.5\gev^2$.} The results are presented in \reffig{fig:case23}. The thin dashed lines indicate the prediction of the approximate one-loop formulae, while the solid lines show the output from \spheno. One observes that \mhl\ can be enhanced through the GFV contribution by up to 3 \gev.
We confirm here the results first obtained in Refs.\cite{AranaCatania:2011ak} and\cite{Arana-Catania:2014ooa} where a 	
distinctive ``M-shape" Higgs mass dependence on \dulrbc\ and \durlbc\ was presented.

Finally, it is worth to make one more remark. An explicit proportionality of $\Delta\mhl^2$ to the common supersymmetric mass scale $\tilde{m}$ in Eqs.(\ref{case2}) and (\ref{case3}) is not in contradiction with an expected decoupling property of non-logarithmic finite corrections to the Higgs boson mass. One should keep in mind that the parameters $(\delta_{ij})_{LR}$, defined in \refeq{deltas_def}, scale as $1/\tilde{m}^2$ when the off-diagonal trilinear terms $T^u_{ij}$ are fixed, so the GFV corrections indeed decouple when $\tilde{m}$ increases. As a confirmation, in \reffig{fig:case23}\subref{fig:c} we show a dependence of $\Delta\mhl^2$ on the common SUSY scale for the benchmark points defined in Table~\ref{tab:BP_23}. The results were obtained under the assumption that flavour-violating entries $T^u_{23}$ were fixed at the values allowing for the maximal Higgs boson mass enhancement. As in the previous plots, the thin dashed lines indicate the prediction of the approximate one-loop formulae, while the solid lines show the output from \spheno. In both cases a clear decoupling behaviour of the GFV correction can be observed.

\paragraph{ Case 4:} GFV and non-universal diagonal entries.

An additional effect arises when a mass hierarchy between the diagonal entries of the matrix $(\muu^2)_{LR}$ is observed. Let us consider a kind of inverted hierarchy scenario, assuming that the first two generations of up squarks have the same common mass $\tilde{m}_2$, while the mass of the third generation is given by $\tilde{m}_3$. An approximate analytical formula in the case of the soft mass splitting in the stop sector have been derived in Ref.\cite{Haber:1996fp}. It is straightforward to extend those results for the case of the inter-generation mixing, using as a reference \refeq{case2} . Assuming that only $\dulrbc\ne 0$ one obtains
\bea\label{case4}
\Delta\mhl^2(\dulrbc)&=&\frac{3}{4\pi^2}\left\{\dulrbc^2\frac{\tilde{m}^2_2\tilde{m}^2_3}{\tilde{m}^2_2-\tilde{m}^2_3}\left[\frac{1}{2}Y_t^2\sin^2\beta\ln\left(\frac{\tilde{m}^2_2}{\tilde{m}^2_3}\right)\right.\right.\nonumber  \\
&+&\left.\left.\dulrbc^2\frac{\tilde{m}^2_2\tilde{m}^2_3}{v^2(\tilde{m}^2_2-\tilde{m}^2_3)}\left(2-\frac{\tilde{m}^2_2+\tilde{m}^2_3}{\tilde{m}^2_2-\tilde{m}^2_3}\ln\left(\frac{\tilde{m}^2_2}{\tilde{m}^2_3}\right)\right)\right]\right\}.
\eea
An immediate consequence of \refeq{case4} is that the Higgs mass correction due to \dulrbc\ can be strongly enhanced by the mass splitting between the first/second and the third generation. 

\begin{table}[t]
\begin{center}
\begin{tabular}{|c|c|c|c|c|c|c|c|c|c|c|c|}
\hline 
    & $M_1$ & $M_2$ & $M_3$ & $A_{t,b,\tau}$ & $\mu$ & \ma & \tanb & $m_{\tilde{L},\tilde{e}}$ & $m_{\tilde{Q}_{1,2},\tilde{u}_{1,2},\tilde{d}}$ & $m_{\tilde{Q}_3,\tilde{u}_3}$\\
\hline 
BP4 &  300 & 160 & 1600 & 0 & 236 & 1665 & 39 & 3000 & 5000 & 756\\
BP5 &  300 & 160 & 1600 & 0 & 189 & 1310 & 49 & 3000 & 5000 & 842 \\
BP6 &  300 & 160 & 1600 & 0 & 236 & 1410 & 49 & 3000 & 5000 & 990\\
\hline 
\end{tabular}
\caption{\footnotesize Input parameters for benchmark points BP4 - BP6 at \msusy. The masses and trilinear terms are in \gev.}
\label{tab:BP_4}
\end{center}
\end{table}

\begin{figure}[b]
\centering
\subfloat[]{
\label{fig:a}
\includegraphics[width=0.45\textwidth]{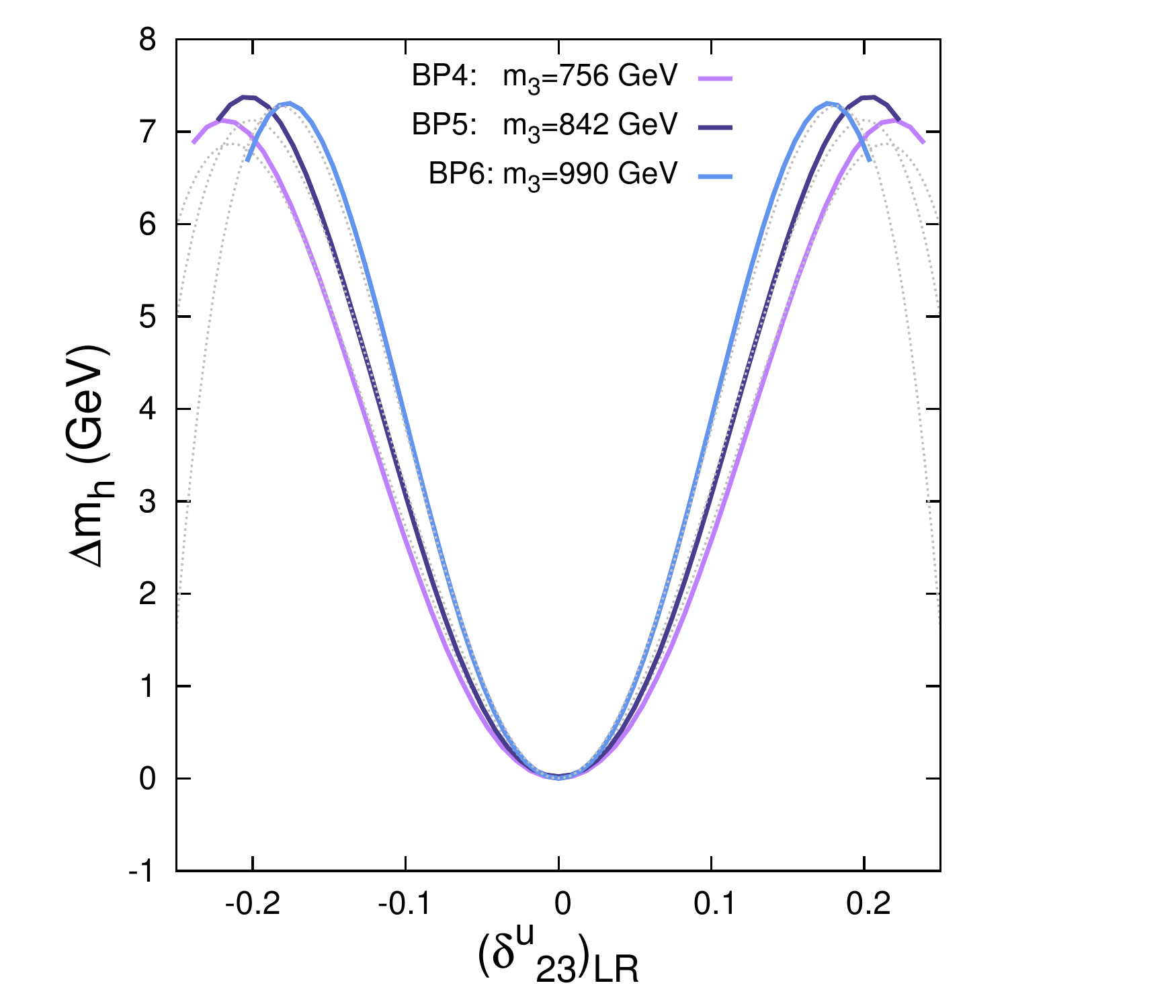}}
\subfloat[]{
\label{fig:b}
\includegraphics[width=0.45\textwidth]{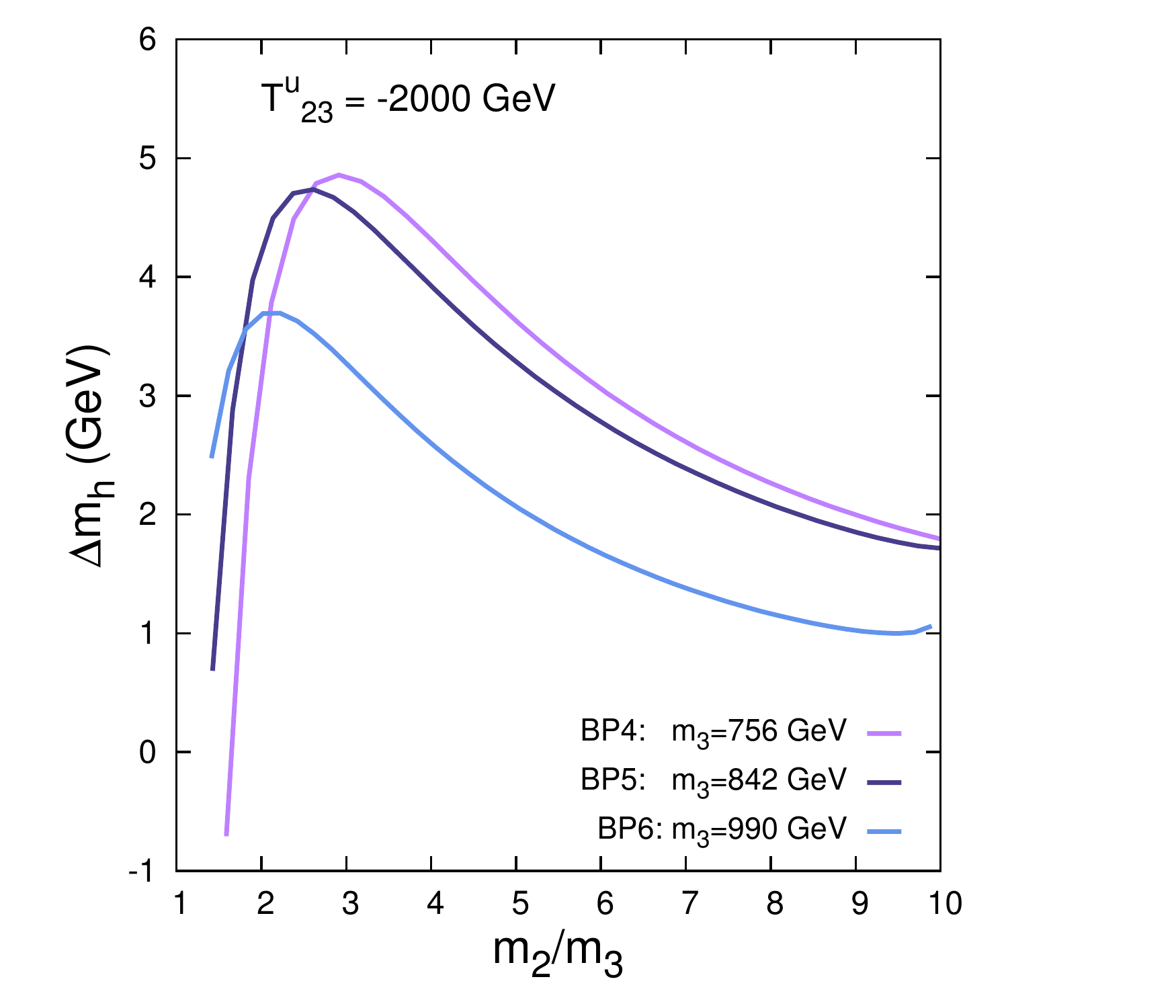}}
\caption{\footnotesize \protect\subref{fig:a} The enhancement of the Higgs boson mass \mhl\ due to flavour violating parameter \dulrbc. Thick solid lines indicate the results form \spheno\ while the thin gray ones show the predictions of analytical formula (\ref{case4}). \protect\subref {fig:b} Dependence of $\Delta\mhl$ on the mass splitting $\tilde{m}_2/\tilde{m}_3$. The benchmark points are defined in Table~\ref{tab:BP_4}. }
\label{fig:case4}
\end{figure}

In \reffig{fig:case4}\subref{fig:a} we show the dependence of \mhl\ on \dulrbc\ for three benchmark points defined in Table~\ref{tab:BP_4}. The soft masses of the first two generations were moved to 5\tev\ to account for the effects of large mass splitting, while the heaviness of right-handed sbottom further reduces the impact from the sbottom mixing, which is now of the order of $0.02\gev^2$. Once more the analytical results (thin grey lines) are compared with the output from \spheno\ (thick lines). As was expected, in this case the enhancement of the Higgs boson mass can be more than twice as large as in the universal mass case, reaching up to $7-8\gev$.

In \reffig{fig:case4}\subref{fig:b} we show the dependence of $\Delta\mhl$ on the mass difference between scharm and stop for our three benchmark points. In all the cases we fixed the value of the off-diagonal entry $(T_u)_{23}$ in the SCKM-basis, $(T_u)_{23}=-2000\gev$. One can observe that the effect is maximalised for the mass ratio $\tilde{m}_2/\tilde{m}_3$ at around $2-4$, while further splitting tends to reduce it.

\paragraph{ Case 5:}Impact of $\tilde{X}_t\ne0$.

\begin{figure}[b]
\centering
\subfloat[]{
\label{fig:a}
\includegraphics[width=0.45\textwidth]{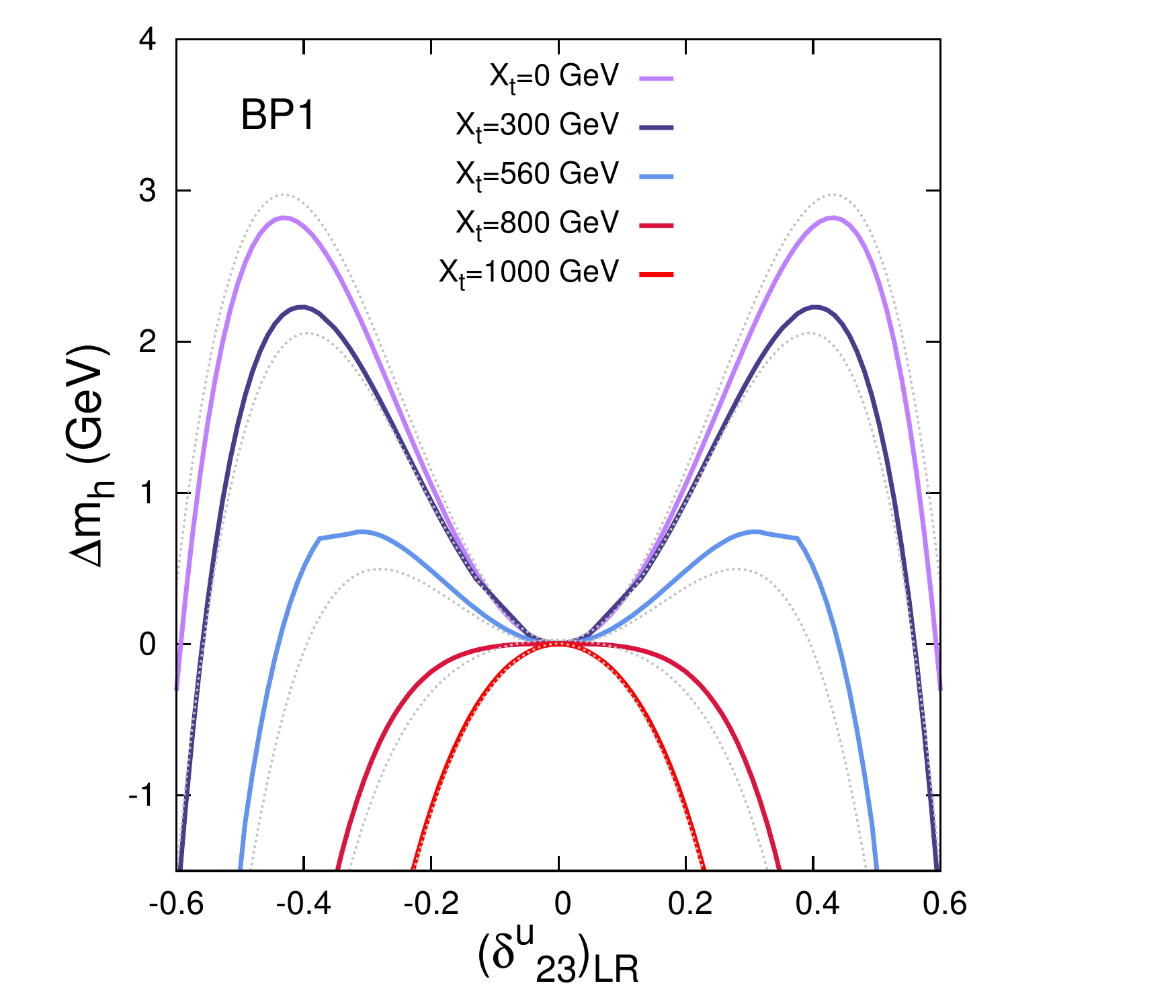}}
\subfloat[]{
\label{fig:b}
\includegraphics[width=0.45\textwidth]{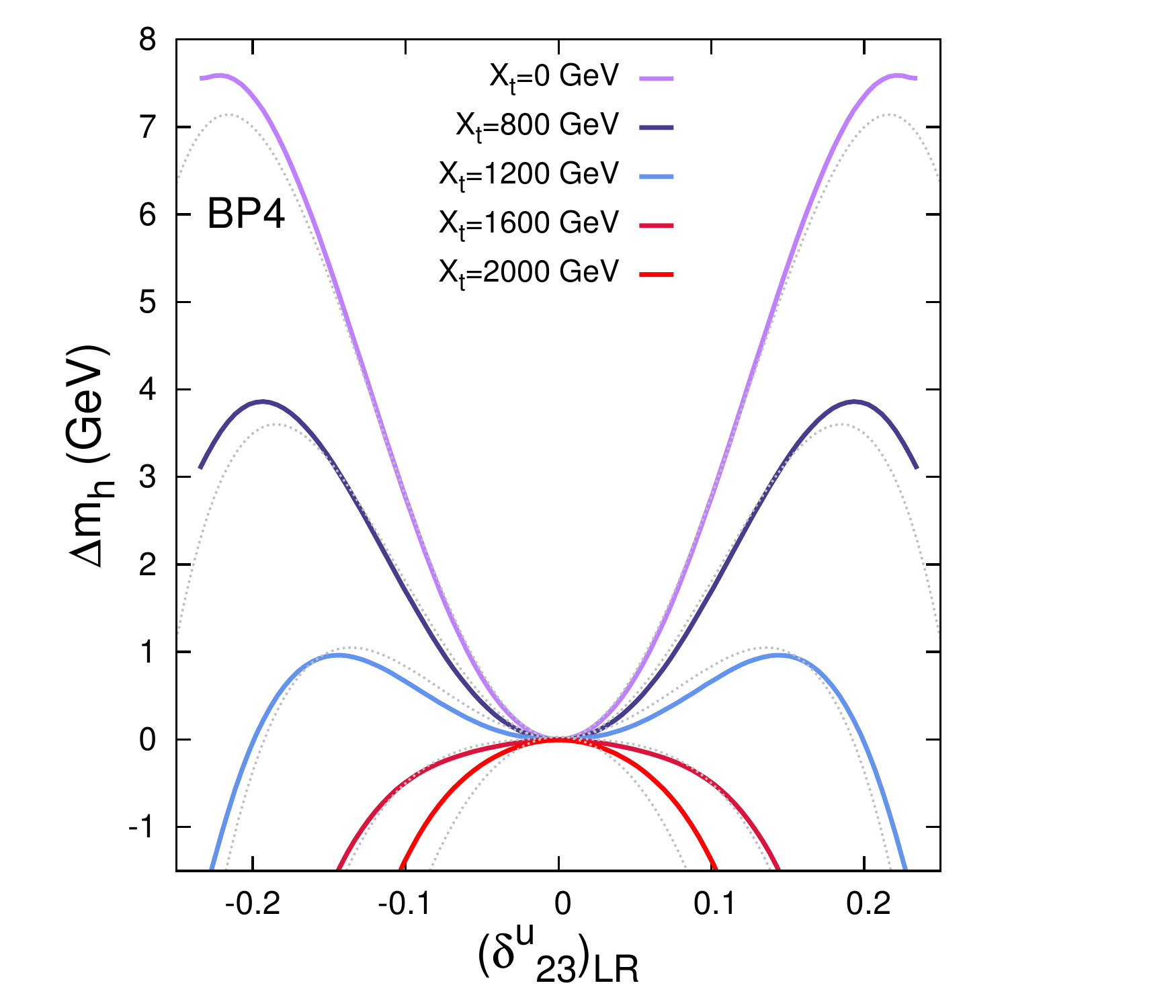}}
\caption{\footnotesize The enhancement of the Higgs boson mass \mhl\ due to flavour violating parameter \dulrbc\ and stop mixing term $\tilde{X}_t$ for benchmark points \protect\subref{fig:a} BP1, and \protect\subref{fig:b} BP4. Thick solid lines indicate the results form \spheno\ while the thin ones show the predictions of analytical formulae (\ref{case7a}) and (\ref{case7b}).}
\label{fig:case7}
\end{figure}

So far we have limited our discussion to the case with no mixing in the stop sector, $\tilde{X}_t=0$. In the presence of non-zero $\tilde{X}_t$ term, the Higgs mass correction takes the form
\be\label{case7a}
\Delta\mhl^2=\frac{3}{4\pi^2}\tilde{m}^2\left[Y_t^2\sin^2\beta\Big((\delta^u_{33})^2+\frac{1}{2}\dulrbc^2\Big)-\frac{\tilde{m}^2}{6v^2}\Big((\delta^u_{33})^4+\dulrbc^4+2(\delta^u_{33})^2\dulrbc^2\Big)\right]
\ee
for the universal sfermion masses, and
\bea\label{case7b}
\Delta\mhl^2(\dulrbc)&=&\frac{3}{4\pi^2}\left\{Y_t^2\sin^2\beta\left[(\delta^u_{33})^2\tilde{m}^2_3+\frac{1}{2}\dulrbc^2\frac{\tilde{m}^2_2\tilde{m}^2_3}{\tilde{m}^2_2-\tilde{m}^2_3}\ln\left(\frac{\tilde{m}^2_2}{\tilde{m}^2_3}\right)\right]\right.\nonumber  \\
&+&\dulrbc^4\frac{\tilde{m}^4_2\tilde{m}^4_3}{v^2(\tilde{m}^2_2-\tilde{m}^2_3)^2}\left(2-\frac{\tilde{m}^2_2+\tilde{m}^2_3}{\tilde{m}^2_2-\tilde{m}^2_3}\ln\left(\frac{\tilde{m}^2_2}{\tilde{m}^2_3}\right)\right)-\frac{\tilde{m}^4_3}{6v^2}(\delta^u_{33})^4\nonumber \\
&-&\left.\frac{(\delta^u_{33})^2\dulrbc^2}{v^2}\left[\frac{2}{\sqrt{6}}\frac{\tilde{m}^2_2\tilde{m}^4_3}{\tilde{m}^2_2-\tilde{m}^2_3}\left|2-\frac{\tilde{m}^2_2+\tilde{m}^2_3}{\tilde{m}^2_2-\tilde{m}^2_3}\ln\left(\frac{\tilde{m}^2_2}{\tilde{m}^2_3}\right)\right|^{1/2}\right]\right\}
\eea
for the hierarchical case. We remind the reader that $\delta^u_{33}=\tilde{m}\frac{\tilde{X}v_2}{\sqrt{2}}$. The main qualitative difference with respect to the cases discussed above is the presence of a negative term $\sim (\delta^u_{33})^2\dulrbc^2$. This new contribution can easily dilute the Higgs mass enhancement obatined through \dulrbc\ if $\tilde{X}_t$ becomes large enough with respect to $\tilde{m}$ or $\tilde{m}_3$. We show this effect in \reffig{fig:case7} for benchmark points BP1 (panel \subref{fig:a}) and BP4 (panel \subref{fig:b}) and different choices of $\tilde{X}_t$ (corresponding to different colours of the solid lines). In particular one can observe that when the ratio $\tilde{X}_t/\tilde{m}$ becomes greater that $\sim 1.5$, it is no longer possible to increase \mhl\ through the non-zero parameter \dulrbc.

\paragraph{ Case 6:} Off-diagonal entries in $(\muu^2)_{LL}$ and $(\muu^2)_{RR}$.

We will now evaluate the contribution to \mhl\ due to parameters $(\delta^{u}_{ij})_{LL}$ and $(\delta^{u}_{ij})_{RR}$.  Note that in the GFV framework large off-diagonal elements $((\muu^2)_{LL})_{23}\sim \lambda^2(\tilde{m}_2^2-\tilde{m}_3^2)$, where $\lambda\sim0.22$ is the Cabibbo angle, can naturally appear due to disalignement between the quark and squark fields if the mass difference between the diagonal elements of the matrix $(m_Q^2)_{LL}$, $\tilde{m}_2^2-\tilde{m}_3^2$, is large. 

Let us consider for simplicity the universal case. The masses of the mixed left-handed eigenstates are given by $m^2_{1L}=\frac{1}{2}\mtop^2+\tilde{m}^2+\tilde{m}^2\dulrbc$ and $m^2_{2L}=\frac{1}{2}\mtop^2+\tilde{m}^2-\tilde{m}^2\dulrbc$ (assuming $\dullbc>\frac{\mtop^2}{2\tilde{m}^2}$). Since \dullbc\ does not depend on $v_2$, it only results in a shift of the soft mass $\tilde{m}$ that lifts the degeneracy between previously universal soft masses of the third generation. In such a situation the Higgs mass is corrected by a factor\cite{Haber:1996fp}
\be
\Delta\mhl^2(\dullbc)=\frac{3}{8\pi^2}\frac{\mz^4}{v^2}\cos2\beta(1-\frac{8}{3}\ssqthw)(\frac{\mtop^2}{\mz^2}+\frac{1}{6}\cos2\beta)\ln(1+\dullbc).
\ee
This contribution is proportional to the mass scale of the EW sector and therefore negligible, unless the parameter \dullbc\ becomes very large. The same effect is observed for \dullac.

On the other hand, the non-zero value of the parameter \dullab\ does not influence \mhl\ at one-loop level, since it does not trigger the mixing with the third generation. 

\paragraph{ Case 7:} More than one $(\delta^u_{ij/ji})_{LR}\ne0$.

\begin{figure}[t]
\centering
\subfloat[]{
\label{fig:a}
\includegraphics[width=0.45\textwidth]{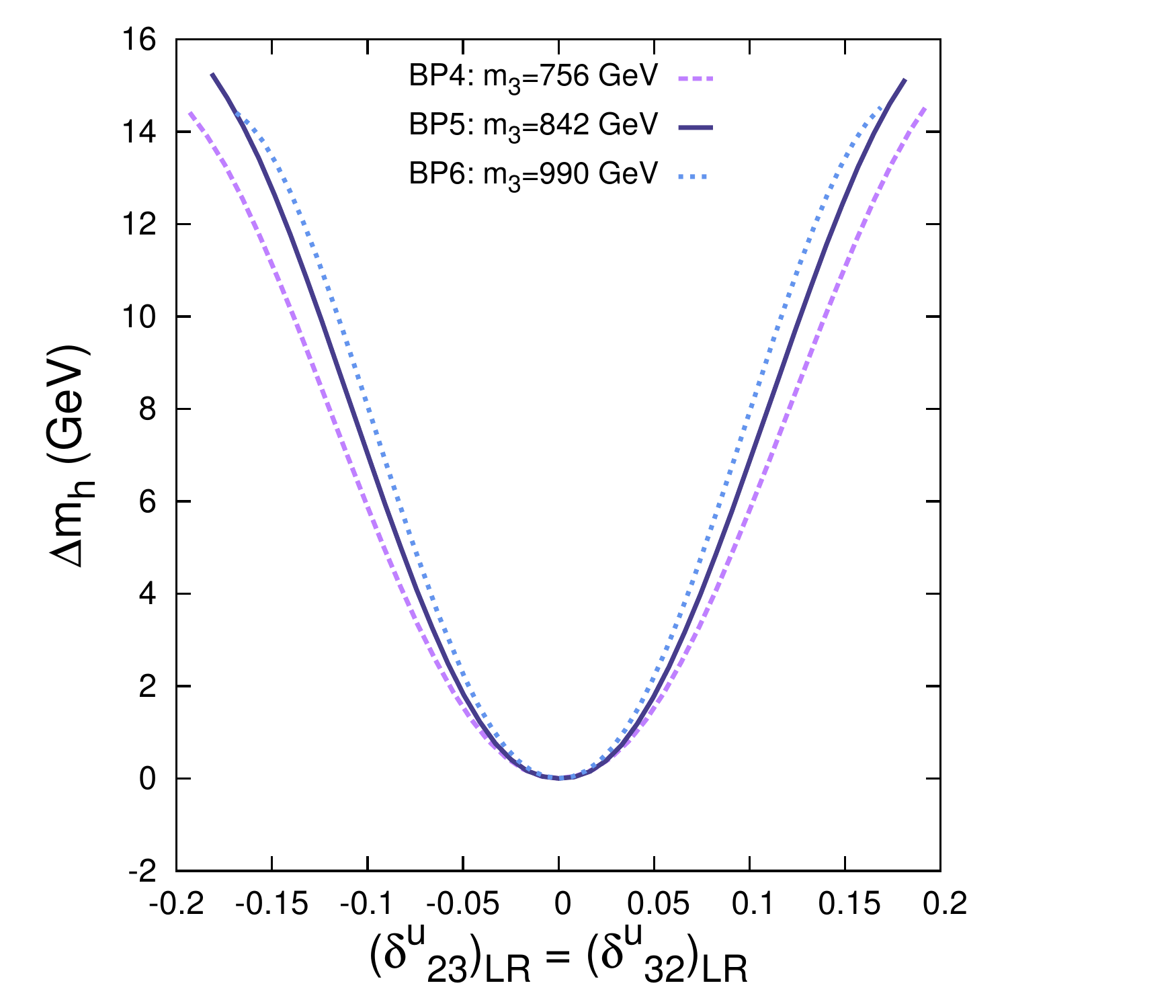}}
\subfloat[]{
\label{fig:b}
\includegraphics[width=0.45\textwidth]{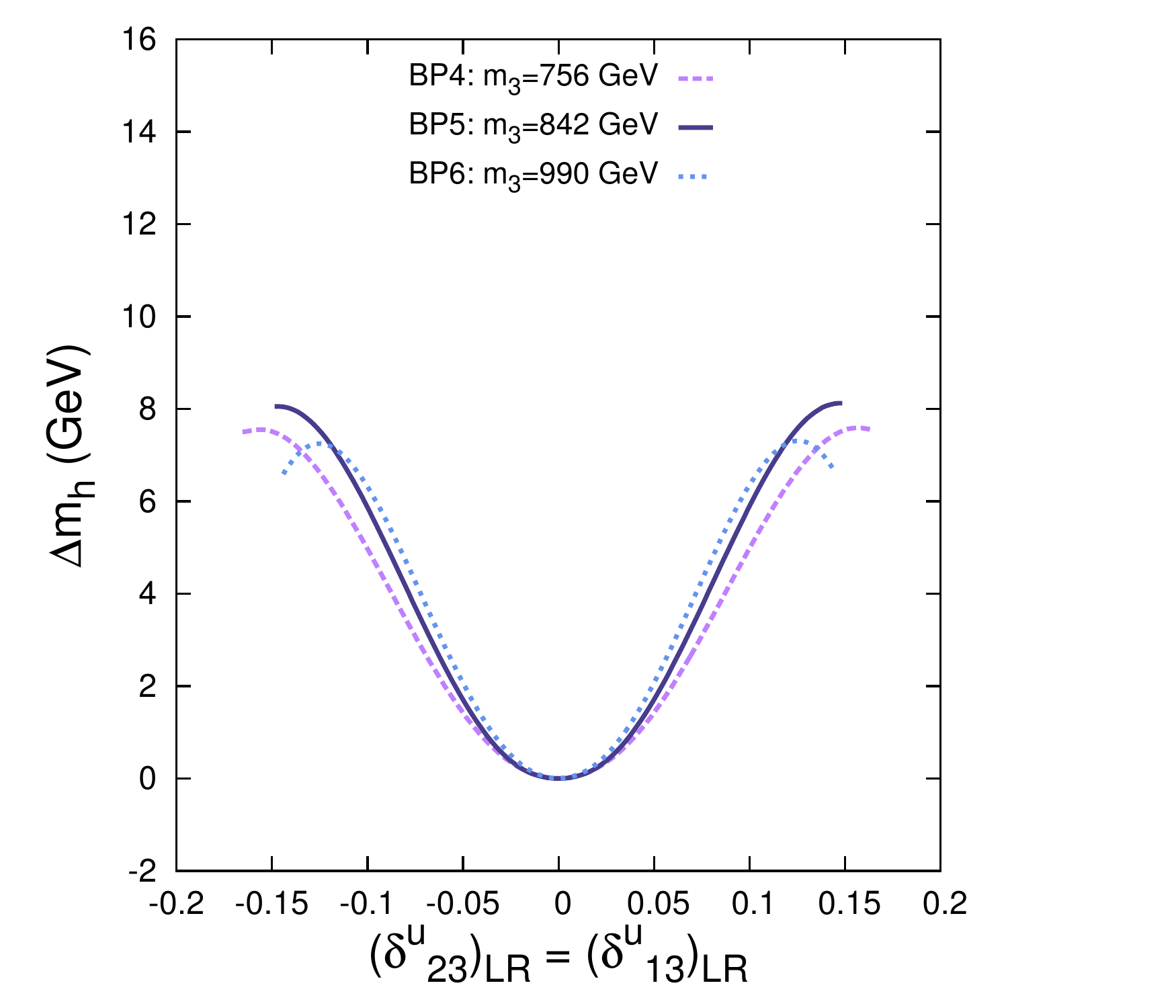}}
\caption{\footnotesize The enhancement of the Higgs boson mass \mhl\ due to flavour violating parameters \protect\subref{fig:a} $\dulrbc=\durlbc$, and \protect\subref{fig:b} $(\delta^u_{23})_{LR}=(\delta^u_{13})_{LR}$. The benchmark points defined in Table~\ref{tab:BP_4}.}
\label{fig:case6}
\end{figure}

Finally, we will analyse the impact of two simultaneously non-zero parameters $(\delta^u_{ij/ji})_{LR}$. The net effect strongly depends on which pairs are considered. If the GFV entries are present in two separate blocks of the matrix $(\muu^2)_{LR}$, namely upper and lower triangle, the individual contributions to $\Delta\mhl$ will sum up allowing for much stronger enhancement of the Higgs boson mass. In \reffig{fig:case6}\subref{fig:a} we show this effect for $\dulrbc=\durlbc$ for three benchmark points defined in Table~\ref{tab:BP_4}. It turns out that \mhl\ can be lifted up even by $14-15\gev$. The same effect would be observed in the case of  $\dulrac=\durlac$ and $\dulrbc=\durlac$. On the contrary, if one considers two GFV entries in the same block of $(\muu^2)_{LR}$, the net effect will be limited as the parameter $\dulrbc$ in Eq.\ref{case2} and Eq.\ref{case4} will be replaced by $\delta_{\textrm{eff}}=\sqrt{(\delta^u_{ij})_{LR}^2+(\delta^u_{ik})_{LR}^2}$. This can be seen in \reffig{fig:case6}\subref{fig:b} for $(\delta^u_{23})_{LR}=(\delta^u_{13})_{LR}$. 

Note that in this case the diluting effect, observed for $\tilde{X}_t$, is absent as a negative mixed term $\sim\dulrbc\durlbc$ does not appear.

\section{Limits on \dulrac,  \durlac, \dulrbc\ and \durlbc}\label{sec:fcnc}

In this section we will discuss the consistency of the GFV scenarios analysed above with both the experimental data and a theoretical requirement of the vacuum stability. In the former case, the strongest limits on the allowed values of the off-diagonal entries in the SSB terms come from the measurements of the FCNC transitions. The influence of various flavour changing processes on the corresponding parameters $\delta_{ij}$ have been thoroughly discussed in the literature, see for example Ref.\cite{Gabbiani:1996hi,Misiak:1997ei}. Interestingly, the four most important parameters from the point of view of the Higgs boson mass enhancement, namely \dulrac,  \durlac, \dulrbc, and \durlbc, are very weakly bounded by flavour constraints. In fact, there are only a few ways of constraining the $(1,3)$ and $(2,3)$ sectors of the left-right mixing blocks in $\mathcal{M}^2_{\tilde{u}}$.  

The chargino-squark loop contributions to rare decays $b\to s\gamma$, $b\to d\gamma$, \bsmumu, as well as to \bsbs\ and \bdbd\ mixing, can provide upper bounds on the allowed values of the chirality flipping deltas. This possibility has been investigated in Refs.\cite{AranaCatania:2011ak} and\cite{Arana-Catania:2014ooa}, where the limits on \dulrbc\ and \durlbc\ have been derived. The actual strength of the bounds clearly depends on the choice of the parameter space, though \durlbc\ as large as $0.22$ can be excluded in certain cases, mainly through the measurement of \brbxsgamma. 

A chargino-loop contribution to the semileptonic decay $\bar{B}\to K^{*}l^{+}l^{-}$ has been recently analysed in Ref.\cite{Behring:2012mv}. It was shown that a strong bound $\dulrbc<0.1$ can be derived for $m_{\tilde{u}_3}\sim 300\gev$ and large stop mixing. A similar study has been performed in Ref.\cite{Colangelo:1998pm,Altmannshofer:2009ma} for the $K \to \pi \nu \bar{\nu}$ decay and \kk\ mixing, providing bounds on \dulrbc\ and \dulrac. However, it should be emphasised that the actual limits depend on the flavour-diagonal MSSM masses  and in general become effective only for relatively light up-squarks with masses below several hundred GeV.

The other way of constraining the chiralilty changing entries of $\mathcal{M}^2_{\tilde{u}}$ would be through rare decays of the top quark, \thc\ and \thu. A recent analysis by CMS, based on $19.5\invfb$ of data at \eight, puts the upper limit on the \thc\ branching ratio as $\textrm{BR}(\thc)<5.6\times10^{-3}$\cite{CMS-PAS-HIG-13-034}. This result, however, is still almost two orders of magnitude larger than the corresponding branching ratios calculated in the MSSM, $6\times 10^{-5}$\cite{Guasch:1999jp,Cao:2007dk,Cao:2006xb}, and henceforth does not provide constraints on the GFV parameters \dulrbc\ and \durlbc.

\begin{table}[t]\footnotesize
\begin{center}
\begin{tabular}{|c|c|c|c|c|c|c|c|}
\hline 
 & BP1 & BP2 & BP3 & BP4 & BP5 & BP6 & FCNC process\\
\hline 
\dulrbc &  $[-0.87:0.55]$ & $-$  & $-$  & $-$  & $-$  & $-$ & \bsgamma\ \\
\durlbc &  $[-0.2:0.52]$ & $-$  & $-$  & $[-0.15:0.3]$ & $[-0.16:0.19]$ & $[-0.2:-]$ & \bsgamma\ (\bsmumu) \\
\dulrac &  $-$ & $-$  & $-$  & $-$ & $-$  & $-$ & $-$ \\
\durlac &  $[-0.45:0.55]$  & $-$  & $-$  & $-$  & $-$  & $-$ & $K^+ \to \pi^+ \nu \bar{\nu}$\\
\hline 
\end{tabular}
\caption{\footnotesize Limits on the parameters \dulrac, \durlac, \dulrbc\ and \durlbc\ from the FCNC processes discussed in the text. The benchmark points are defined in Table~\ref{tab:BP_23} and Table~\ref{tab:BP_4}. In the last column a process that provides the strongest bound is shown.}
\label{tab:FCNC}
\end{center}
\end{table}

In Table~\ref{tab:FCNC} we present the limits on the flavour-violating parameters \dulrac, \durlac, \dulrbc\ and \durlbc\ for six benchmark points defined in Table~\ref{tab:BP_23} and Table~\ref{tab:BP_4}, obtained by considering the following FCNC processes: \bsgamma, \bsmumu, \bdmumu, \delmbs, $K^0 \to \pi^0 \nu \bar{\nu}$ and $K^+ \to \pi^+ \nu \bar{\nu}$.  The flavour observables were calculated with the public code \texttt{\susyflav\ v2.10}\cite{Crivellin:2012jv}, and the following experimental measurements were applied: $\brbxsgamma=3.43\pm 0.3\times 10^{-4}$\cite{bsgamma}, $\brbsmumu=2.9\pm 0.76\times 10^{-9}$\cite{Aaij:2013aka,Chatrchyan:2013bka}, $\brbdmumu< 7.4\times 10^{-10}$\cite{Aaij:2013aka,Chatrchyan:2013bka}, $\delmbs= 1.166\pm0.158\times 10^{-11}\gev$\cite{Beringer:1900zz}, $\textrm{BR}(K^0 \to \pi^0 \nu \bar{\nu})<2.6\times 10^{-8}$\cite{Beringer:1900zz}, $\textrm{BR}(K^+\to \pi^+ \nu \bar{\nu})=1.73\pm1.15\times 10^{-10}$\cite{Beringer:1900zz}, where theoretical and experimental errors were added in quadrature. When deriving the limits on $\delta_{ij}$ , a $2\sigma$ deviation from the experimentally measured central value was allowed. 

One can observe that the benchmark point BP1 is the only one that can be significantly constrained by the FCNC processes. This had to be expected as BP1 is characterised by the lightest SUSY spectrum. More importantly, however, even after imposing the FCNC constraints the maximal Higgs mass enhancement through all parameters $\delta_{ij}$ discussed in the previous section is still possibile. One can also notice that \bsgamma\ seems to be the strongest experimental constraint, which is in agreement with the findings of Refs.\cite{AranaCatania:2011ak} and\cite{Arana-Catania:2014ooa}. Interestingly, it can provide stringent bounds on the allowed values of parameter \durlbc\ for the points with the hierarchical up-squark matrices (mainly due to the large values of \tanb), which however remain consistent with the values required to increase \mhl.

The off-diagonal entries in the SSB matrices can also be bounded by requiring that the radiative corrections to the CKM elements generated through the squark-gluino loops do not exceed the experimental values, as discussed in Ref.\cite{Crivellin:2008mq,Crivellin:2009sd,Crivellin:2011jt}. In fact, such limits can be stronger than the FCNC ones if SUSY spectrum is heavier than 500\gev\ and they become even more effective when the squark mass increases. The bound on \durlac\ is particularly important in that context as it was shown to be very strong\cite{Crivellin:2008mq,Crivellin:2009sd,Crivellin:2011jt}. 

In Table~\ref{tab:CKM} we present the limits on the flavour-violating parameters \dulrac, \durlac, \dulrbc, and \durlbc\ after the inclusion of chirally enhanced corrections to CKM matrix elements. The corrections were calculated with \texttt{\susyflav\ v2.10} and the following experimental values were assumed: $V_{ts}=0.0429\pm 0.0026$, $V_{cb}=0.0409\pm 0.0011$, $V_{ub}=0.00415\pm 0.00049$\cite{Beringer:1900zz}. One can immediately see that the impact of those additional constraints is dramatic. The non-zero values of \durlac\ are now totally excluded by demanding that a radiative correction to $V_{ub}$ does not exceed the experimentally measured value. Also \durlbc\ becomes very strongly constrained. However, the Higgs mass enhancement is still perfectly possible with $\dulrac\ne0$ as this parameter remains virtually unconstrained.

\begin{table}[t]\footnotesize
\begin{center}
\begin{tabular}{|c|c|c|c|c|c|c|}
\hline 
    & BP1 & BP2 & BP3 & BP4 & BP5 & BP6  \\
\hline 
\dulrbc &  $[-0.45:0.35]$ & $[-0.15:0.12]$  & $[-0.14:0.12]$  & excl.  & excl.  & excl.\\
\durlbc &  $[-0.05:0.04]$ & $[-0.02:0.02]$  & $[-0.05:0.02]$  &  $[-0.04:0.01]$  & $[-0.04:0.01]$ & $[-0.02:0.01]$  \\
\dulrac &  $-$ & $-$  & $-$  & $-$ & $-$  & $-$ \\
\durlac &  excl.  & excl.  & excl.  & excl.  & excl.  & excl. \\
\hline 
\end{tabular}
\caption{\footnotesize \footnotesize Limits on the parameters \dulrac, \durlac, \dulrbc\ and \durlbc\ from the FCNC processes discussed in the text and the chirally enhanced corrections to CKM matrix elements. The benchmark points are defined in Table~\ref{tab:BP_23} and Table~\ref{tab:BP_4}.}
\label{tab:CKM}
\end{center}
\end{table}

Additional bounds on the flavour-violating parameters \dulrac, \durlac, \dulrbc\ and \durlbc\ arise from the requirement that the EW vacuum is stable. The scalar potential of the MSSM can develop a charge/color breaking (CCB) minimum lower than the EW one, or it can become unbounded from below (UFB), if the trilinear couplings are too large\cite{Frere:1983ag,AlvarezGaume:1983gj,Derendinger:1983bz,Kounnas:1983td,Casas:1996de}. More importantly, unlike the FCNC constraints, these kind of bounds do not weaken when \msusy\ increases. In the case of the up-squark sector the corresponding formulae for the CCB limits are given by\cite{Casas:1996de}
\bea
(\delta_{i3}^u)_{LR}&\leq& \mtop\frac{[(\muu^2)^{ii}_{LL}+(\muu^2)^{33}_{RR}+\mhu^2+\mu^2]^{1/2}}{(\muu)^{ii}_{RR}(\muu)^{33}_{LL}},\qquad i=1,2,
\eea
and the bounds on $(\delta_{3i}^u)_{LR}$ are obtained by switching the indices $3$ and $i$. Similarly, the UFB limits read\cite{Casas:1996de}
\bea
(\delta_{i3}^u)_{LR}&\leq& \mtop\frac{[(\muu^2)^{ii}_{LL}+(\muu^2)^{33}_{RR}+(\mll^2)^{ii}_{LL}+(\mll^2)^{33}_{RR}]^{1/2}}{(\muu)^{ii}_{RR}(\muu)^{33}_{LL}},\qquad i=1,2.
\eea
Note that the denominator in the above expressions differ from the one shown in\cite{Casas:1996de} due to different normalization of the parameters $\delta_{ij}$ in \refeq{deltas_def}. 

\begin{table}[b]\footnotesize
\begin{center}
\begin{tabular}{|c|c|c|c|c|c|c|}
\hline 
    & BP1 & BP2 & BP3 & BP4 & BP5 & BP6  \\
\hline 
CCB &  $\leq 0.47$ & $\leq 0.28$  & $\leq 0.20$ & $\leq 0.23$ & $\leq 0.21$ & $\leq 0.18$\\
UFB &  $\leq 4.90$ & $\leq 0.52$  & $\leq 0.44$ & $\leq 0.30$ & $\leq 0.27$ & $\leq 0.23$\\
\hline 
\end{tabular}
\caption{\footnotesize \footnotesize CCB and UFB vacuum stability bounds on the parameters \dulrac, \durlac, \dulrbc\ and \durlbc. The benchmark points are defined in Table~\ref{tab:BP_23} and Table~\ref{tab:BP_4}.}
\label{tab:CCB}
\end{center}
\end{table}

In Table~\ref{tab:CCB} we present the vacuum stability bounds on the flavour-violating parameters \dulrac, \durlac, \dulrbc, and \durlbc\ for our six benchmark points. The limits are the same for all considered deltas due to the fact that two first generations of sfermions are degenerate. The CCB bounds are much more constraining than the UFB ones, mainly due to the presence of heavy sleptons in the spectrum. This effect is particularly strong for the benchmark points BP1-BP3, characterized by relatively light squarks. One can also observe that the bounds become stronger when the masses of the up-squark increase. It should be emphasised, however, that the derived limts are consistent with the value of parameters $\delta_{ij}$ necessary for the maximal or nearly maximal enhacement of the Higgs boson mass.
 
On the other hand, the appearance of the CCB minimum does not necessarily need to be a problem for a model, as long as the lifetime of the EW vacuum is longer than the age of the universe. The derivation of the metastability limits is quite complex and involves numerical calculation of the bounce action for a given scalar potential. Such an analysis was performed in Ref.\cite{Park:2010wf} in the context of the off-diagonal trilinear terms. It was shown that the resulting metastability bounds do not depend on the Yukawa couplings and therefore are in general much less stringent than the CCB ones. The effect is, however, least significant for \dulrac, \durlac, \dulrbc, and \durlbc, where the upper limit from vacuum stability requirement can be weakened by only $10-15\%$.

\section{GFV effects in GUT-constrained scenarios}\label{sec:guthiggs}

In the previous section we showed that the mass of the lightest Higgs boson can be significantly enhanced by non-zero chirality changing mixing between the third and second/first generations, triggered by the presence of non-zero off-diagonal entries in the trilinear coupling matrix $(\muu^2)_{LR}$. In this section we will analyse the corresponding GFV effects in the framework of GUT-constrained scenarios. Here the dependence of \mhl\ on the parameters $\delta_{ij}$ is slightly more complicated, as one must take into account the RGE evolution of the input parameters between the scales \mgut\ and \msusy. 

While working in the framework of General Flavour Violation, it is convenient to rewrite the corresponding RGEs in the SCKM-basis. Then the input parameters at the GUT-scale are defined in the basis in which the Yukawa matrices are diagonal and the flavour violating effects manifest themselves through the presence of the CKM-matrix elements in the RG equations. From now on we will always assume that the soft terms at \mgut\ are given in the SCKM-basis.

Let us first analyse the impact of the non-zero element $(T_u)_{23}$ in the up-squark trilinear matrix $T_u$. The only RGEs directly affected by its presence (neglecting the contribution from $g_1$ and $g_2$ and all the Yukawa couplings but $Y_t$) are:
\bea
\frac{d(T_u)_{23}}{dt}&=&\frac{1}{16\pi^2}(T_u)_{23}\left(8Y_t^2-\frac{16}{3}g_3^2\right),\nonumber\\
\frac{\mhu^2}{dt}&=&\frac{6}{16\pi^2}\left(Y_t^2\mhu^2+Y_t^2\mqthree+Y_t^2\muthree+\sum_{i}(T_u)^2_{ii}+(T_{u})^2_{23}\right),
\eea 
where we used the formulae given in\cite{Martin:1993zk}. 
Therefore, two effects are present while evolving the soft SUSY-breaking parameters down from the GUT scale: a)  reduction of the soft squark masses due to a positive contribution to \mhu\ from $(T_u)_{23}$ and, b) increase of the term $(T_u)_{23}$ driven by top-Yukawa. Both factors make \dulrbc\ at \msusy\ increase, thus facilitating the \mhl\ enhancement. On the other hand, lower squark masses mean lowering of the Higgs boson mass with respect to the MFV case. Therefore, in order to increase the scalar mass \mhl\ in a GUT-constrained scenario, the former effect should dominate the latter.

\begin{table}[t]
\begin{center}
\begin{tabular}{|c|c|c|c|c|c|c|c|c|c|c|c|}
\hline 
    & C1 & C2 & C3 & C4 & C5 &  I1 & I2 & I3 & I4 & I5 \\
\hline 
$\mzero(1,2)$ & 2320 & 1320 & 1320 & 900 & 900 & 8000 & 5000 & 5000 & 3000 & 3000\\
\mhalf &   1380 & 1080 & 1080 & 508 & 508& 1380 & 1080 & 1080 & 508 & 508 \\
\azero &   0 & 0 & 2000 & 0 & 1000 & 0 &  0 & 1000 & 0 & 1000\\
\tanb &   31 & 28 & 28 & 18 & 18 &31 & 28 & 28 & 18 & 18\\
\hline 
$\mzero(3)$  & 2320 & 1320 & 1320 & 900 & 900 &  2320 & 1320 & 1320 & 900 & 900\\
\hline 
\end{tabular}
\caption{\footnotesize Input parameters for benchmark points at \mgut. Letter C refers to the CMSSM boundary conditions while I to the inverted hierarchy scenario. The masses and trilinear terms are in \gev.}
\label{tab:BP_GUT}
\end{center}
\end{table}
\begin{figure}[b]
\centering
\subfloat[]{
\label{fig:a}
\includegraphics[width=0.45\textwidth]{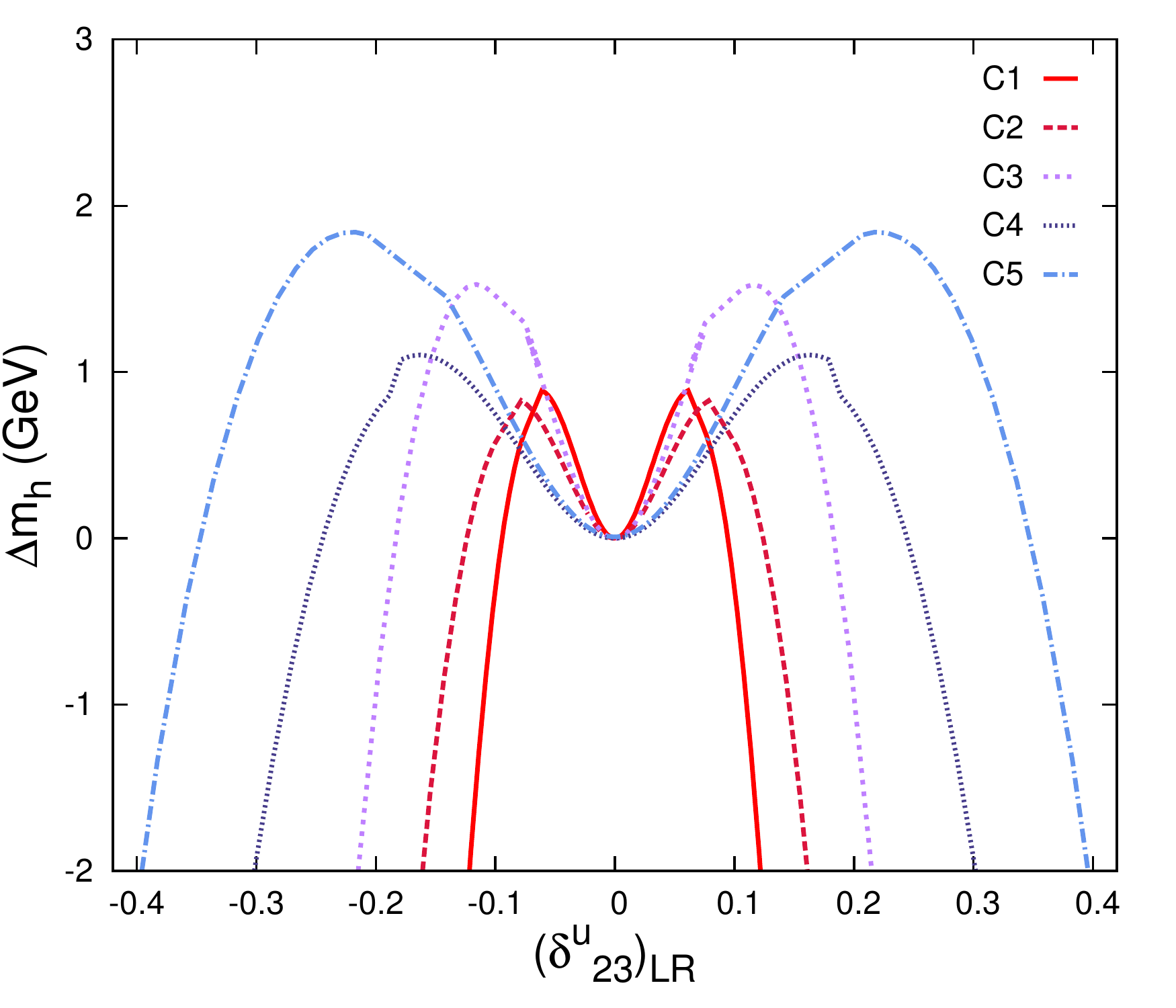}
}
\subfloat[]{
\label{fig:b}
\includegraphics[width=0.45\textwidth]{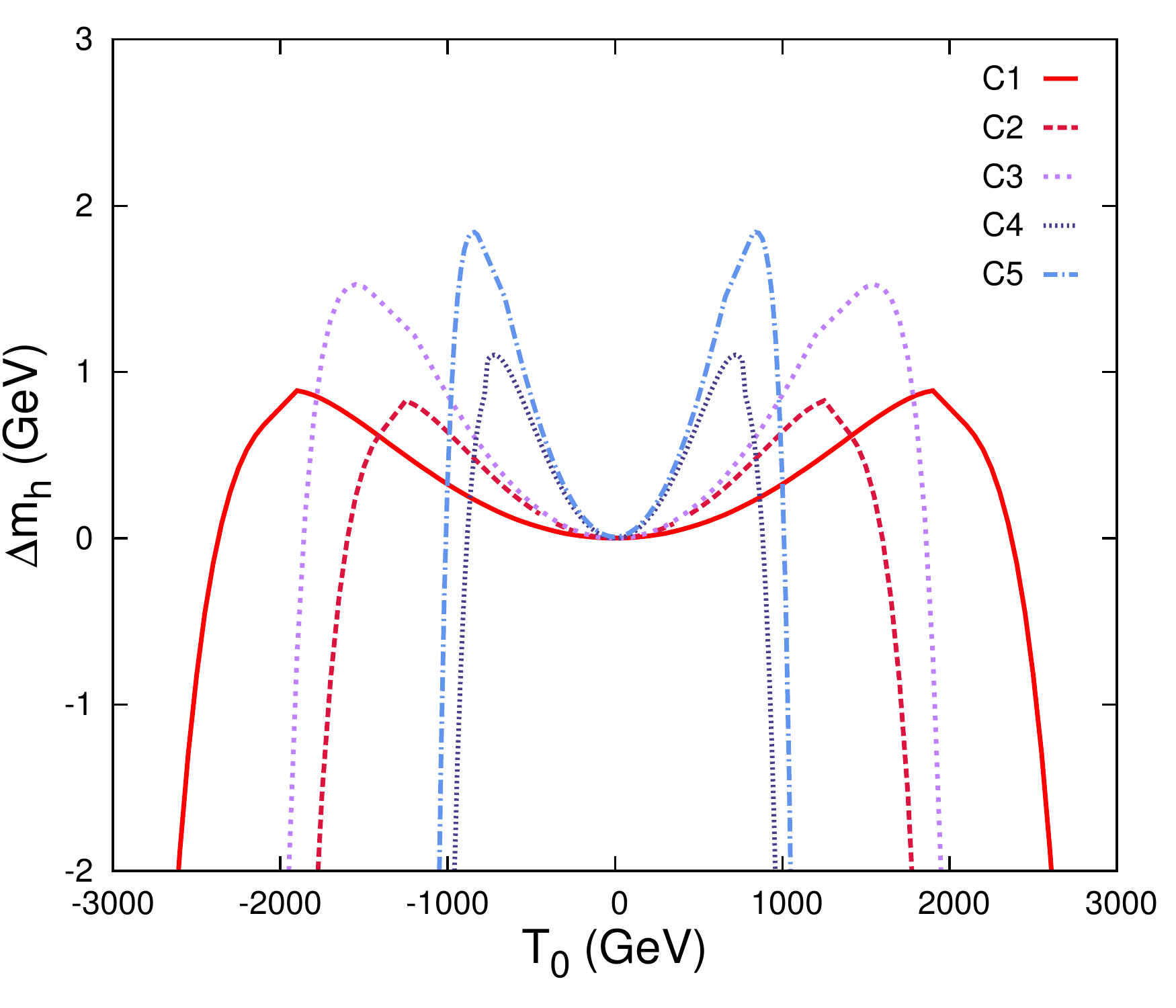}
}
\caption{\footnotesize The enhancement of the Higgs boson mass \mhl\ due to parameter $T_0$ as a function of \protect\subref{fig:a} \dulrbc, and \protect\subref{fig:b} $T_0$ in the CMSSM. The benchmark points are defined in Table~\ref{tab:BP_GUT}.}
\label{fig:gut_cmssm}
\end{figure}

In \reffig{fig:gut_cmssm}\subref{fig:a} we show the dependence $\Delta\mhl$ vs \dulrbc\ for five benchmark points with mSUGRA inspired universal boundary conditions at the \gut\ scale. The input values of the soft parameters are given in Table~\ref{tab:BP_GUT} (labelled with C) and the off-diagonal entry $(T_u)_{23}(\mgut)$, denoted as $T_0$, is varied in the range $[-3000:3000]$ \gev. The reference value of the Higgs boson mass in the case of Minimal Flavour Violation corresponds to $T_0=0$. In \reffig{fig:gut_cmssm}\subref{fig:b} we present the same mass dependence as a function of $T_0$.  As one can see, the enhancement of the Higgs boson mass with respect to the MVF case is still possible, but its magnitude is mitigated by the RG running so it does not exceed 2\gev. The effect is distinctly strongest for the benchmark points C3 and C5, which are characterised by the ratio $\azero/\mhalf\simeq2$.

To better understand the mechanism underlying this behaviour, let us write down approximate formulae that quantify the dependence of the relevant soft SUSY-breaking terms at the EW scale on the input \gut-scale parameters:
\bea\label{rges_cmssm}
(\mql^2)_{33}(\mew)&\simeq&2.96\,\mhalf^2-0.07\,\azero^2-0.31\,T^2_0+0.14\,\azero\mhalf+0.73\,m_0^2,\nonumber\\
(\muu^2)_{22}(\mew)&\simeq&3.56\,\mhalf^2+0.1\,\azero^2-0.96\,T^2_0-0.03\,\azero\mhalf+0.97\,m_0^2,\nonumber\\
(T_u)_{23}(\mew)&\simeq&-0.04\,\mhalf+1.34\,T_0.
\eea
The expansion coefficients multiplying the input parameters in the above polynomial have been calculated for the benchmark point C3 with \spheno\_v.3.2.4. Since they only depend on the RGE running of the masses and trilinear terms, they do not change over the parameter space by more than 10--20\% and the qualitative conclusions derived for a sample point are quite universal. From \refeq{rges_cmssm} it is easy to see that with all other parameters fixed, larger \mzero\ allows for larger values of $T_0$ before any of the soft mass $(\mql^2)_{33}$ or $(\muu^2)_{22}$ becomes negative at \mew. That is confirmed by \reffig{fig:gut_cmssm}\subref{fig:b} while comparing the maximal allowed $T_0$ for the points C1, C3 and C5. Of course, for the points with larger \mzero\ the maximal $T_0$ corresponds to smaller \dulrbc, as the suppression of its value due to soft masses is stronger than the enhancement through $T_0$. Therefore in \reffig{fig:gut_cmssm}\subref{fig:a} the pattern of benchmark points is inverted.

The impact of \azero\ is somehow more subtle. With all other parameters fixed, larger \azero\ also makes it easier to obtain larger $T_0$. This effect is, however, much weaker than the one due to \mzero, since $(\muu^2)_{22}$ essentially does not depend on \azero\ and is quickly driven below zero when $T_0$ increases. This behaviour is particularly clear when both \mzero\ and \mhalf\ are small, as can be seen for the points C4 and C5 for which the maximum $T_0$ is almost the same. On the other hand, exactly the same feature makes it possible to enhance the Higgs mass more than in the case with $A_0=0$. A closer look at \refeq{rges_cmssm} shows that for a given set of input parameters there exists a range of $T_0$ when even a small change in its value results in a quick drop of $(\muu^2)_{22}(\mew)$ and, subsequently, a quick rise of $\dulrbc$ (this tends to happen for smaller $T_0$ when \mzero\ becomes smaller). Therefore, larger \azero\ allows for slightly larger $T_0$ and that is enough to enhance \mhl.

\begin{figure}[t]
\centering
\subfloat[]{
\label{fig:a}
\includegraphics[width=0.45\textwidth]{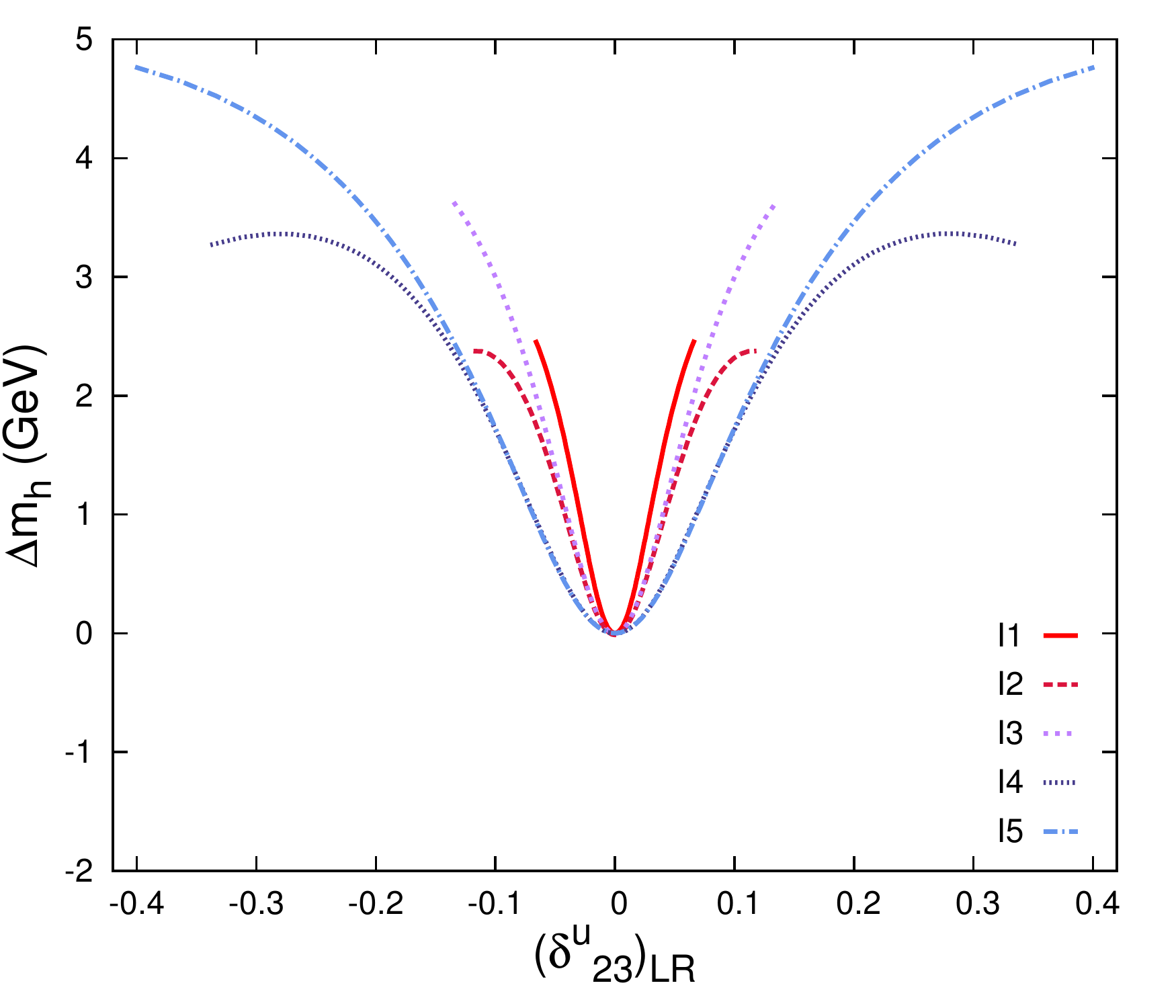}
}
\subfloat[]{
\label{fig:b}
\includegraphics[width=0.45\textwidth]{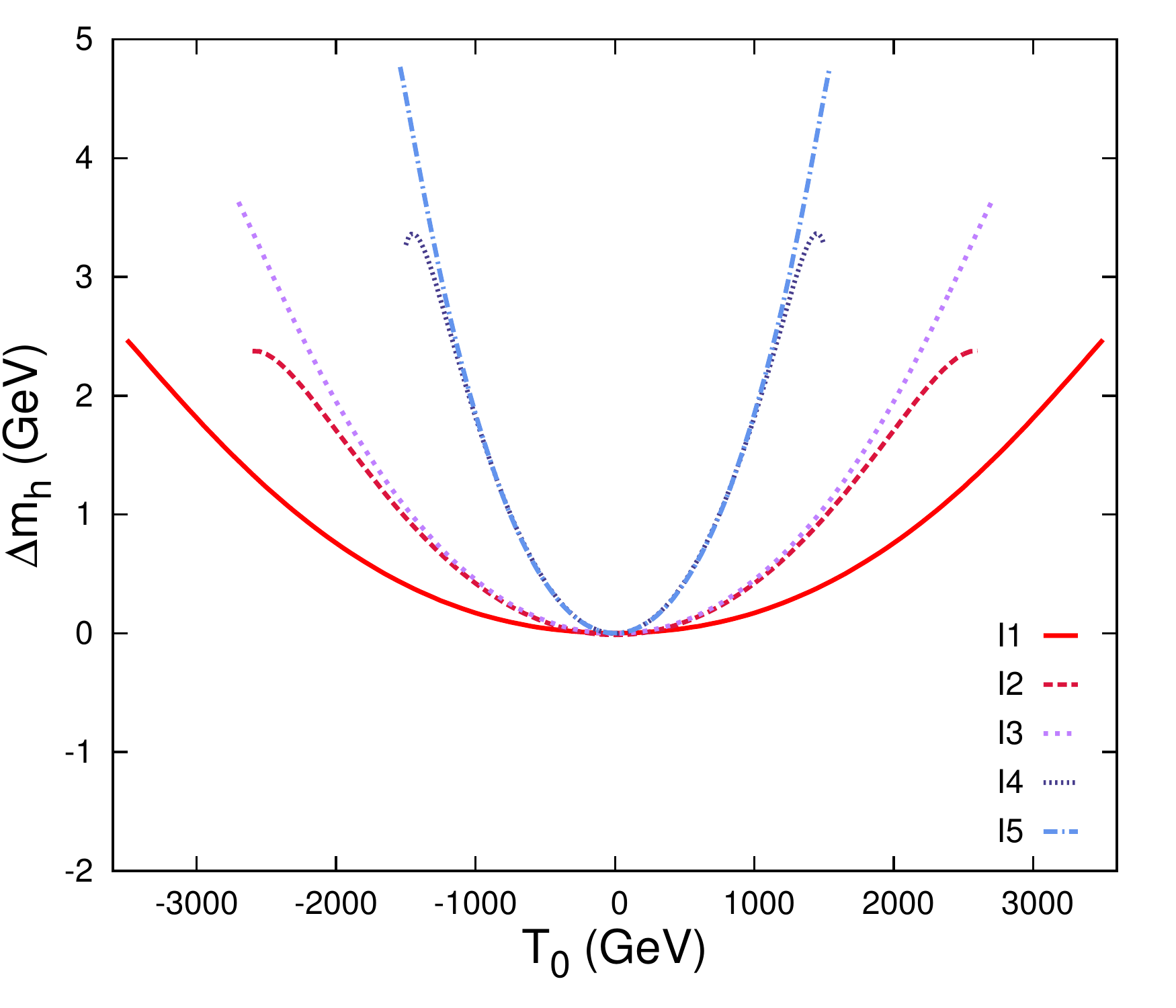}
}
\caption{\footnotesize The enhancement of the Higgs boson mass \mhl\ due to parameter $T_0$ as a function of \protect\subref{fig:a} \dulrbc, and \protect\subref{fig:b} $T_0$ in the inverted hierarchy scenario. The benchmark points are defined in Table~\ref{tab:BP_GUT}.}
\label{fig:gut_ih}
\end{figure}

On the other hand, it was pointed out in Sec.\ref{sec:higgs} that the effects due to inter-generation mixing can be stronger in the case of  large mass splitting between stops and scharms of different chiralities. In \reffig{fig:gut_ih}\subref{fig:a} we show the dependence $\Delta\mhl$ vs \dulrbc\ for five benchmark points defined in Table~\ref{tab:BP_GUT} (labelled with I), and in \reffig{fig:gut_ih}\subref{fig:b} the dependence $\Delta\mhl$ vs $T_0$. All the input parameters but $m_0(1,2)$ are equal to the ones analysed in the case with universal boundary conditions. The amount of mass splitting between the first/second and third generations have been chosen to maximalise the impact of \dulrbc, as discussed in Sec.\ref{sec:higgs}. One can see that in this scenario the Higgs boson mass can be enhanced more than in the CMSSM, up to $5\gev$. This is a direct consequence of the mass splitting between the up-squarks. The EW-scale value of $(\muu^2)_{22}$ is now modified as
\be\label{rges_ih}
(\muu^2)_{22}(\mew)\simeq4.6\,\mhalf^2+1.16\,\azero^2-1.05\,T^2_0+0.01\,\azero\mhalf-0.03\,\mzero^2(3)+0.24\,\mzero^2(1,2),
\ee
so its dependence on $T_0$ is weaker with respect to the dependence on other parameters. As a result, the drop of the soft masses due to $T_0$ is now much slower and $\dulrbc$ can be largely increased before $(\mql^2)_{33}$ becomes negative.

\begin{figure}[b]
\centering
\subfloat[]{
\label{fig:a}
\includegraphics[width=0.45\textwidth]{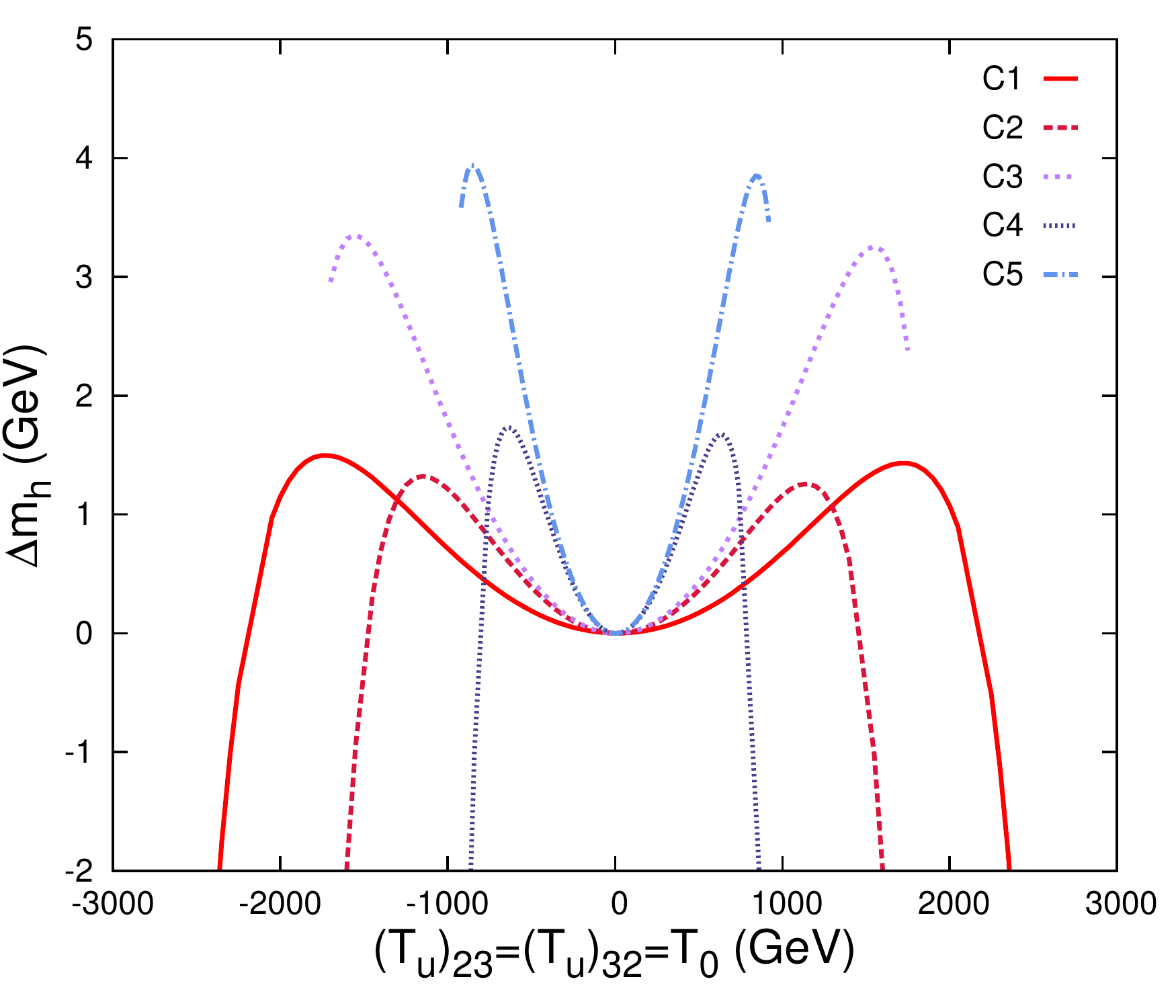}
}
\subfloat[]{
\label{fig:b}
\includegraphics[width=0.45\textwidth]{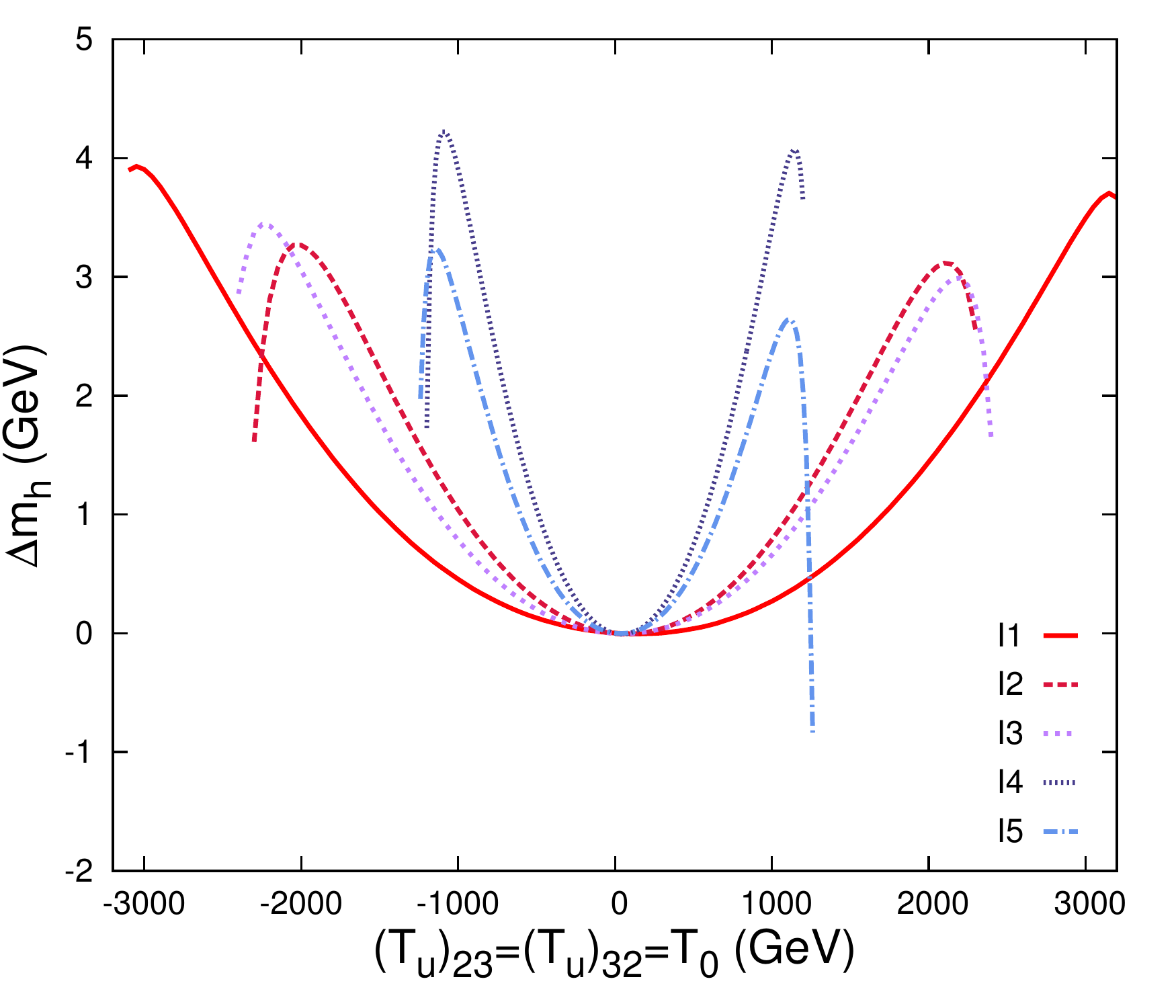}
}
\caption{\footnotesize The enhancement of the Higgs boson mass \mhl\ due to parameter $(T_u)_{32}=(T_u)_{23}=T_0$ for \protect\subref{fig:a} CMSSM, and \protect\subref{fig:b} inverted hierarchy scenario. The benchmark points are defined in Table~\ref{tab:BP_GUT}. }
\label{fig:gut_2d}
\end{figure}

Finally we will analyse the case with both $(T_u)_{23}$ and $(T_u)_{32}$  different from zero and equal to $T_0$ at the GUT-scale. After the discussion in Sec.\ref{sec:higgs} one could expect that the enhancement of the Higgs boson mass will be now much stronger. \reffig{fig:gut_2d}  shows that while it is in fact possible for the CMSSM (panel \subref{fig:a}), it is totally not the case in the inverted hierarchy scenario (panel \subref{fig:b}). The reason is once more the RGE running. For the IH mass pattern the renormalization of the soft term $(\mql^2)_{33}$ due to both off-diagonal entries is so strong that it quickly dominates the effect of the increased $T_0$.

\begin{figure}[t]
\centering
\subfloat[]{
\label{fig:a}
\includegraphics[width=0.47\textwidth]{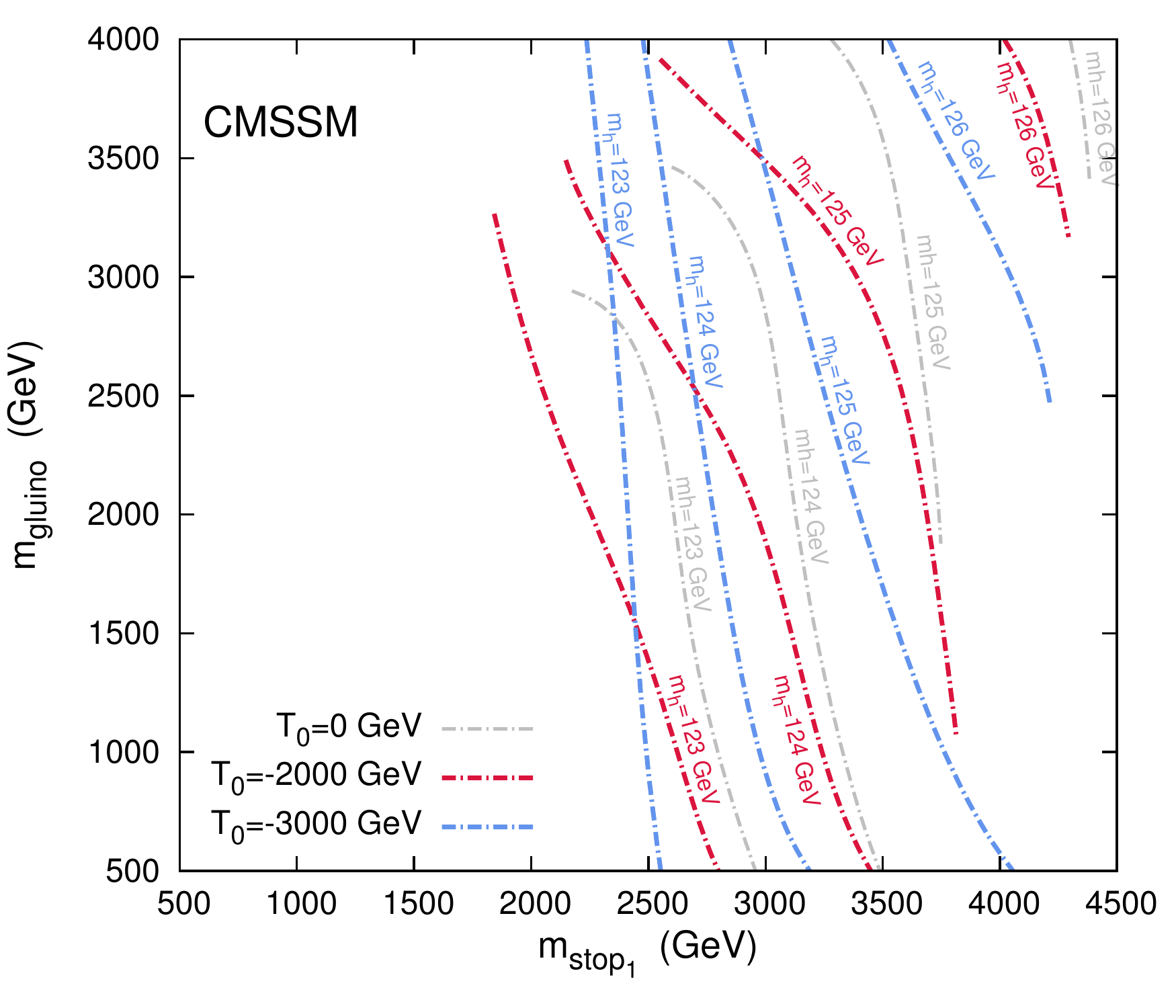}
}
\subfloat[]{
\label{fig:b}
\includegraphics[width=0.47\textwidth]{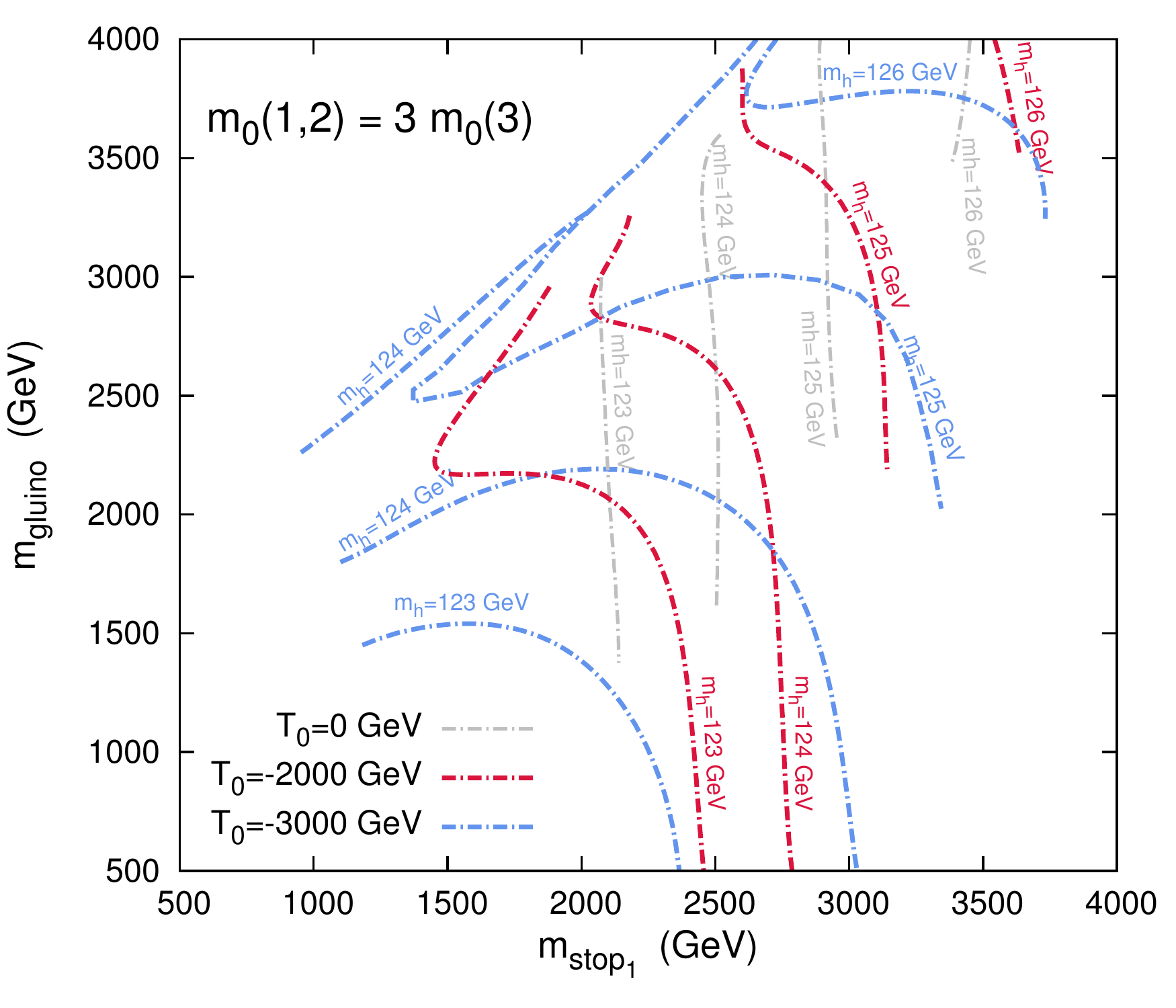}
}
\caption{\footnotesize Iso-contours of \mhl\ in the $(m_{\tilde{t}_1}, \mgluino)$ plain for \protect\subref{fig:a} CMSSM, and \protect\subref{fig:a} inverted hierarchy with $\mzero(1,2)=3\mzero(3)$. In both plots $\azero=0$, $\tanb=30$, while the off-diagonal entry $(T_u)_{23}=T_0$ is set at  $T_0=0$ (gray), $T_0=-2000\gev$ (dark red), and $T_0=-3000\gev$ (light blue).}
\label{fig:gut_grid}
\end{figure}

To summarise the findings of this section, we will briefly comment on possible implications of the GFV contributions to the Higgs boson mass on the phenomenology of two scenarios discussed above, and in particular on the allowed masses of stops and gluinos (the full phenomenological analysis is a subject of the next section).
In \reffig{fig:gut_grid} we show the iso-contours of \mhl\ in the $(m_{\tilde{t}_1}, \mgluino)$ plain for \subref{fig:a} CMSSM, and \subref{fig:b} inverted hierarchy scenario with $\mzero(1,2)=3\mzero(3)$. The universal scalar and gaugino masses where scanned freely, while the remaining parameters were kept fixed at $\azero=0$ and $\tanb=30$. The colours correspond to different values of the off-diagonal entry $(T_u)_{23}=T_0$ at the GUT scale: $T_0=0$ (gray), $T_0=-2000\gev$ (dark red), $T_0=-3000\gev$ (light blue). The lines break when reaching the region of the parameter space corresponding either to no EWSB, or to tachyonic stop, or to the sequence of stop and gluino masses that cannot be obtained given the input parameter range. We do not show those regions explicitly on the plots as their position depends on the choice of $T_0$. 

From \reffig{fig:gut_grid} one can see that the impact of $T_0$ is particularly important in the inverted hierarchy case. If $T_0=-3000\gev$, the Higgs boson with the mass at 125 \gev\ can be obtained for $m_{\tilde{t}_1}=1370\gev$, $\mgluino=2500\gev$ and a small stop-sector mixing $|X_t/\msusy|=0.82$. On the other hand, if $T_0=0$ the corresponding value of \mhl\ requires $m_{\tilde{t}_1}=2850\gev$ and $\mgluino=3600\gev$.

On the contrary, the impact of non-zero $T_0$ in the case of the CMSSM is not particularly strong.

\section{Global analysis}\label{sec:scans}

So far we have discussed the possibility of enhancing the Higgs boson mass through the GFV corrections to the scalar potential without questioning the validity of such scenarios when confronted with the experimental data. To address this issue, in this section we will study the phenomenology of the GUT-constrained inverted hierarchy scenario in the context of its predictions about the relic density of dark matter, EW precision observables and flavour physics. 

\begin{table}[b]\footnotesize
\begin{center}
\begin{tabular}{|l|l|l|l|l|l|}
\hline
Measurement & Mean or range & Error:~exp.,~th. & Ref.\\
\hline
\mhl\ (by CMS) & $125.7\gev$ & $0.4\gev, 3\gev$ & \cite{CMS-PAS-HIG-13-005} \\
\abundchi      & $0.1199$ 	& $0.0027$,~$10\%$ 		& \cite{Ade:2013zuv}\\
\brbxsgamma $\times 10^{4}$ 		& $3.43$   	& $0.22$,~$0.21$ & \cite{bsgamma}\\
$\brbsmumu\times 10^9$			& 2.9 &  0.7, 10\% &  \cite{Aaij:2013aka,Chatrchyan:2013bka}\\
\brbutaunu $\times 10^{4}$          & $0.72$  	& $0.27$,~$0.38$ & \cite{Adachi:2012mm}\\
$\Delta M_{B_s}\times 10^{11}$ & $1.166\gev$ & $0.008\gev$, $0.158\gev$ & \cite{Beringer:1900zz}\\
\sinsqeff 			& $0.23146$     & $0.00012$, $0.00015$ &  \cite{Beringer:1900zz}\\
$M_W$                     	& $80.385\gev$      & $0.015\gev$, $0.015\gev$ &  \cite{Beringer:1900zz}\\
\hline
\mtpole\ & 173.34\gev & 0.76\gev, 0 &  \cite{Beringer:1900zz} \\
\mbmbmsbar\ & 4.18\gev\ & 0.03\gev, 0 &  \cite{Beringer:1900zz} \\
$\alphaem^{-1}$ & 127.916 & 1.00.015, 0 &  \cite{Beringer:1900zz} \\
\alphas\ & 0.1184 & 0.0007, 0 &  \cite{Beringer:1900zz} \\
\hline
\end{tabular}
\caption{\footnotesize
The experimental constraints applied in the analysis.} 
\label{tab:exp_constraints}
\end{center}
\end{table}


The numerical analysis was performed with the package BayesFITSv3.1,
described in detail in\cite{Fowlie:2012im}. 
The package is linked to \texttt{\multinest\ v2.7}\cite{Feroz:2008xx}
for sampling. Mass spectra are calculated with \texttt{\spheno\ v3.2.4};
the branching ratios \brbxsgamma, \brbsmumu\ and \brbutaunu\ as well as mass difference $\Delta M_{B_s}$ and SUSY contribution to anomalous magnetic moment of the muon \deltagmtwomususy\ with \texttt{\susyflav\ v2.10};
the relic density and spin-independent neutralino-proton cross section \sigsip\ 
with \texttt{\dsusy\ v5.0.6}\cite{Gondolo:2004sc}; and 
EW precision constraints with  \texttt{\feynhiggs\ v2.10.0}.
To include the exclusion limits from Higgs boson searches at LEP, Tevatron, and the LHC and the \chisq\ contributions from the Higgs boson signal rates from Tevatron and the LHC we use \texttt{\higgsbounds\ v1.0.0}\cite{Bechtle:2008jh,Bechtle:2011sb,Bechtle:2013wla} interfaced with \texttt{\higgssignals\ v1.0.0}\cite{Bechtle:2013xfa}.
The SM parameters (\mtpole, \mbmbmsbar, \alphaem\ and \alphas) where treated as nuisance parameters and randomly drawn from a Gaussian distribution around the central value. The experimental constraints applied in the analysis are listed in Table~\ref{tab:exp_constraints}. Note that we do not assign any statistical interpretation to the presented results.

We scanned the input parameters of the model in the following ranges:
\begin{gather}\label{ranges1}
100\gev\leq\mzero(3)\leq 6000\gev\, \qquad 1 \leq\mzero(1,2)/\mzero(3)\leq 10,\qquad 100\gev\leq\mhalf\leq 3000\gev\,\nonumber\\
3\leq\tanb\leq 62,\qquad -3000\gev\leq T_0\leq -1000\gev.
\end{gather}
The universal trilinear coupling was set as $\azero=0$. This choice comes from the fact that our main interest is to investigate the possible enhancement of \mhl\ in the situation when the mixing in the stop sector is far from its maximal value. For the same reason we scanned the parameter $T_0$ far from zero.

\begin{figure}[t]
\centering
\subfloat[]{
\label{fig:a}
\includegraphics[width=0.42\textwidth]{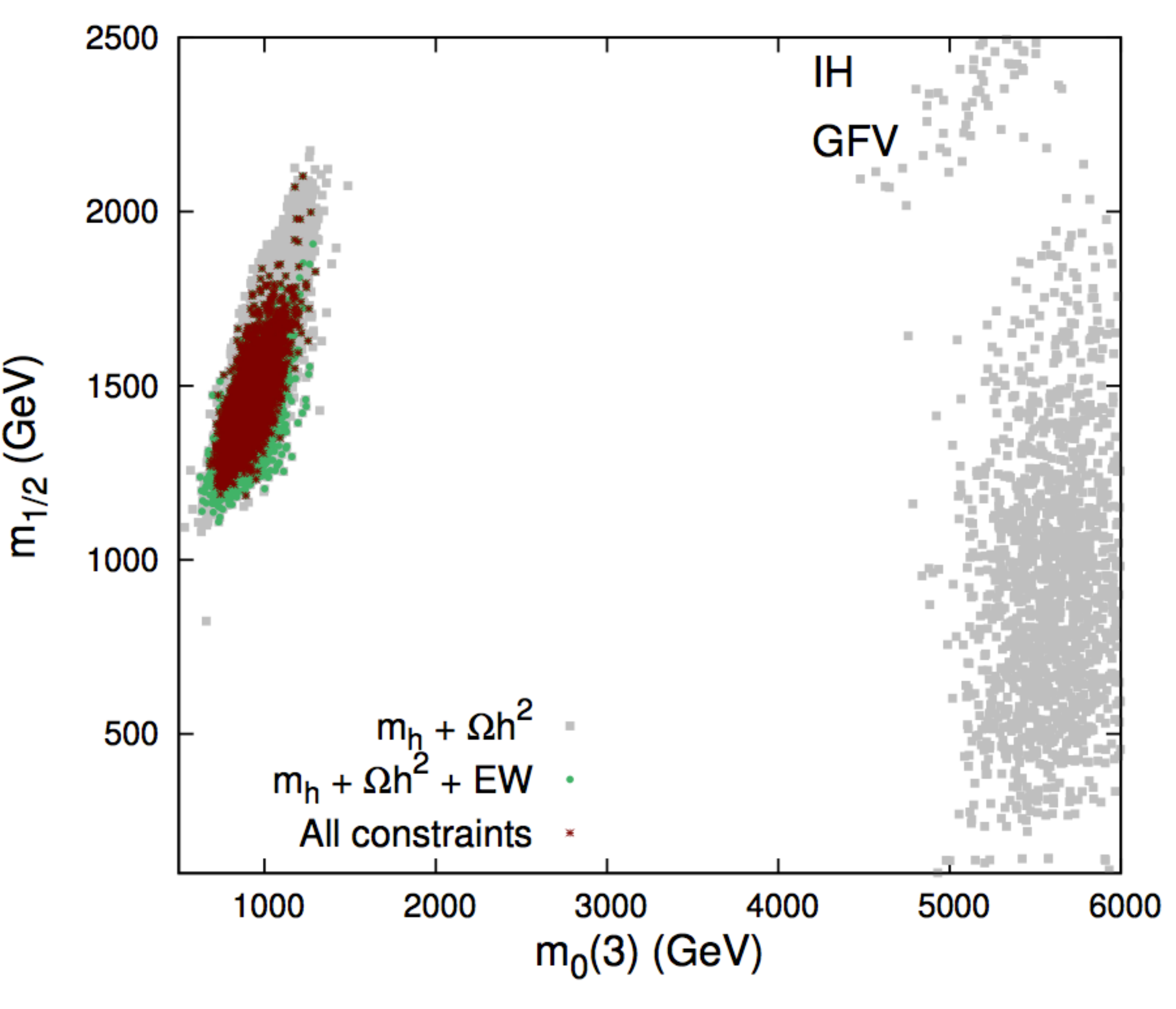}
}
\subfloat[]{
\label{fig:b}
\includegraphics[width=0.42\textwidth]{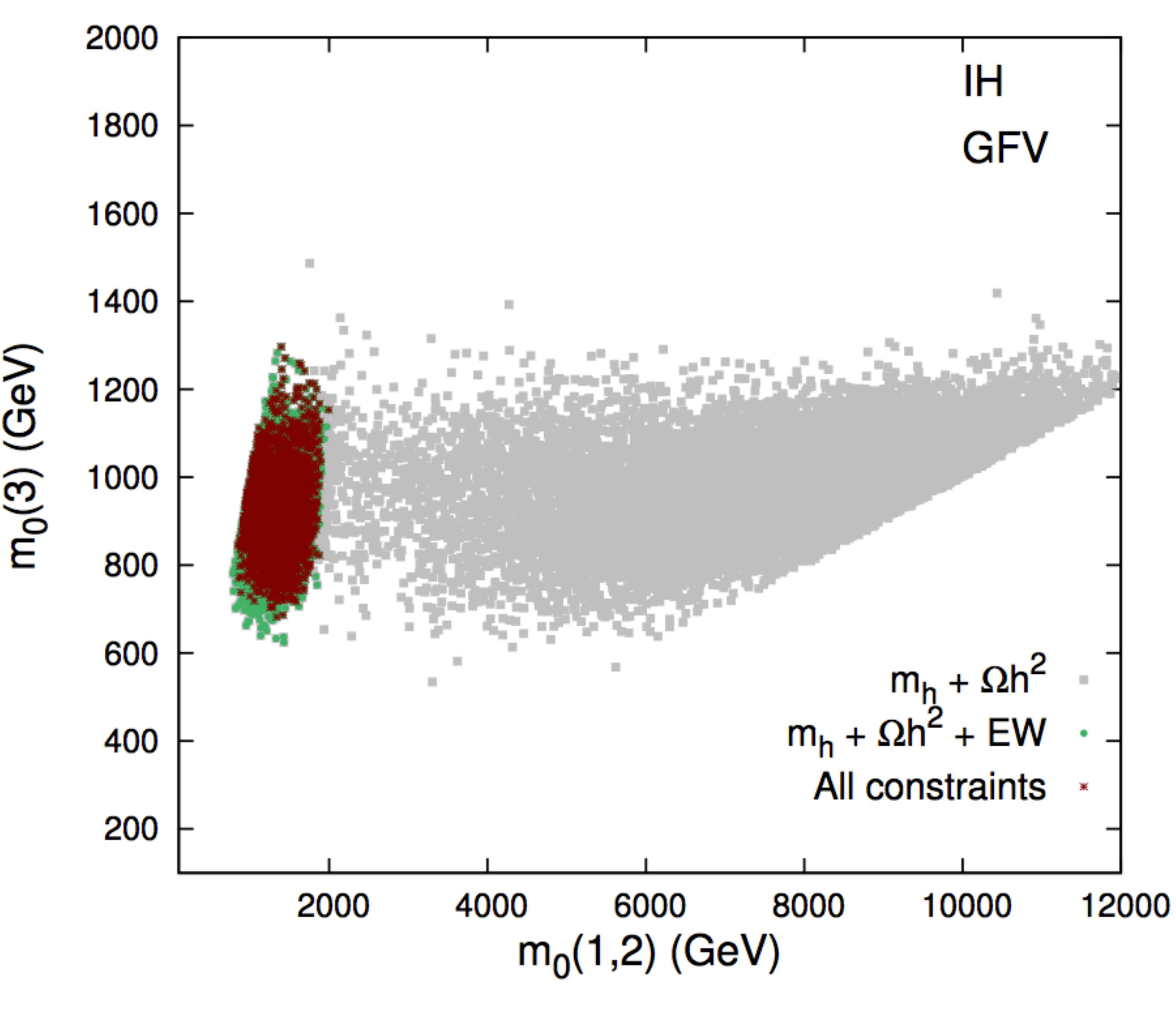}
}\\
\subfloat[]{
\label{fig:c}
\includegraphics[width=0.42\textwidth]{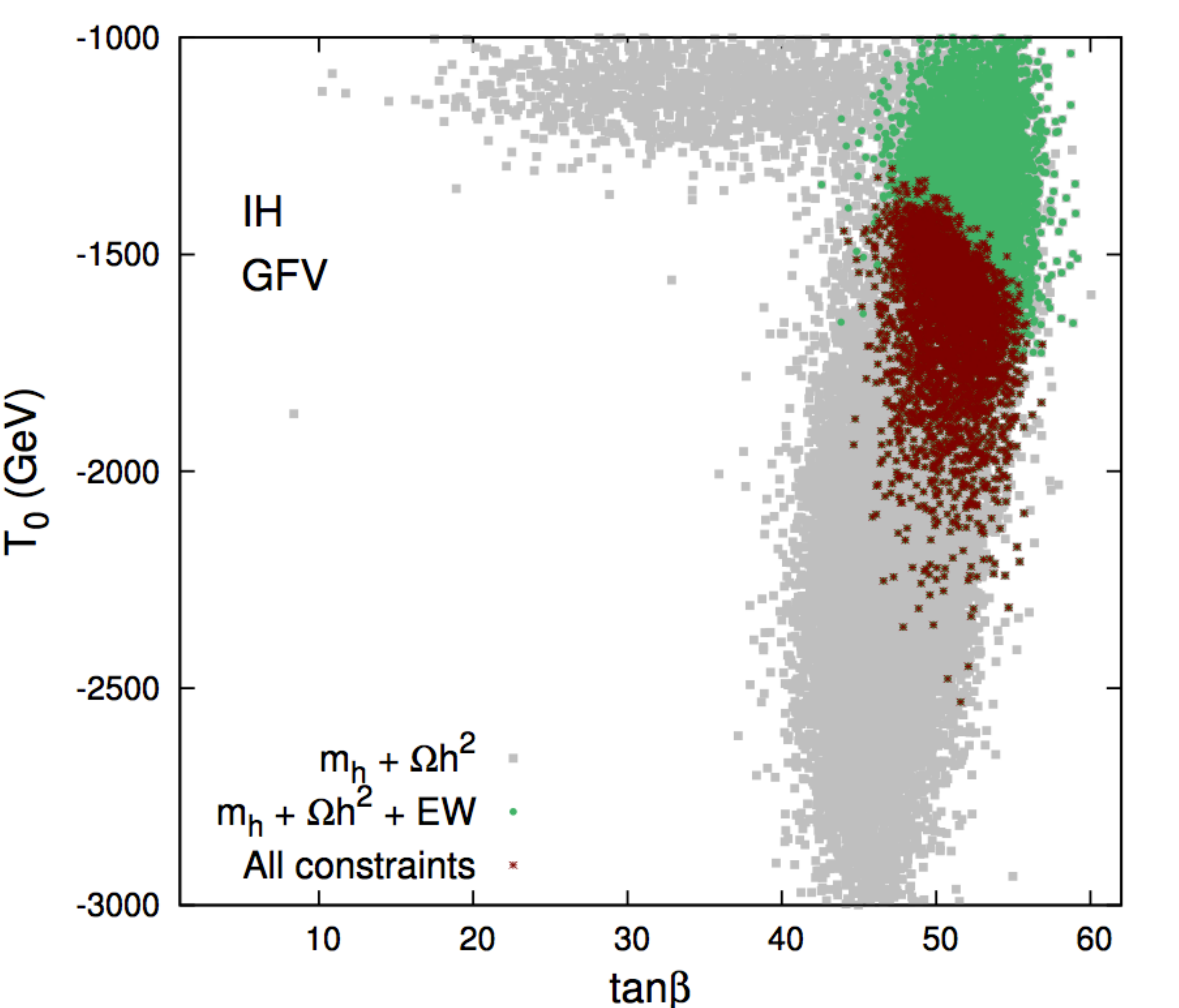}
}
\caption{\footnotesize Scatter plots of the IH scenario points in the planes of \protect\subref{fig:a} ($m_0(3)$, $\mhalf$), \protect\subref{fig:b} ($m_0(1,2)$, $m_0(3)$), and \protect\subref{fig:c} ($\tanb$, $T_0$). The grey squares correspond to the points satisfying the $2\sigma$ bounds on the Higgs mass and the relic density, the green circles additionally satisfy the $2\sigma$ bound on the EW precision observables, while the brown crosses all the constraints listed in Table~\ref{tab:exp_constraints}.
}
\label{fig:scan_1}
\end{figure}

In \reffig{fig:scan_1} we show the distribution of the model points in the planes corresponding to the input parameters: \subref{fig:a} ($m_0(3)$, $\mhalf$), \subref{fig:b} ($m_0(1,2)$, $m_0(3)$), \subref{fig:c} ($\tanb$, $T_0$). The points presented as grey squares satisfy $2\sigma$ bounds on the Higgs mass and the relic density, while the ones depicted as green circles additionally belong to $2\sigma$ acceptance region for the EW precision observables \mw\ and \sineff. Finally, the brown stars correspond to those points that satisfy at $2\sigma$ all the experimental constraints listed in Table~\ref{tab:exp_constraints}.

The requirement of satisfying the bound on the relic density is usually the most stringent one. The only dark matter candidate in the analysed scenario is the lightest neutralino. In the narrow region spanned around $\mzero\approx1\tev$ it is predominantly bino-like and the proper value of the relic density is obtained through neutralino co-annihilation with the lightest slepton (stau). The corresponding value of \tanb\ is limited to $40-60$, while the off-diagonal entry $T_0$ remains unconstrained. On the other hand, in a vast region with \mzero\ between $5-6\tev$ neutralino is a mixture of bino and higgsinos. Subsequently, the annihilation cross-section is enhanced by a non-zero higgsino component that opens a possibility of efficient annihilation into gauge bosons through a t-channel exchange of higgsino-like \charone\ and \neuttwo. In \reffig{fig:scan_1}\subref{fig:c} this region corresponds to a narrow strip around $T_0=-1200\gev$ spanned over the wide range of \tanb, and it is not shown in \reffig{fig:scan_1}\subref{fig:b}. 

The situation severely changes after imposing constraints from the EW precision measurements. The whole part of the parameter space corresponding to $\mzero(3)>1500\gev$ and $\mzero(1,2)>2000\gev$ is now disfavoured at $2\sigma$ level. This is a direct consequence of abandoning the MFV assumption. After the matrix \mqlij\ is rotated to the SCKM-basis, an off-diagonal element $((\muu^2)_{LL})_{23}\sim \lambda^2(\tilde{m}_2^2-\tilde{m}_3^2)$ is generated. Since in the inverted hierarchy scenario the corresponding mass difference between the diagonal entries is large, the flavour violating term $((\muu^2)_{LL})_{23}$ is substantial and induces large 1-loop corrections to the electroweak parameter $\rho$, and consequently to \mw\ and \sineff, as discussed in details in Ref.\cite{Heinemeyer:2004by}. One can also observe an upper bound on the gaugino mass term $\mhalf<1800\gev$ in the \stauc\ region that translates into a lower bound on $T_0 >-2500\gev$. It results from the fact that as long as $\mzero(3)$ remains low, the squark soft masses become much more strongly renormalised when \mhalf\ increases. In effect, \msusy\ also increases and the impact of the off-diagonal entry $((\muu^2)_{LL})_{23}$ is stronger. 

\begin{figure}[t]
\centering
\subfloat[]{
\label{fig:a}
\includegraphics[width=0.42\textwidth]{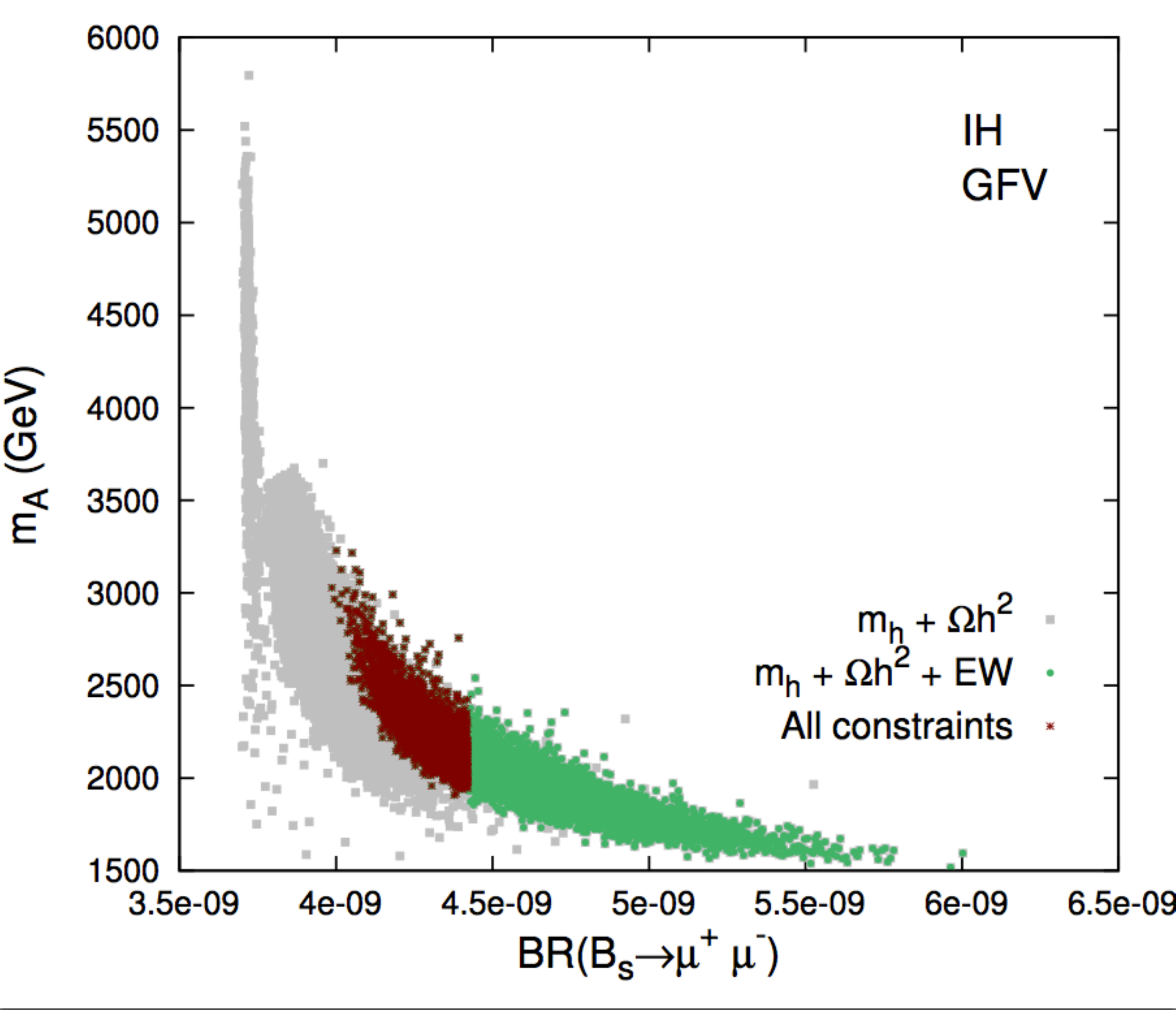}
}
\subfloat[]{
\label{fig:b}
\includegraphics[width=0.42\textwidth]{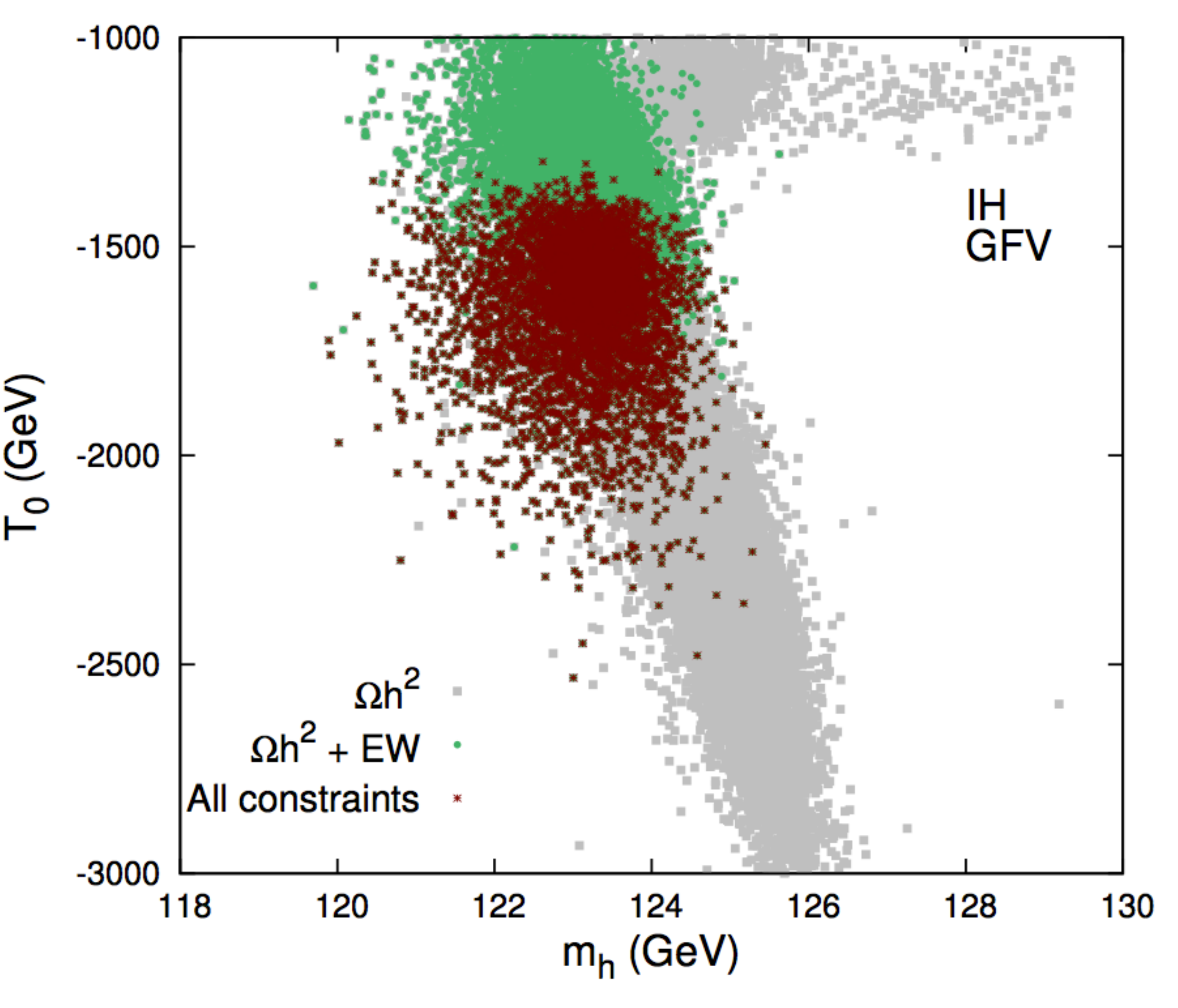}
}\\
\subfloat[]{
\label{fig:c}
\includegraphics[width=0.42\textwidth]{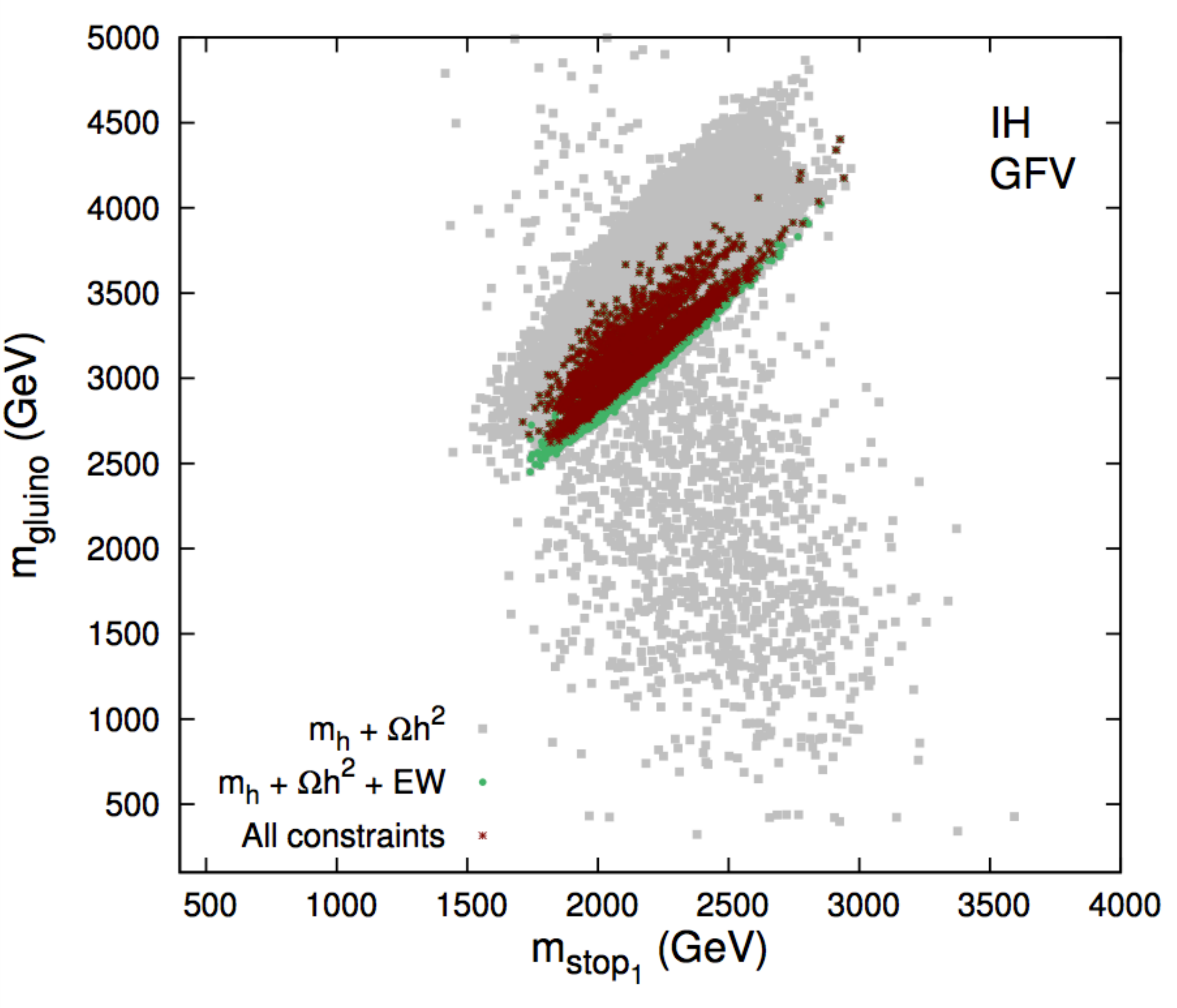}
}
\subfloat[]{
\label{fig:d}
\includegraphics[width=0.42\textwidth]{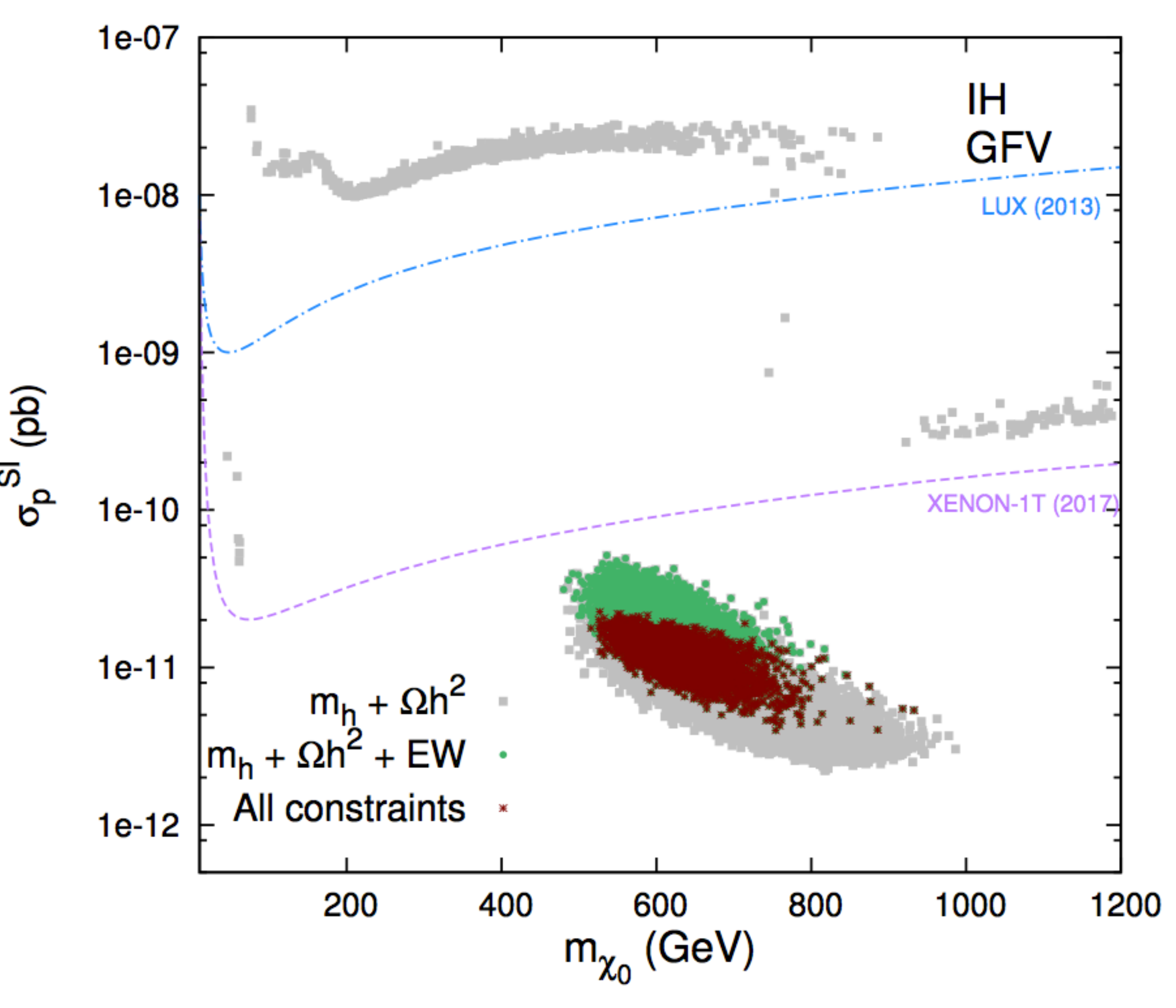}
}
\caption{\footnotesize Scatter plots of the IH scenario points in the planes of \protect\subref{fig:a} (\brbsmumu,\ma), \protect\subref{fig:b} ($T_0$,\mhl), \protect\subref{fig:c} ($m_\tone$,\mgluino), \protect\subref{fig:d} ($m_{\neutone}$,\sigsip). The colour code is the same as in \reffig{fig:scan_1}.}
\label{fig:scan_2}
\end{figure}

Finally, after all the experimental constraints listed in Table~\ref{tab:exp_constraints} are taken into account, the only parameter that becomes further affected is $T_0$, now disfavoured above $-1400\gev$. The main reason here is a tension with \brbsmumu\ measurement. A supersymmetric contribution to this branching ratio is driven by a factor $\brbsmumu\sim\frac{\tanb^6}{\ma^4}$ \cite{Bobeth:2001sq}. 
In \reffig{fig:scan_2}\subref{fig:a} we show a distribution of the analysed points in the $(\brbsmumu,\ma)$ plane. Comparing this plot with \reffig{fig:scan_1}\subref{fig:c} one can observe that in the \stauc\ region mass of the pseudoscalar $A$ significantly decreases when $|T_0|$ becomes smaller, while in the mixed bino-higgsino region it stays always above $2500\gev$. Additionally, in the former case \tanb\ is bound to be large. Remembering that $\ma^2\sim -\mhu^2$ for large \tanb, this behaviour can be qualitatively understood from an approximate relation between the value of $\mhu^2$ at the EW-scale and the \gut-scale input parameters:
\bea\label{rges_mhu}
(\mhu^2)(\mew)\simeq& -&0.96\,\mhalf^2-0.40\,\azero^2-0.38\,T^2_0+0.61\,\azero\mhalf \nonumber\\
&+&0.02\,\mzero^2(3)+0.003\,\mzero^2(1,2),
\eea
which was derived for a sample point $\mhalf=1500\gev$, $\mzero(3)=1000\gev$, $\mzero(1,2)=1000\gev$, $T_0=-1500\gev$, $\tanb=50$ and $\azero=0$. It is now clear from \refeq{rges_mhu} that the mass of a CP-odd scalar is enhanced when the value of GFV entry $T_0$ increases, making it easier to accommodate predictions of the model within the experimental bounds. In particular, we checked in a supplementary scan that when $T_0=0$, the whole parameter space of the analysed scenario is disfavoured by the  \brbsmumu\ constraint.  

We would like to emphasise the importance of this effect. When the MFV assumption is abandoned, the experimental measurement of the branching ratio for rare decay \bsmumu\ strongly disfavours inverted hierarchy scenarios unless large $T_0$ is introduced. Interestingly, the same parameter can be responsible for enhancement of the Higgs boson mass without the need of introducing large stop mixing. 

We conclude this section with a brief discussion of the most relevant phenomenological features characterising the considered model. In \reffig{fig:scan_2} we present distributions of points in the planes of \subref{fig:a} (\brbsmumu,\ma), \subref{fig:b} ($T_0$,\mhl), \subref{fig:c} ($m_\tone$,\mgluino), and \subref{fig:d} ($m_{\neutone}$,\sigsip). The colour code is the same as in \reffig{fig:scan_1}. From  \reffig{fig:scan_2}\subref{fig:b} one can see that for $\mzero(3)$ fixed at $\sim~1000\gev$ the Higgs mass in the ballpark of $126\gev$ can be easily obtained by increasing $T_0$, as discussed in the previous sections. However, the bounds from the EW precision measurements strongly disfavour the region of $\mhl>126\gev$. 

The allowed spectrum of SUSY particles is characterised by the lightest up-squarks heavier than $1600\gev$ and gluinos heavier than $2500\gev$. While far beyond the reach of the LHC 8\tev, spectra of that kind might be still accessible by the LHC 14\tev\ with 3000\invfb\ of collected data. Bino-like neutralino can be as light as 500\gev\ in this scenario, therefore multi-jet signatures with large amount of missing energy seem to be a promising tool for testing such spectra in the proton colliders. On the other hand, they will probably remain beyond the reach of direct dark matter detection experiments, as the corresponding spin-independent proton-neutralino cross-section is in general too low to be tested. This can be inferred from \reffig{fig:scan_2}\subref{fig:d} where the dashed lines show the 90\% \cl\ exclusion bound by LUX\cite{Akerib:2013tjd} (blue line) and a projected sensitivity by XENON1T\cite{Aprile:2012zx} (purple line). 

\section{Conclusions}\label{sec:concl}

In this study we analysed the phenomenology of SUSY scenarios beyond the Minimal Flavour Violation. As the general topic is very broad, we concentrated on the effects from the up-squark sector only. 

Firstly, we investigated the possibility of enhancing the Higgs boson mass through non-zero entries in the SSB squark matrices that generate additional contributions to the one-loop scalar potential. The largest effect was achieved through non-zero (2,3), (3,2), (1,3) and (3,1) entries of the up-squark trilinear coupling in the absence of mixing in the stop sector. Interestingly, the GFV parameters responsible for the \mhl\ enhancement are not very strongly constrained neither by the FCNC transitions nor by the vacuum stability bounds, and therefore can be kept relatively large. We also showed that the effect can be stronger in a class of models where the first/second generation of the up squarks is heavier than the third one.

Secondly, we analysed the phenomenology of a class of GUT-constrained models that assume inverted mass hierarchy in the squark sector. Such scenarios are particularly sensitive to GFV effects as non-degenerate squarks of the second and third generations gives rise to a large flavour violating entry $((\muu^2)_{LL})_{23}$ after the rotation to the SCKM-basis. Although it does not affect the value of the Higgs boson mass, the presence of such an entry has other important consequences. First of all, it induces large loop contributions to the EW precision observables \mw\ and \sineff. Consequently, the allowed parameter space of the model is strongly limited to the values of $\mzero(3)$ below $1500\gev$. Secondly, the same entry strongly reduces the mass of the CP-odd Higgs boson in the \stauc\ strip, leading to dangerously large enhancement of \brbsmumu. This effect, however, can be counterbalanced by the presence of a large non-zero (2,3) entry in the SSB trilinear matrix of the up squarks. 

The latter observation suggests that once the MFV assumption is lifted, a no-trivial GFV structure in the up squark sector is required for all the experimental constraints to be satisfied. Moreover, it was shown in Ref.\cite{Blanke:2013uia} that mixing between the right-handed charm and top squarks can reduce the experimental LHC bound on the stop mass, explaining the hitherto null result of the SUSY direct searches.

This might be a hint that supersymmetry, if realised in nature, should be considered in a more general, flavour violating framework.
\bigskip
\bigskip
\begin{center}
\textbf{ACKNOWLEDGMENTS}
\end{center}

  I would like to thank Werner Porod for claryfing some issues related to usage of \spheno\ in the GFV framework.
  I am also very grateful to Enrico Maria Sessolo for many helpful discussions and comments on this manuscript. 
  This work has been supported by the EU and MSHE Grant No. POIG.02.03.00-00-013/09.
  The use of the CIS computer cluster at the National Centre for Nuclear Research is gratefully acknowledged.
\bigskip
\bigskip
\appendix
\section{Derivation of the formulae for $\Delta\mhl$}\label{sec:appen}
In this appendix we present the details of our derivation of the analytical formulae for the GFV contributions to the lightest Higgs scalar mass shown in Sec.\ref{sec:higgs}. The approach is based on Section 6 of Ref.\cite{Haber:1993an}.

We start with defining a one-loop corrected scalar potential of the MSSM as
\be
V_1=V_0+\Delta V_1,
\ee
where $V_0$ is the tree-level potential and the one-loop radiative contribution in the dimensional reduction ($\overline{DR}$) scheme is given by a well know Coleman-Weinberg formula\cite{Coleman:1973jx}
\be\label{deltav1}
\Delta V_1=\frac{1}{64\pi^2}\textrm{Str}\left[\mathcal{M}^4\left(\ln\frac{\mathcal{M}^2}{Q^2}-\frac{3}{2}\right)\right],
\ee
where $Q$ denotes the renormalization scale, the supertrace is defined as\linebreak  $\textrm{Str} F(\mathcal{M})=\sum_i C_i(-1)^{2s_i}(2s_i+1)F(m_i^2)$, and the sum goes over all the states in the theory. For a particle $i$, $C_i$ denotes the colour degrees of freedom and $s_i$ the spin.
The mass matrix of the CP-even neutral Higgs bosons is given by the second derivatives of the one-loop corrected effective potential in its minimum,
\be\label{mhmatrix}
\mathcal{M}^2_{h^0,H^0}=\frac{1}{2}\left( \begin{array}{cc} \frac{\partial^2V_1}{\partial H_2\partial H_2^*} &  \frac{\partial^2V_1}{\partial H_2\partial H_1^*} \\  \frac{\partial^2V_1}{\partial H_1\partial H_2^*} &  \frac{\partial^2V_1}{\partial H_1\partial H_1^*}  \end{array}\right)\Bigg|_{\textrm{vev}}.
\ee
In the decouplig limit, $\ma\gg \mz$ where $A$ is the CP-odd scalar, the mass of the lightest Higgs boson is given by:
\be\label{mhmass}
\mhl^2=(\mathcal{M}^2_{h^0,H^0})_{11}\cos^2\beta+(\mathcal{M}^2_{h^0,H^0})_{22}\sin^2\beta+(\mathcal{M}^2_{h^0,H^0})_{12}\sin2\beta.
\ee

Let us only consider the contribution to the scalar potential from the Higgs field and the up-squark sector. In such a simple case the only mass matrix entering \refeq{deltav1} is $\mathcal{M}^2_{\tilde{u}}$ defined by \refeq{mu2approx},
\be\label{v1approx}
\Delta V_1=\frac{3}{64\pi^2}\textrm{tr}\left[\mathcal{M}_{\tilde{u}}^4\left(\ln\frac{\mathcal{M}_{\tilde{u}}^2}{Q^2}-\frac{3}{2}\right)-2\mtop^4\left(\ln\frac{\mtop^2}{Q^2}-\frac{3}{2}\right)\right].
\ee
It is now convenient to decompose the matrix $\mathcal{M}^2_{\tilde{u}}$ as $\mathcal{M}^2_{\tilde{u}}=\tilde{m}^2\mathbb{I}+\mathcal{M}^2_T+\mathcal{M}^2_F$ (where the first term refers to the SSB mass contribution, $\mathcal{M}^2_T$ to the trilinear term contribution and $\mathcal{M}^2_F$ to the F-term superpotential contribution, respectively) and expand \refeq{v1approx} up to the fourth power of $H_2$. Keeping only the quartic terms one gets:
\be\label{v1expand}
\Delta V_1=\frac{3}{64\pi^2}\left[\ln\frac{\tilde{m}^2}{Q^2}\textrm{tr}(\mathcal{M}^2_F)^2+\frac{1}{\tilde{m}^2}\textrm{tr}((\mathcal{M}^2_T)^2\mathcal{M}^2_F)-\frac{1}{12\tilde{m}^4}\textrm{tr}(\mathcal{M}^2_T)^4-2\mtop^4\left(\ln\frac{\mtop^2}{Q^2}-\frac{3}{2}\right)\right].
\ee
Since $\mathcal{M}^2_F=\mtop^2\mathbb{I}_2$, the first and the fourth term of \refeq{v1expand} combine to reproduce the one-loop leading logarithm term (where we used the minimalization condition $\partial \Delta V_1/\partial H_2=0$):
\be
\Delta(\mhl^2)_{LL}=\frac{3}{8\pi^2 v^2}Y_t^4v_2^4\ln\frac{\tilde{m}^2}{\mtop^2}.
\ee
The remaining terms define the non-logarithmic finite corrections to the scalar potential which in general depends on the full structure of the mass matrix (\ref{mu2approx}):
\be\label{finite}
(\Delta V_1)_{\textrm{finite}}=\frac{3}{64\pi^2}\frac{1}{\tilde{m}^2}\left[\textrm{tr}((\mathcal{M}^2_T)^2\mathcal{M}^2_F)-\frac{1}{12\tilde{m}^2}\textrm{tr}(\mathcal{M}^2_T)^4\right].
\ee
As a first example let us consider the case when $\dulrbc\ne 0$. The terms in \refeq{finite} are then equal $\textrm{tr}(\mathcal{M}^2_T)^4=2(T_u)^4_{23}H_2^4$ and $\textrm{tr}((\mathcal{M}^2_T)^2\mathcal{M}^2_F)=(T_u)^2_{23}H_2^4(Y_t^2+Y_c^2)$. Using \refeq{mhmatrix} and \refeq{mhmass} and neglecting the charm Yukawa coupling the GFV contribution to the Higgs mass takes the form 
\bea
\Delta\mhl^2(\dulrbc)&=&\frac{3v^4_2}{8\pi^2v^2}\left[\frac{(T_u)^{2}_{23}}{\tilde{m}^2}\left(\frac{1}{2}Y_t^2-\frac{(T_u)^{2}_{23}}{12\tilde{m}^2}\right)\right]\nonumber\\
&=&\frac{3}{4\pi^2}\sin^2\beta\left[\tilde{m}^2\dulrbc^2\left(\frac{1}{2}Y_t^2-\frac{\tilde{m}^2\dulrbc^2}{6v^2_2}\right)\right].
\eea

As a second example let us evaluate the impact of non-zero stop mixing parameter $\tilde{X}_t=(T_{u})_{33}$. In such a case the non-zero eigenvalues of the matrix $(\mathcal{M}^2_T)^4$ are again degenerate and read $\lambda_{1,2}=((T_u)^{2}_{23}+(T_u)^{2}_{33})^2H_2^4$. On the other hand, for the matrix $(\mathcal{M}^2_T)^2\mathcal{M}^2_F$ one obtains $\lambda_{1}=((T_u)^{2}_{23}+(T_u)^{2}_{33})H_2^4Y_t^2$, $\lambda_{2}=(T_u)^{2}_{33}H_2^4Y_t^2$. Therefore the traces give $\textrm{tr}(\mathcal{M}^2_T)^4=2((T_u)^{2}_{23}+(T_u)^{2}_{33})^2H_2^4$ and $\textrm{tr}((\mathcal{M}^2_T)^2\mathcal{M}^2_F)=(2(T_u)^2_{33}+(T_u)^2_{23})H_2^4Y_t^2$, and the one-loop correction to the Higgs boson mass takes the form
\bea
\Delta\mhl^2&=&\frac{3v^4_2}{8\pi^2v^2\tilde{m}^2}\left[Y_t^2\Big((T_u)^2_{33}+\frac{1}{2}(T_u)^2_{23}\Big)-\frac{1}{12\tilde{m}^2}\Big((T_u)^4_{33}+(T_u)^4_{23}+2(T_u)^2_{33}(T_u)^2_{23}\Big)\right]\\
&=&\frac{3}{4\pi^2}\tilde{m}^2\left[Y_t^2\sin^2\beta\Big((\delta^u_{33})^2+\frac{1}{2}\dulrbc^2\Big)-\frac{\tilde{m}^2}{6v^2}\Big((\delta^u_{33})^4+\dulrbc^4+2(\delta^u_{33})^2\dulrbc^2\Big)\right].\nonumber
\eea

\bigskip
\bigskip

\bibliographystyle{JHEP}

\bibliography{myref}

\providecommand{\href}[2]{#2}\begingroup\raggedright\begin{thebibliography}{10}

\bibitem{Chatrchyan:2012ufa}
{\bf CMS Collaboration} Collaboration, S.~Chatrchyan et~al., {\it {Observation
  of a new boson at a mass of 125 GeV with the CMS experiment at the LHC}},
  {\em Phys.Lett.} {\bf B716} (2012) 30--61,
  [\href{http://xxx.lanl.gov/abs/1207.7235}{{\tt arXiv:1207.7235}}].

\bibitem{Aad:2012tfa}
{\bf ATLAS Collaboration} Collaboration, G.~Aad et~al., {\it {Observation of a
  new particle in the search for the Standard Model Higgs boson with the ATLAS
  detector at the LHC}},  {\em Phys.Lett.} {\bf B716} (2012) 1--29,
  [\href{http://xxx.lanl.gov/abs/1207.7214}{{\tt arXiv:1207.7214}}].

\bibitem{CMS-PAS-HIG-13-005}
{\bf CMS} Collaboration, {\it {Combination of standard model Higgs boson
  searches and measurements of the properties of the new boson with a mass near
  125 GeV}},  Tech. Rep. CMS-PAS-HIG-13-005, CERN, Geneva, 2013.

\bibitem{Haber:1996fp}
H.~E. Haber, R.~Hempfling, and A.~H. Hoang, {\it {Approximating the radiatively
  corrected Higgs mass in the minimal supersymmetric model}},  {\em Z.Phys.}
  {\bf C75} (1997) 539--554,
  [\href{http://xxx.lanl.gov/abs/hep-ph/9609331}{{\tt hep-ph/9609331}}].

\bibitem{Fowlie:2012im}
A.~Fowlie, M.~Kazana, K.~Kowalska, S.~Munir, L.~Roszkowski, et~al., {\it {The
  CMSSM Favoring New Territories: The Impact of New LHC Limits and a 125 GeV
  Higgs}},  {\em Phys.Rev.} {\bf D86} (2012) 075010,
  [\href{http://xxx.lanl.gov/abs/1206.0264}{{\tt arXiv:1206.0264}}].

\bibitem{Akula:2012kk}
S.~Akula, P.~Nath, and G.~Peim, {\it {Implications of the Higgs Boson Discovery
  for mSUGRA}},  {\em Phys.Lett.} {\bf B717} (2012) 188--192,
  [\href{http://xxx.lanl.gov/abs/1207.1839}{{\tt arXiv:1207.1839}}].

\bibitem{Beskidt:2012sk}
C.~Beskidt, W.~de~Boer, D.~Kazakov, and F.~Ratnikov, {\it {Constraints on
  Supersymmetry from LHC data on SUSY searches and Higgs bosons combined with
  cosmology and direct dark matter searches}},  {\em Eur.Phys.J.} {\bf C72}
  (2012) 2166, [\href{http://xxx.lanl.gov/abs/1207.3185}{{\tt
  arXiv:1207.3185}}].

\bibitem{Buchmueller:2012hv}
O.~Buchmueller, R.~Cavanaugh, M.~Citron, A.~De~Roeck, M.~Dolan, et~al., {\it
  {The CMSSM and NUHM1 in Light of 7 TeV LHC, Bs to mu+mu- and XENON100 Data}},
   {\em Eur.Phys.J.} {\bf C72} (2012) 2243,
  [\href{http://xxx.lanl.gov/abs/1207.7315}{{\tt arXiv:1207.7315}}].

\bibitem{Altmannshofer:2012ks}
W.~Altmannshofer, M.~Carena, N.~R. Shah, and F.~Yu, {\it {Indirect Probes of
  the MSSM after the Higgs Discovery}},  {\em JHEP} {\bf 1301} (2013) 160,
  [\href{http://xxx.lanl.gov/abs/1211.1976}{{\tt arXiv:1211.1976}}].

\bibitem{Strege:2012bt}
C.~Strege, G.~Bertone, F.~Feroz, M.~Fornasa, R.~Ruiz~de Austri, et~al., {\it
  {Global Fits of the cMSSM and NUHM including the LHC Higgs discovery and new
  XENON100 constraints}},  {\em JCAP} {\bf 1304} (2013) 013,
  [\href{http://xxx.lanl.gov/abs/1212.2636}{{\tt arXiv:1212.2636}}].

\bibitem{Cabrera:2012vu}
M.~E. Cabrera, J.~A. Casas, and R.~R. de~Austri, {\it {The health of SUSY after
  the Higgs discovery and the XENON100 data}},  {\em JHEP} {\bf 1307} (2013)
  182, [\href{http://xxx.lanl.gov/abs/1212.4821}{{\tt arXiv:1212.4821}}].

\bibitem{Kowalska:2013hha}
K.~Kowalska, L.~Roszkowski, and E.~M. Sessolo, {\it {Two ultimate tests of
  constrained supersymmetry}},  {\em JHEP} {\bf 1306} (2013) 078,
  [\href{http://xxx.lanl.gov/abs/1302.5956}{{\tt arXiv:1302.5956}}].

\bibitem{Badziak:2012rf}
M.~Badziak, E.~Dudas, M.~Olechowski, and S.~Pokorski, {\it {Inverted Sfermion
  Mass Hierarchy and the Higgs Boson Mass in the MSSM}},  {\em JHEP} {\bf 1207}
  (2012) 155, [\href{http://xxx.lanl.gov/abs/1205.1675}{{\tt
  arXiv:1205.1675}}].

\bibitem{Arbey:2012dq}
A.~Arbey, M.~Battaglia, A.~Djouadi, and F.~Mahmoudi, {\it {The Higgs sector of
  the phenomenological MSSM in the light of the Higgs boson discovery}},  {\em
  JHEP} {\bf 1209} (2012) 107, [\href{http://xxx.lanl.gov/abs/1207.1348}{{\tt
  arXiv:1207.1348}}].

\bibitem{Ellis:1986yg}
J.~R. Ellis, K.~Enqvist, D.~V. Nanopoulos, and F.~Zwirner, {\it {Observables in
  Low-Energy Superstring Models}},  {\em Mod.Phys.Lett.} {\bf A1} (1986) 57.

\bibitem{Barbieri:1987fn}
R.~Barbieri and G.~Giudice, {\it {Upper Bounds on Supersymmetric Particle
  Masses}},  {\em Nucl.Phys.} {\bf B306} (1988) 63.

\bibitem{Cohen:1996vb}
A.~G. Cohen, D.~Kaplan, and A.~Nelson, {\it {The More minimal supersymmetric
  standard model}},  {\em Phys.Lett.} {\bf B388} (1996) 588--598,
  [\href{http://xxx.lanl.gov/abs/hep-ph/9607394}{{\tt hep-ph/9607394}}].

\bibitem{Heinemeyer:2004by}
S.~Heinemeyer, W.~Hollik, F.~Merz, and S.~Penaranda, {\it {Electroweak
  precision observables in the MSSM with nonminimal flavor violation}},  {\em
  Eur.Phys.J.} {\bf C37} (2004) 481--493,
  [\href{http://xxx.lanl.gov/abs/hep-ph/0403228}{{\tt hep-ph/0403228}}].

\bibitem{Herrmann:2011xe}
B.~Herrmann, M.~Klasen, and Q.~Le~Boulc'h, {\it {Impact of squark flavour
  violation on neutralino dark matter}},  {\em Phys.Rev.} {\bf D84} (2011)
  095007, [\href{http://xxx.lanl.gov/abs/1106.6229}{{\tt arXiv:1106.6229}}].

\bibitem{AranaCatania:2011ak}
M.~Arana-Catania, S.~Heinemeyer, M.~Herrero, and S.~Penaranda, {\it {Higgs
  Boson masses and B-Physics Constraints in Non-Minimal Flavor Violating SUSY
  scenarios}},  {\em JHEP} {\bf 1205} (2012) 015,
  [\href{http://xxx.lanl.gov/abs/1109.6232}{{\tt arXiv:1109.6232}}].

\bibitem{Arana-Catania:2014ooa}
M.~Arana-Catania, S.~Heinemeyer, and M.~Herrero, {\it {Updated Constraints on
  General Squark Flavor Mixing}},
  \href{http://xxx.lanl.gov/abs/1405.6960}{{\tt arXiv:1405.6960}}.

\bibitem{Cao:2006xb}
J.~Cao, G.~Eilam, K.-i. Hikasa, and J.~M. Yang, {\it {Experimental constraints
  on stop-scharm flavor mixing and implications in top-quark FCNC processes}},
  {\em Phys.Rev.} {\bf D74} (2006) 031701,
  [\href{http://xxx.lanl.gov/abs/hep-ph/0604163}{{\tt hep-ph/0604163}}].

\bibitem{Ade:2013zuv}
{\bf Planck Collaboration} Collaboration, P.~Ade et~al., {\it {Planck 2013
  results. XVI. Cosmological parameters}},  {\em Astron.Astrophys.} (2014)
  [\href{http://xxx.lanl.gov/abs/1303.5076}{{\tt arXiv:1303.5076}}].

\bibitem{Beringer:1900zz}
{\bf Particle Data Group} Collaboration, J.~Beringer et~al., {\it {Review of
  Particle Physics (RPP)}},  {\em Phys.Rev.} {\bf D86} (2012) 010001.

\bibitem{Allanach:2008qq}
B.~Allanach, C.~Balazs, G.~Belanger, M.~Bernhardt, F.~Boudjema, et~al., {\it
  {SUSY Les Houches Accord 2}},  {\em Comput.Phys.Commun.} {\bf 180} (2009)
  8--25, [\href{http://xxx.lanl.gov/abs/0801.0045}{{\tt arXiv:0801.0045}}].

\bibitem{feynhiggs:99}
S.~Heinemeyer, W.~Hollik, and G.~Weiglein, {\it {The Masses of the neutral CP -
  even Higgs bosons in the MSSM: Accurate analysis at the two loop level}},
  {\em Eur.Phys.J.} {\bf C9} (1999) 343--366,
  [\href{http://xxx.lanl.gov/abs/hep-ph/9812472}{{\tt hep-ph/9812472}}].

\bibitem{feynhiggs:00}
S.~Heinemeyer, W.~Hollik, and G.~Weiglein, {\it {FeynHiggs: A Program for the
  calculation of the masses of the neutral CP even Higgs bosons in the MSSM}},
  {\em Comput.Phys.Commun.} {\bf 124} (2000) 76--89,
  [\href{http://xxx.lanl.gov/abs/hep-ph/9812320}{{\tt hep-ph/9812320}}].

\bibitem{feynhiggs:03}
G.~Degrassi, S.~Heinemeyer, W.~Hollik, P.~Slavich, and G.~Weiglein, {\it
  {Towards high precision predictions for the MSSM Higgs sector}},  {\em
  Eur.Phys.J.} {\bf C28} (2003) 133--143,
  [\href{http://xxx.lanl.gov/abs/hep-ph/0212020}{{\tt hep-ph/0212020}}].

\bibitem{feynhiggs:06}
M.~Frank, T.~Hahn, S.~Heinemeyer, W.~Hollik, H.~Rzehak, et~al., {\it {The Higgs
  Boson Masses and Mixings of the Complex MSSM in the Feynman-Diagrammatic
  Approach}},  {\em JHEP} {\bf 0702} (2007) 047,
  [\href{http://xxx.lanl.gov/abs/hep-ph/0611326}{{\tt hep-ph/0611326}}].

\bibitem{Porod:2003um}
W.~Porod, {\it {SPheno, a program for calculating supersymmetric spectra, SUSY
  particle decays and SUSY particle production at e+ e- colliders}},  {\em
  Comput.Phys.Commun.} {\bf 153} (2003) 275--315,
  [\href{http://xxx.lanl.gov/abs/hep-ph/0301101}{{\tt hep-ph/0301101}}].

\bibitem{Porod:2011nf}
W.~Porod and F.~Staub, {\it {SPheno 3.1: Extensions including flavour,
  CP-phases and models beyond the MSSM}},  {\em Comput.Phys.Commun.} {\bf 183}
  (2012) 2458--2469, [\href{http://xxx.lanl.gov/abs/1104.1573}{{\tt
  arXiv:1104.1573}}].

\bibitem{Haber:1993an}
H.~E. Haber and R.~Hempfling, {\it {The Renormalization group improved Higgs
  sector of the minimal supersymmetric model}},  {\em Phys.Rev.} {\bf D48}
  (1993) 4280--4309, [\href{http://xxx.lanl.gov/abs/hep-ph/9307201}{{\tt
  hep-ph/9307201}}].

\bibitem{Ellis:1990nz}
J.~R. Ellis, G.~Ridolfi, and F.~Zwirner, {\it {Radiative corrections to the
  masses of supersymmetric Higgs bosons}},  {\em Phys.Lett.} {\bf B257} (1991)
  83--91.

\bibitem{Ellis:1991zd}
J.~R. Ellis, G.~Ridolfi, and F.~Zwirner, {\it {On radiative corrections to
  supersymmetric Higgs boson masses and their implications for LEP searches}},
  {\em Phys.Lett.} {\bf B262} (1991) 477--484.

\bibitem{Gabbiani:1996hi}
F.~Gabbiani, E.~Gabrielli, A.~Masiero, and L.~Silvestrini, {\it {A Complete
  analysis of FCNC and CP constraints in general SUSY extensions of the
  standard model}},  {\em Nucl.Phys.} {\bf B477} (1996) 321--352,
  [\href{http://xxx.lanl.gov/abs/hep-ph/9604387}{{\tt hep-ph/9604387}}].

\bibitem{Misiak:1997ei}
M.~Misiak, S.~Pokorski, and J.~Rosiek, {\it {Supersymmetry and FCNC effects}},
  {\em Adv.Ser.Direct.High Energy Phys.} {\bf 15} (1998) 795--828,
  [\href{http://xxx.lanl.gov/abs/hep-ph/9703442}{{\tt hep-ph/9703442}}].

\bibitem{Behring:2012mv}
A.~Behring, C.~Gross, G.~Hiller, and S.~Schacht, {\it {Squark Flavor
  Implications from B --> K* l+ l-}},  {\em JHEP} {\bf 1208} (2012) 152,
  [\href{http://xxx.lanl.gov/abs/1205.1500}{{\tt arXiv:1205.1500}}].

\bibitem{Colangelo:1998pm}
G.~Colangelo and G.~Isidori, {\it {Supersymmetric contributions to rare kaon
  decays: Beyond the single mass insertion approximation}},  {\em JHEP} {\bf
  9809} (1998) 009, [\href{http://xxx.lanl.gov/abs/hep-ph/9808487}{{\tt
  hep-ph/9808487}}].

\bibitem{Altmannshofer:2009ma}
W.~Altmannshofer, A.~J. Buras, D.~M. Straub, and M.~Wick, {\it {New strategies
  for New Physics search in $B \to K^{*} \nu \bar{\nu}$, $B \to K \nu
  \bar{\nu}$ and $B \to X_{s} \nu \bar{\nu}$ decays}},  {\em JHEP} {\bf 0904}
  (2009) 022, [\href{http://xxx.lanl.gov/abs/0902.0160}{{\tt
  arXiv:0902.0160}}].

\bibitem{CMS-PAS-HIG-13-034}
{\bf CMS Collaboration} Collaboration, {\it {Combined multilepton and diphoton
  limit on t to cH}},  Tech. Rep. CMS-PAS-HIG-13-034, CERN, Geneva, 2014.

\bibitem{Guasch:1999jp}
J.~Guasch and J.~Sola, {\it {FCNC top quark decays: A Door to SUSY physics in
  high luminosity colliders?}},  {\em Nucl.Phys.} {\bf B562} (1999) 3--28,
  [\href{http://xxx.lanl.gov/abs/hep-ph/9906268}{{\tt hep-ph/9906268}}].

\bibitem{Cao:2007dk}
J.~Cao, G.~Eilam, M.~Frank, K.~Hikasa, G.~Liu, et~al., {\it {SUSY-induced FCNC
  top-quark processes at the large hadron collider}},  {\em Phys.Rev.} {\bf
  D75} (2007) 075021, [\href{http://xxx.lanl.gov/abs/hep-ph/0702264}{{\tt
  hep-ph/0702264}}].

\bibitem{Crivellin:2012jv}
A.~Crivellin, J.~Rosiek, P.~Chankowski, A.~Dedes, S.~Jaeger, et~al., {\it
  {SUSY\textunderscore FLAVOR v2: A Computational tool for FCNC and
  CP-violating processes in the MSSM}},  {\em Comput.Phys.Commun.} {\bf 184}
  (2013) 1004--1032, [\href{http://xxx.lanl.gov/abs/1203.5023}{{\tt
  arXiv:1203.5023}}].

\bibitem{bsgamma}
{\url{http://www.slac.stanford.edu/xorg/hfag/rare/2012/radll/index.html}}.

\bibitem{Aaij:2013aka}
{\bf LHCb} Collaboration, R.~Aaij et~al., {\it {Measurement of the $B^0_s \to
  \mu^+ \mu^-$ branching fraction and search for $B^0 \to \mu^+ \mu^-$ decays
  at the LHCb experiment}},  {\em Phys.Rev.Lett.} {\bf 111} (2013) 101805,
  [\href{http://xxx.lanl.gov/abs/1307.5024}{{\tt arXiv:1307.5024}}].

\bibitem{Chatrchyan:2013bka}
{\bf CMS} Collaboration, S.~Chatrchyan et~al., {\it {Measurement of the $B_s^0
  \to \mu^+ \mu^-$ branching fraction and search for $B^0 \to \mu^+ \mu^-$ with
  the CMS Experiment}},  {\em Phys.Rev.Lett.} {\bf 111} (2013) 101804,
  [\href{http://xxx.lanl.gov/abs/1307.5025}{{\tt arXiv:1307.5025}}].

\bibitem{Crivellin:2008mq}
A.~Crivellin and U.~Nierste, {\it {Supersymmetric renormalisation of the CKM
  matrix and new constraints on the squark mass matrices}},  {\em Phys.Rev.}
  {\bf D79} (2009) 035018, [\href{http://xxx.lanl.gov/abs/0810.1613}{{\tt
  arXiv:0810.1613}}].

\bibitem{Crivellin:2009sd}
A.~Crivellin, {\it {Effects of right-handed charged currents on the
  determinations of V(ub) and V(cb)}},  {\em Phys.Rev.} {\bf D81} (2010)
  031301, [\href{http://xxx.lanl.gov/abs/0907.2461}{{\tt arXiv:0907.2461}}].

\bibitem{Crivellin:2011jt}
A.~Crivellin, L.~Hofer, and J.~Rosiek, {\it {Complete resummation of
  chirally-enhanced loop-effects in the MSSM with non-minimal sources of
  flavor-violation}},  {\em JHEP} {\bf 1107} (2011) 017,
  [\href{http://xxx.lanl.gov/abs/1103.4272}{{\tt arXiv:1103.4272}}].

\bibitem{Frere:1983ag}
J.~Frere, D.~Jones, and S.~Raby, {\it {Fermion Masses and Induction of the Weak
  Scale by Supergravity}},  {\em Nucl.Phys.} {\bf B222} (1983) 11.

\bibitem{AlvarezGaume:1983gj}
L.~Alvarez-Gaume, J.~Polchinski, and M.~B. Wise, {\it {Minimal Low-Energy
  Supergravity}},  {\em Nucl.Phys.} {\bf B221} (1983) 495.

\bibitem{Derendinger:1983bz}
J.~Derendinger and C.~A. Savoy, {\it {Quantum Effects and SU(2) x U(1) Breaking
  in Supergravity Gauge Theories}},  {\em Nucl.Phys.} {\bf B237} (1984) 307.

\bibitem{Kounnas:1983td}
C.~Kounnas, A.~Lahanas, D.~V. Nanopoulos, and M.~Quiros, {\it {Low-Energy
  Behavior of Realistic Locally Supersymmetric Grand Unified Theories}},  {\em
  Nucl.Phys.} {\bf B236} (1984) 438.

\bibitem{Casas:1996de}
J.~Casas and S.~Dimopoulos, {\it {Stability bounds on flavor violating
  trilinear soft terms in the MSSM}},  {\em Phys.Lett.} {\bf B387} (1996)
  107--112, [\href{http://xxx.lanl.gov/abs/hep-ph/9606237}{{\tt
  hep-ph/9606237}}].

\bibitem{Park:2010wf}
J.-h. Park, {\it {Metastability bounds on flavour-violating trilinear soft
  terms in the MSSM}},  {\em Phys.Rev.} {\bf D83} (2011) 055015,
  [\href{http://xxx.lanl.gov/abs/1011.4939}{{\tt arXiv:1011.4939}}].

\bibitem{Martin:1993zk}
S.~P. Martin and M.~T. Vaughn, {\it {Two loop renormalization group equations
  for soft supersymmetry breaking couplings}},  {\em Phys.Rev.} {\bf D50}
  (1994) 2282, [\href{http://xxx.lanl.gov/abs/hep-ph/9311340}{{\tt
  hep-ph/9311340}}].

\bibitem{Adachi:2012mm}
{\bf Belle} Collaboration, I.~Adachi et~al., {\it {Measurement of $B^- \to
  \tau^- \bar{\nu}_\tau$ with a Hadronic Tagging Method Using the Full Data
  Sample of Belle}},  {\em Phys.Rev.Lett.} {\bf 110} (2013) 131801,
  [\href{http://xxx.lanl.gov/abs/1208.4678}{{\tt arXiv:1208.4678}}].

\bibitem{Feroz:2008xx}
F.~Feroz, M.~Hobson, and M.~Bridges, {\it {MultiNest: an efficient and robust
  Bayesian inference tool for cosmology and particle physics}},  {\em
  Mon.Not.Roy.Astron.Soc.} {\bf 398} (2009) 1601--1614,
  [\href{http://xxx.lanl.gov/abs/0809.3437}{{\tt arXiv:0809.3437}}].

\bibitem{Gondolo:2004sc}
P.~Gondolo, J.~Edsjo, P.~Ullio, L.~Bergstrom, M.~Schelke, et~al., {\it
  {DarkSUSY: Computing supersymmetric dark matter properties numerically}},
  {\em JCAP} {\bf 0407} (2004) 008,
  [\href{http://xxx.lanl.gov/abs/astro-ph/0406204}{{\tt astro-ph/0406204}}].

\bibitem{Bechtle:2008jh}
P.~Bechtle, O.~Brein, S.~Heinemeyer, G.~Weiglein, and K.~E. Williams, {\it
  {HiggsBounds: Confronting Arbitrary Higgs Sectors with Exclusion Bounds from
  LEP and the Tevatron}},  {\em Comput.Phys.Commun.} {\bf 181} (2010) 138--167,
  [\href{http://xxx.lanl.gov/abs/0811.4169}{{\tt arXiv:0811.4169}}].

\bibitem{Bechtle:2011sb}
P.~Bechtle, O.~Brein, S.~Heinemeyer, G.~Weiglein, and K.~E. Williams, {\it
  {HiggsBounds 2.0.0: Confronting Neutral and Charged Higgs Sector Predictions
  with Exclusion Bounds from LEP and the Tevatron}},  {\em Comput.Phys.Commun.}
  {\bf 182} (2011) 2605--2631, [\href{http://xxx.lanl.gov/abs/1102.1898}{{\tt
  arXiv:1102.1898}}].

\bibitem{Bechtle:2013wla}
P.~Bechtle, O.~Brein, S.~Heinemeyer, O.~Stål, T.~Stefaniak, et~al., {\it
  {$\mathsf{HiggsBounds}-4$: Improved Tests of Extended Higgs Sectors against
  Exclusion Bounds from LEP, the Tevatron and the LHC}},  {\em Eur.Phys.J.}
  {\bf C74} (2014) 2693, [\href{http://xxx.lanl.gov/abs/1311.0055}{{\tt
  arXiv:1311.0055}}].

\bibitem{Bechtle:2013xfa}
P.~Bechtle, S.~Heinemeyer, O.~Stål, T.~Stefaniak, and G.~Weiglein, {\it
  {$HiggsSignals$: Confronting arbitrary Higgs sectors with measurements at the
  Tevatron and the LHC}},  {\em Eur.Phys.J.} {\bf C74} (2014) 2711,
  [\href{http://xxx.lanl.gov/abs/1305.1933}{{\tt arXiv:1305.1933}}].

\bibitem{Bobeth:2001sq}
C.~Bobeth, T.~Ewerth, F.~Kruger, and J.~Urban, {\it {Analysis of neutral Higgs
  boson contributions to the decays $\bar{B}$( $s^{)} \to \ell^{+} \ell^{-}$
  and $\bar{B} \to K \ell^{+} \ell^{-}$}},  {\em Phys.Rev.} {\bf D64} (2001)
  074014, [\href{http://xxx.lanl.gov/abs/hep-ph/0104284}{{\tt
  hep-ph/0104284}}].

\bibitem{Akerib:2013tjd}
{\bf LUX Collaboration} Collaboration, D.~Akerib et~al., {\it {First results
  from the LUX dark matter experiment at the Sanford Underground Research
  Facility}},  {\em Phys.Rev.Lett.} {\bf 112} (2014) 091303,
  [\href{http://xxx.lanl.gov/abs/1310.8214}{{\tt arXiv:1310.8214}}].

\bibitem{Aprile:2012zx}
{\bf XENON1T collaboration} Collaboration, E.~Aprile, {\it {The XENON1T Dark
  Matter Search Experiment}},  {\em Springer Proc.Phys.} {\bf C12-02-22} (2013)
  93--96, [\href{http://xxx.lanl.gov/abs/1206.6288}{{\tt arXiv:1206.6288}}].

\bibitem{Blanke:2013uia}
M.~Blanke, G.~F. Giudice, P.~Paradisi, G.~Perez, and J.~Zupan, {\it {Flavoured
  Naturalness}},  {\em JHEP} {\bf 1306} (2013) 022,
  [\href{http://xxx.lanl.gov/abs/1302.7232}{{\tt arXiv:1302.7232}}].

\bibitem{Coleman:1973jx}
S.~R. Coleman and E.~J. Weinberg, {\it {Radiative Corrections as the Origin of
  Spontaneous Symmetry Breaking}},  {\em Phys.Rev.} {\bf D7} (1973) 1888--1910.

\end{thebibliography}\endgroup

\end{document}